\documentclass[12pt]{article}
\usepackage{amsmath,amssymb,bm,epsfig,euscript,psfrag}
\usepackage[square,comma,sort&compress]{natbib}

\def\nslash{\rlap{\hspace{0.02cm}/}{n}}
\def\nbslash{\rlap{\hspace{0.02cm}/}{\bar n}}
\def\delslash{\rlap{\hspace{0.02cm}/}{\partial}}
\def\Dslash{\rlap{\hspace{0.07cm}/}{D}}
\def\Aslash{\rlap{\hspace{0.08cm}/}{A}}
\def\calAslash{\rlap{\hspace{0.08cm}/}{{\EuScript A}}}
\def\vslash{\rlap{\hspace{0.02cm}/}{v}}
\def\Ahc{{\cal X}}
\def\hc{\xi_{hc}}
\def\ahc{\xi_{\overline{hc}}}
\def\hcbar{\bar\xi_{hc}}
\def\ahcbar{\bar\xi_{\overline{hc}}}
\def\Ahc{{\EuScript A}_{hc}}
\def\Aahc{{\EuScript A}_{\overline{hc}}}
\def\bsg{$\bar B\to X_s\gamma$}
\def\bulv{$\bar B\to X_u l\,\bar\nu$}

\begin{document}

\begin{titlepage}

\begin{flushright}
MZ-TH/09-20\\
HD-THEP-10-6\\
EFI 10-7\\
March 25, 2010
\end{flushright}

\vspace{0.1cm}
\begin{center}
\Large\bf\boldmath
Factorization at Subleading Power and Irreducible 
Uncertainties in $\bar B\to X_s\gamma$ Decay
\unboldmath
\end{center}

\vspace{0.2cm}
\begin{center}
{\sc Michael Benzke$^a$, Seung J.~Lee$^b$, Matthias Neubert$^{a,c}$, and Gil Paz$^d$}\\
\vspace{0.4cm}
{\sl 
${}^a$Institut f\"ur Physik (THEP), Johannes Gutenberg-Universit\"at\\
D-55099 Mainz, Germany\\[0.3cm]
${}^b$Department of Particle Physics, Weizmann Institute of Science\\ 
Rehovot 76100, Israel\\[0.3cm]
${}^c$Institut f\"ur Theoretische Physik, Ruprecht-Karls-Universit\"at Heidelberg\\
Philosophenweg 16, D-69120 Heidelberg, Germany\\[0.3cm]
${}^d$Enrico Fermi Institute, University of Chicago\\ 
Chicago, IL 60637, U.S.A.}
\end{center}

\vspace{0.2cm}
\begin{abstract}\noindent
Using methods from soft-collinear and heavy-quark effective theory, a systematic factorization analysis is performed for the \bsg\ photon spectrum in the endpoint region $m_b-2E_\gamma={\cal O}(\Lambda_{\rm QCD})$. It is proposed that, to all orders in $1/m_b$, the spectrum obeys a novel factorization formula, which besides terms with the structure $H\,J\otimes S$ familiar from inclusive \bulv\ decay distributions contains ``resolved photon" contributions of the form $H\,J\otimes S\otimes\bar J$ and $H\,J\otimes S\otimes\bar J\otimes\bar J$. Here $S$ and $\bar J$ are new soft and jet functions, whose form is derived. These contributions arise whenever the photon couples to light partons instead of coupling directly to the effective weak interaction. The new contributions appear first at order $1/m_b$ and are related to operators other than $Q_{7\gamma}$ in the effective weak Hamiltonian. They give rise to non-vanishing $1/m_b$ corrections to the total decay rate, which cannot be described using a local operator product expansion. A systematic analysis of these effects is performed at tree level in hard and hard-collinear interactions. The resulting uncertainty on the decay rate defined with a cut $E_\gamma>1.6$\,GeV is estimated to be approximately $\pm 5\%$. It could be reduced by an improved measurement of the isospin asymmetry $\Delta_{0-}$ to the level of $\pm 4\%$. We see no possibility to reduce this uncertainty further using reliable theoretical methods.
\end{abstract}
\vfil

\end{titlepage}

\section{Introduction and outline}

The radiative decay \bsg\ plays an important role in testing the Standard Model and constraining its possible extensions at or beyond the TeV scale. Comparing the predictions for the branching ratio of this decay obtained in extensions of the Standard Model with experiment provides powerful constraints on the parameter space of many new-physics models (see e.g.\ \cite{Carena:2007aq,Ellis:2007fu,Domingo:2007dx} for analyses in the context of the MSSM, and \cite{Haisch:2008ar} for an overview of several other models). The calculation of the \bsg\ branching ratio in the Standard Model has been pushed to the next-to-next-to-leading order in renormalization-group improved perturbation theory \cite{Misiak:2006zs}, leading to the prediction ${\rm Br} (\bar B\to X_s\gamma)=(3.15\pm 0.23)\cdot 10^{-4}$ for a cut $E_\gamma>1.6$\,GeV on the photon energy measured in the $B$-meson rest frame. A dedicated analysis of cut-related effects and uncertainties gives the slightly lower value ${\rm Br}(\bar B\to X_s\gamma)=(2.98\pm 0.26)\cdot 10^{-4}$ \cite{Becher:2006pu}. These theoretical estimates are in good agreement with the current experimental world average ${\rm Br}(\bar B\to X_s\gamma)=(3.52\pm 0.23\pm 0.09)\cdot 10^{-4}$ \cite{Barberio:2008fa,Antonelli:2009ws}.

The shape of the photon energy spectrum in \bsg\ decay is sensitive to non-perturbative hadronic physics. At lowest order in the heavy-quark expansion, it is related to a universal shape function describing the momentum distribution of the $b$ quark inside the $B$ meson \cite{Neubert:1993ch,Neubert:1993um,Bigi:1993ex,Kagan:1998ym}. The same shape function parameterizes the leading bound-state effects in the inclusive semileptonic decay \bulv. As a result, a precise measurement of the photon spectrum can be used to derive useful hadronic input for the analysis of \bulv\ decay spectra, and in this way enable a precise determination of $|V_{ub}|$ \cite{Lange:2005yw,Gambino:2007rp}. One goal of the present paper is to complete the analysis of non-perturbative effects on the \bsg\ photon spectrum at subleading order in the heavy-quark expansion. This will allow us to estimate the irreducible theoretical uncertainties in the calculation of the \bsg\ branching ratio computed with a cut on photon energy, and it will also have implications for the extraction of $|V_{ub}|$. In the process, we will discuss that certain terms in the standard formulae for the \bsg\ decay rate and photon spectrum result from an incorrect matching procedure and thus carry unphysical sensitivity to long-distance physics. A second goal of this paper is to properly factorize the short- and long-distance contributions into perturbatively calculable functions and non-perturbative matrix elements, using methods of effective field theory. 

The \bsg\ decay rate and photon spectrum can be calculated using the optical theorem, which relates them to a restricted discontinuity of the forward $B$-meson matrix element of the product of two effective weak Hamiltonians,
\begin{equation}\label{THH}
   d\Gamma(\bar B\to X_s\gamma)\propto \mbox{Disc}_{\rm \,restr.}\,
   \Big[ i\int d^4 x\,\langle\bar B| {\cal H}_{\rm eff}^\dagger(x)\,
   {\cal H}_{\rm eff}(0) |\bar B\rangle \Big] \,.
\end{equation}
The discontinuity is restricted by the requirement that the cut propagators must include that of the photon and a strange quark. The effective weak Hamiltonian ${\cal H}_{\rm eff}$ consists of a sum of local operators, whose definitions are collected in Appendix~\ref{app:Heff}. The most important ones are the electromagnetic and chromomagnetic dipole operators $Q_{7\gamma}$ and $Q_{8g}$ as well as the current-current operator $Q_1^c$. At the lowest order in $\alpha_s$ and $1/m_b$, only the dipole operator $Q_{7\gamma}$ contributes to the decay rate.

Note the important fact that, unlike for semileptonic inclusive $B$-meson decays, the \bsg\ decay rate cannot be written as the discontinuity of a forward matrix elements of time-ordered products of fields. The reason is that not all cuts of the relevant Feynman graphs correspond to the \bsg\ process. For example, diagrams with penguin contractions of the four-quark operator $Q_1^c$ contain cuts corresponding to the decay $b\to c\bar c s$ without a photon in the final state, which clearly do not contribute to the decay rate in (\ref{THH}). As a result, the fields belonging to the $\bar B\to X_s\gamma$ amplitude are time-ordered, while those belonging to the complex conjugate amplitude are anti-time-ordered. A path-integral method for the evaluation of the cut diagrams contributing to expressions such as (\ref{THH}) is the Keldysh (or time-loop) formalism \cite{Schwinger:1960qe,Keldysh:1964ud}. Here we will not expose the technical details of this approach (see \cite{Becher:2007ty} for a concise recent discussion), but we will mention at the appropriate places in our discussion where the anti-time-ordering of fields is important.

Theoretical calculations of the forward scattering amplitude utilize the fact that $\Lambda_{\rm QCD}\ll m_b$ to express the decay rate and the photon spectrum as a series of operator matrix elements suppressed by powers of $1/m_b$ \cite{Bigi:1993fe,Manohar:1993qn,Falk:1993dh}. The photon spectrum has been measured accurately for energies $E_\gamma>2$\,GeV, and some less accurate data is available in the range between 1.7 and 2\,GeV \cite{Chen:2001fja,Koppenburg:2004fz,Aubert:2005cua,Aubert:2006gg,Abe:2008sxa}. The partially inclusive rates obtained experimentally are defined as integrals over the endpoint region $E_0<E_\gamma<M_B/2$. The shape of the photon spectrum in the region above 2\,GeV is most useful for extracting information that can be used to determine $|V_{ub}|$ from \bulv\ decay distributions \cite{Lange:2005yw,Gambino:2007rp}. Note that in the relevant region of phase space the variable $\Delta=m_b-2E_0$ is a hadronic scale of order $\Lambda_{\rm QCD}$, which is much smaller than the hard scale $m_b$ of the process. In this ``endpoint region", the hadronic final state $X_s$ has large energy $E_X\sim m_b$ but small invariant mass $M_X\sim\sqrt{m_b\Delta}\sim\sqrt{m_b\Lambda_{\rm QCD}}$. This follows from the fact that $M_X^2=M_B(M_B-2E_\gamma)$, which implies that the photon energy spectrum contains the same information as the hadronic invariant mass distribution in \bsg\ decay. In this case the appropriate theoretical description of hadronic effects involves an expansion of the forward scattering amplitude in non-local operator matrix elements called shape functions \cite{Neubert:1993ch,Neubert:1993um}. If in the future it will be possible to lower the photon cut to a value such that $m_b\gg\Delta\gg\Lambda_{\rm QCD}$ (this will require $E_0<1.6$\,GeV or so), then many (but not all) of the non-local matrix elements can be expanded in matrix elements of local operators using a multi-scale operator product expansion, consisting of a double expansion in powers of $\Lambda_{\rm QCD}/m_b$ and $\Lambda_{\rm QCD}/\Delta$ \cite{Neubert:2004dd}. However, even in the hypothetical limit $E_0\to 0$ some non-local effects remain, which cannot be described using a local heavy-quark expansion in power of $\Lambda_{\rm QCD}/m_b$.

In this paper we perform a comprehensive study of the factorization properties of the \bsg\ photon spectrum in the endpoint region, which is more systematic than previous analyses. Using methods of effective field theory, we propose a novel factorization formula valid at any order in the $1/m_b$ expansion, which is a generalization of the familiar soft-collinear factorization formula \cite{Korchemsky:1994jb,Bauer:2001yt,Bosch:2004th,Lee:2004ja,Bosch:2004cb,Beneke:2004in}
\begin{equation}\label{simplefact}
   d\Gamma(\bar B\to X_u l\,\bar\nu)
   = \sum_{n=0}^\infty\,\frac{1}{m_b^n}\,
   \sum_i\,H_i^{(n)} J_i^{(n)}\otimes S_i^{(n)}
\end{equation}
for the differential distributions in the inclusive semileptonic decay \bulv. Here $H_i^{(n)}$ are hard functions parameterizing physics at the scale $m_b$, $J_i^{(n)}$ are jet functions describing the physics of the hadronic final state $X_u$ with invariant mass $M_X\sim\sqrt{m_b\Lambda_{\rm QCD}}$, and $S_i^{(n)}$ are soft functions incorporating hadronic physics associated with the scale $\Lambda_{\rm QCD}$. The soft or shape functions are defined in terms of forward matrix elements of non-local HQET operators on the light cone. The symbol $\otimes$ implies a convolution, which arises when the soft and jet functions share some common variables.

\begin{figure}
\begin{center}
\epsfig{file=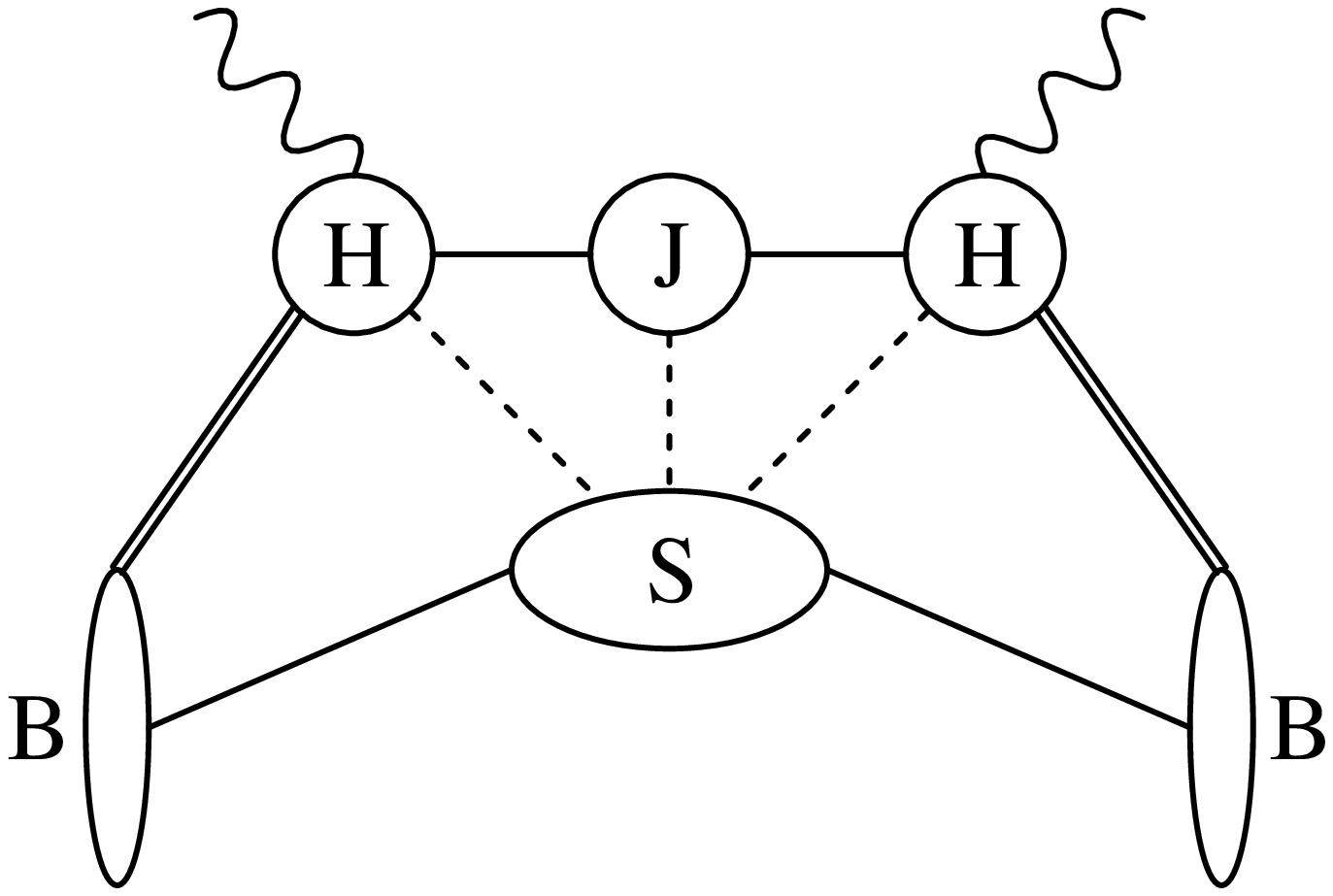,width=4.5cm}\qquad
\epsfig{file=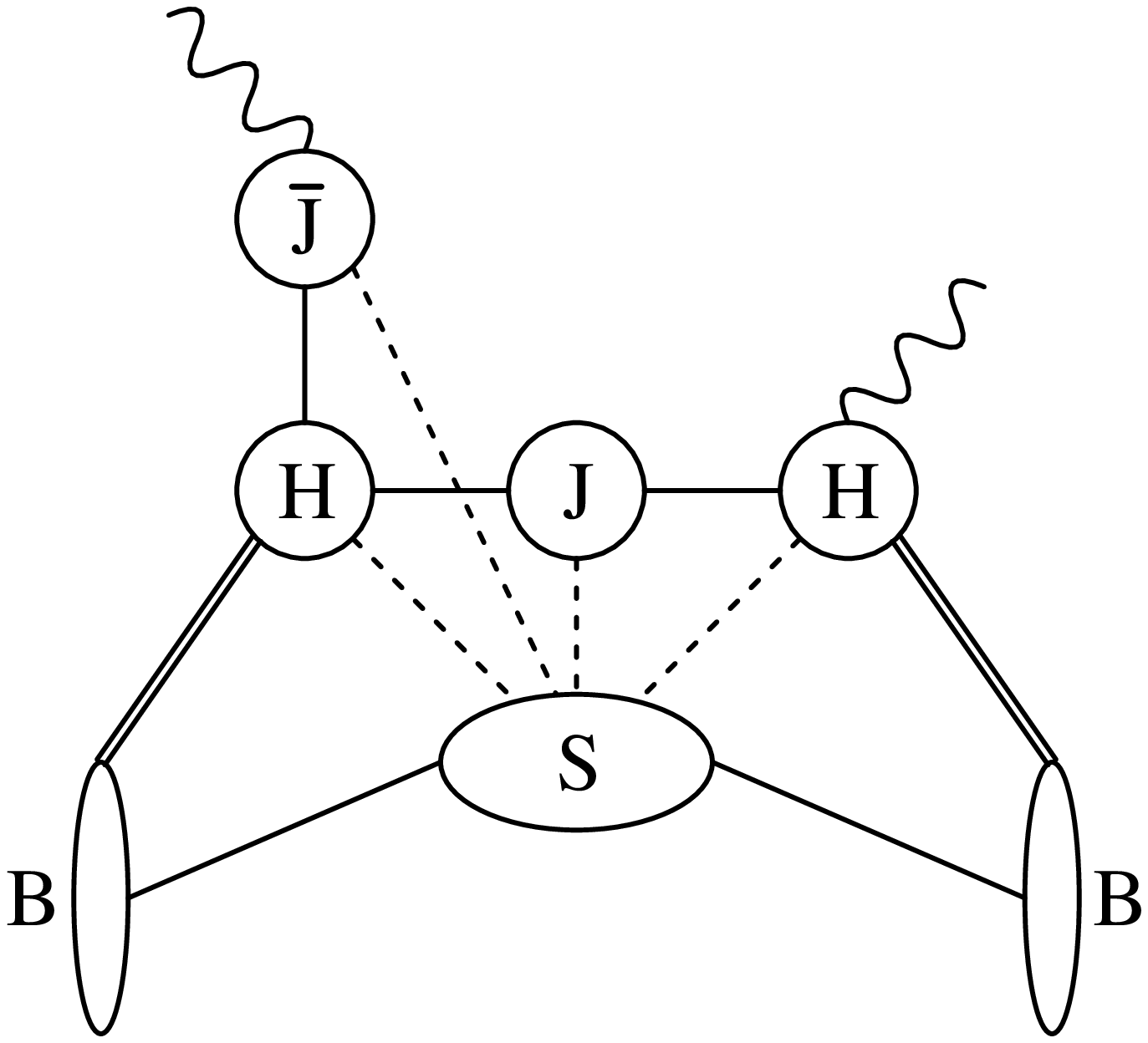,width=4.5cm}\qquad
\epsfig{file=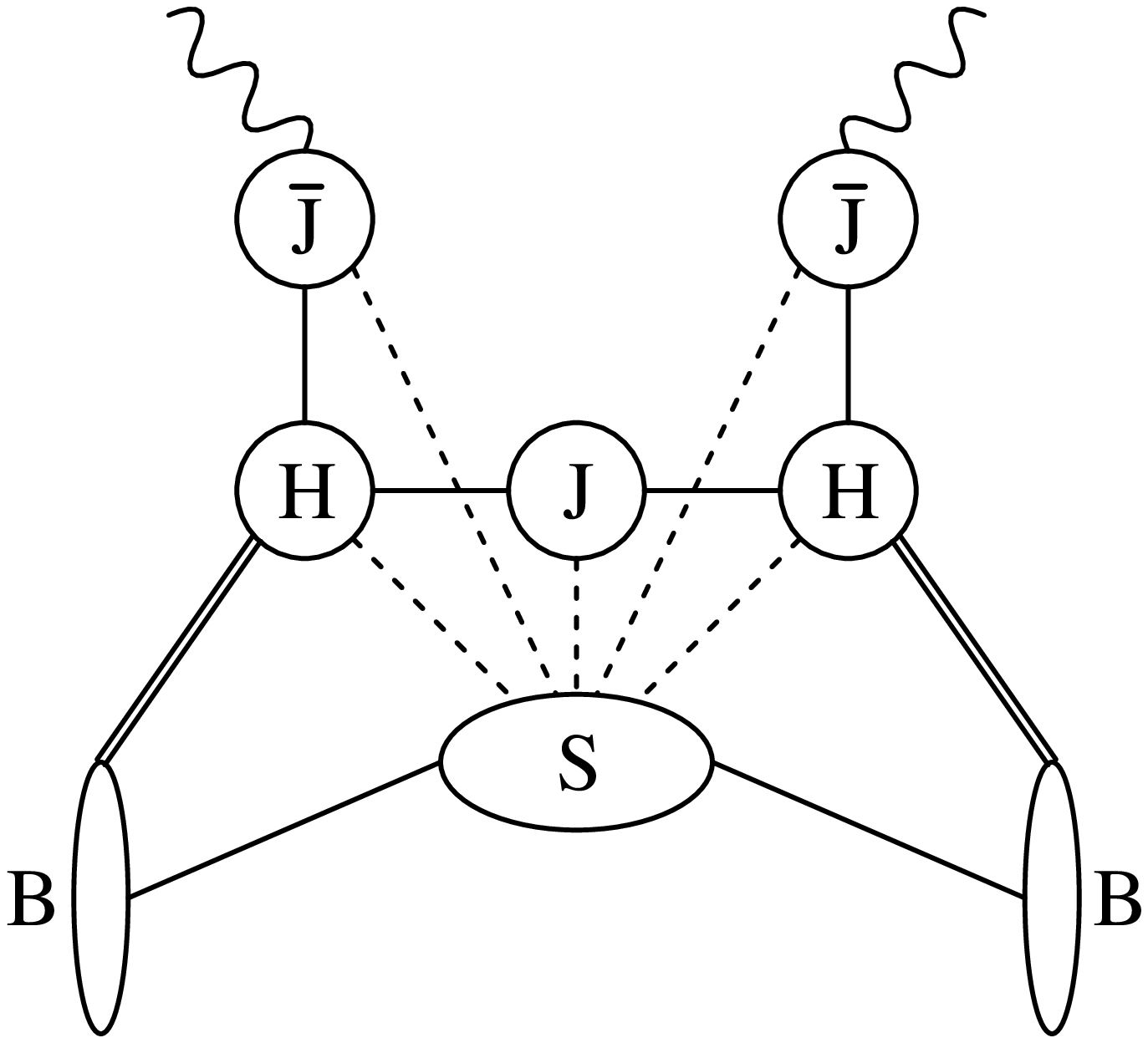,width=4.5cm}
\parbox{15.5cm}{\caption{\label{fig:theorem}
Graphical illustration of the three terms in the QCD factorization theorem (\ref{fact2}) for $\bar B\to X_s\gamma$ decay in the endpoint region. The dashed lines represent soft interactions, which must be power expanded and factored off the remaining building blocks to derive factorization.}}
\end{center}
\end{figure}

The new element, which makes the analysis of \bsg\ decay more involved than that of semileptonic decays, is the presence of ``resolved photon'' contributions, which contain subprocesses in which the photon couples to light partons instead of connecting directly to the effective weak-interaction vertex \cite{Kapustin:1995fk,Voloshin:1996gw,Ligeti:1997tc,Grant:1997ec,Buchalla:1997ky,Lee:2006wn}. As we will show, these subprocesses probe the hadronic substructure of the photon at a scale of order $\sqrt{2E_\gamma\Lambda_{\rm QCD}}$. The corresponding effects can be described by introducing new jet functions $\bar J_i^{(n)}$. There is no analog of this phenomenon in semileptonic decays, because a lepton-neutrino pair can only couple to light partons via $W$-boson exchange. The factorization formula we obtain for the photon spectrum in the endpoint region is
\begin{eqnarray}\label{fact2}
   d\Gamma(\bar B\to X_s\gamma)
   &=& \sum_{n=0}^\infty\,\frac{1}{m_b^n}\,
    \sum_i\,H_i^{(n)} J_i^{(n)}\otimes S_i^{(n)} \\
   &+& \sum_{n=1}^\infty\,\frac{1}{m_b^n}\,\bigg[
    \sum_i\,H_i^{(n)} J_i^{(n)}\otimes S_i^{(n)}\otimes\bar J_i^{(n)}
    + \sum_i\,H_i^{(n)} J_i^{(n)}\otimes S_i^{(n)}
    \otimes\bar J_i^{(n)}\otimes\bar J_i^{(n)} \bigg] . \nonumber
\end{eqnarray}
It contains ``direct photon'' contributions of the same form as (\ref{simplefact}), accompanied by single and double resolved photon contributions that are new. Our notation is symbolic; objects denoted by the same symbol in the various terms refer, in general, to different quantities. Note the important fact that the new contributions appear first at order $1/m_b$ in the heavy-quark expansion. While the jet functions $J_i^{(n)}$ are cut propagator functions dressed by Wilson lines, the jet functions $\bar J_i^{(n)}$ are given in terms of full propagator functions dressed by Wilson lines. A graphical illustration of the factorization formula is shown in Figure~\ref{fig:theorem}.

When the photon spectrum is integrated over an interval much larger than the endpoint region, the direct photon contributions simplify to a series of hard coefficients multiplying forward $B$-meson matrix elements of local operators, in analogy to what happens in semileptonic \bulv\ decay \cite{Bigi:1993fe,Manohar:1993qn,Falk:1993dh}. In particular, it follows that the corrections of first-order in $\Lambda_{\rm QCD}/m_b$ integrate to zero, since there does not exist a local, gauge-invariant operator that could account for such terms. An important result of our analysis is that the resolved photon contributions do not reduce to matrix elements of local operators in that case. Their effects on the total decay rate must still be described in terms of non-local operator matrix elements, as illustrated with a specific example in \cite{Lee:2006wn}.\footnote{The total \bsg\ decay rate is not an infra-red (IR) safe observable. What is usually meant by this term is the rate defined with a very low cut on photon energy, and with a subtraction of duality-violating charmonium resonance contributions \cite{Beneke:2009az}.} 

Resolved photon contributions can only arise from operators in the effective weak Hamiltonian that do not contain the photon field as part of the effective, local weak interactions. The most important such operators are the chromomagnetic dipole operator $Q_{8g}$ and the current-current operator $Q_1^c$. Power-suppressed contributions from other operators can be safely neglected for phenomenological purposes. It follows that double resolved photon contributions can only arise from the operator pairs $Q_{8g}-Q_{8g}$, $Q_1^c-Q_1^c$, and $Q_1^c-Q_{8g}$, while single resolved photon contributions can also arise from the pairs $Q_{8g}-Q_{7\gamma}$ and $Q_1^c-Q_{7\gamma}$. Direct photon contributions can arise from all operator pairings. Some effects involving the conversion of the photon into light partons have been discussed previously in the literature \cite{Kapustin:1995fk,Voloshin:1996gw,Ligeti:1997tc,Grant:1997ec,Buchalla:1997ky,Lee:2006wn}, and it is known (though not widely appreciated) that they fall outside the realm of the local operator product expansion. Let us comment on the various effects one by one:
\begin{itemize}
\item
The perturbative analysis of the $Q_{8g}-Q_{8g}$ contribution gives rise to IR-singular contributions, which at one-loop order can be regularized by introducing a non-zero mass for the strange quark \cite{Ali:1995bi}. It was argued in \cite{Kapustin:1995fk} that these singularities can be absorbed into the photon fragmentation functions of the strange quark and the gluon. We find that this factorization no longer holds in the endpoint region. Instead, the IR singularities must be factored into a subleading four-quark shape function.
\item
The current-current operator $Q_1^c$ can induce penguin-type transitions, in which two charm or up quarks convert into a photon and soft gluon. Previous studies of this effect have focused on its contribution to the total decay rate, which arises from its interference with the matrix element of $Q_{7\gamma}$ \cite{Voloshin:1996gw,Ligeti:1997tc,Grant:1997ec,Buchalla:1997ky}. In the present work we will generalize this analysis to the case of the photon spectrum in the endpoint region.
\item
The square of the charm-penguin amplitude, the $Q_1^c-Q_1^c$ double resolved photon contribution, has not yet been analyzed in the literature, but it is sometimes mentioned as a potentially large source of power corrections due to the fact that the operator $Q_1^c$ has by far the largest Wilson coefficient in the effective weak Hamiltonian. We will show that this contribution arises first at order $1/m_b^2$ in the heavy-quark expansion. Its effects on the decay rate and spectrum are therefore strongly suppressed. The same is true for the $Q_1^c-Q_{8g}$ double resolved photon contribution.
\item
Resolved photon contributions from the $Q_{8g}-Q_{7\gamma}$ interference term were first studied in \cite{Lee:2006wn}, again with regard to their impact on the total decay rate. We will complete this study and generalize it to the case of the photon spectrum.
\end{itemize}

We begin our analysis with a review of known results for the \bsg\ photon spectrum in Section~\ref{sec:review}, indicating a couple of problematic features in the formulae routinely used in the literature. The new factorization formula (\ref{fact2}) will be derived in Section~\ref{sec:fact} using a two-step matching procedure from QCD to soft-collinear effective theory (SCET) \cite{Bauer:2001yt,Bauer:2000yr,Beneke:2002ph} and heavy-quark effective theory (HQET) \cite{Neubert:1993mb}. In Section~\ref{sec:matching} we discuss the factorization properties of the various contributions to the \bsg\ photon spectrum, which arise from different pairs of operators in the effective weak Hamiltonian. This includes, in particular, a detailed discussion of the new subleading shape functions for the contributions from operator pairs other than $Q_{7\gamma}-Q_{7\gamma}$. These have not been considered previously in the literature, except for a particular subleading shape-function contribution to the total \bsg\ decay rate arising from the operator pair $Q_{7\gamma}-Q_{8g}$ \cite{Lee:2006wn}. In Section~\ref{sec:PT} we use the invariance of the strong interaction under the discrete symmetry $PT$ to prove that the subleading soft functions are real, i.e., they do not carry non-trivial strong phases. The implications of our findings for the integrated \bsg\ decay rate are studied in Section~\ref{sec:Gtot}. We show that the resolved photon contributions must still be described in terms of matrix elements of non-local operators, whose effects cannot be reduced by lowering the cutoff on the photon energy. Finally, in Section~\ref{sec:conc} we study the phenomenological implications of our results by estimating the irreducible theoretical uncertainty in the prediction for the \bsg\ decay rate integrated over the range $E_\gamma>1.6$\,GeV. We then summarize our results and give some conclusions. Three Appendices contain our definitions of the operators in the effective weak Hamiltonian, a summary of input parameters, and a detailed exposition of the matching of the effective weak Hamiltonian onto operators in SCET. Readers not interested in the technical details of our derivations should consult Section~\ref{sec:review} and then proceed with Sections~\ref{sec:Gtot} and \ref{sec:conc}.

Even though we only sketch the derivation of the new factorization formula in Section~\ref{sec:fact}, we consider this discussion as solid as that for many other processes discussed in the context of SCET. Still, we do not claim to have a rigorous proof of factorization. Indeed, in our analysis we will encounter one particular contribution to the \bsg\ photon spectrum and decay rate, for which the resulting convolution integrals derived using SCET suffer from an ultra-violet (UV) divergence. In its current formulation, the effective theory does not provide a systematic framework for regularizing this divergence. The problem of divergent convolution integrals in SCET has been encountered previously in the context of heavy-to-light form factors \cite{Beneke:2000ry,Bauer:2002aj,Beneke:2003pa,Lange:2003pk} and power-suppressed contributions to hadronic $B$-meson decays \cite{Beneke:2001ev,Arnesen:2006vb}. It is to some extent still an open question whether these integrals indicate a failure of factorization, or whether they can be cured by a generalization of the theoretical framework of SCET (an attempt in this direction was initiated in \cite{Manohar:2006nz}). An important difference is that in all previous cases these divergences were of IR origin. In our case, the convolution integrals diverge in the UV. Such divergences appear to be rather generic in the description of higher-order power corrections, because the resulting convolution integrals contain higher powers of soft momentum variables. The physical origin of the divergence and its interpretation are entirely transparent. Still, the presence of this effect is problematic for the consistency of SCET as a {\em bona fide\/} effective field theory and calls for a cure. We will discuss a simple treatment of the divergence using a hard cutoff on the convolution integrals. We do not claim, however, to have a systematic procedure that would work at higher orders in perturbation theory and allow for a consistent resummation of large logarithms. In that sense our derivation of factorization is incomplete.

\section{Review of known results and preview of new ones}
\label{sec:review}

Let us briefly summarize what is known in the literature about the various terms in the factorization formula (\ref{fact2}). Separating the contributions from different operators in the effective weak Hamiltonian, we write the heavy-quark expansion of the CP-averaged $\bar B\to X_s\gamma$ photon-energy spectrum in the endpoint region $p_+\equiv m_b-2E_\gamma={\cal O}(\Lambda_{\rm QCD})$ in the form
\begin{equation}\label{HQexp}
\begin{aligned}
   \frac{d\Gamma}{dE_\gamma} 
   &= \frac{G_F^2\alpha|V_{tb} V_{ts}^*|^2}{2\pi^4}\,
    \overline{m}_b^2(\mu)\,E_\gamma^3\,\Bigg[\,
    |H_\gamma(\mu)|^2 \int_{-p_+}^{\bar\Lambda}\!d\omega\,m_b\,
    J\big( m_b(\omega+p_+),\mu \big)\,S(\omega,\mu) \\
   &\hspace{4.9cm}\mbox{}+ \frac{1}{m_b}\,\sum_{i\le j}\,
    \mbox{Re}\big[C_i^*(\mu)\,C_j(\mu)\big]\,F_{ij}(E_\gamma,\mu)
    + \dots \Bigg] \,,
\end{aligned}
\end{equation}
where $\bar\Lambda=M_B-m_b$, and the ellipses represent terms of order $1/m_b^2$ and higher. For convenience we have factored out two powers of the running $b$-quark mass (defined in the $\overline{\rm MS}$ scheme) and three powers of the photon energy, as this is the correct energy dependence of the leading contribution to the spectrum.

The term in the first line of (\ref{HQexp}) is the leading-power contribution and is well understood theoretically. At this order the effective weak Hamiltonian for \bsg\ decay matches onto a unique leading-order current operator in SCET. The hard matching coefficient $H_\gamma(\mu)=C_{7\gamma}(\mu)+{\cal O}(\alpha_s)$ for this current receives contributions from all operators in the effective weak Hamiltonian, not just $Q_{7\gamma}$, as soon as one goes beyond the leading order in perturbation theory \cite{Neubert:2004dd}. The contribution proportional to $C_{7\gamma}$ is known to order $\alpha_s^2$ \cite{Melnikov:2005bx,Asatrian:2006sm}, while the remaining terms are known to order $\alpha_s$. When the effective current operator is further matched onto HQET, a single jet function $J(p^2,\mu)=\delta(p^2)+{\cal O}(\alpha_s)$ arises, which is given by the discontinuity of the quark propagator in light-cone gauge and has been calculated to two-loop order in \cite{Becher:2006qw}. The remaining HQET matrix element defines a single, leading-order shape function via \cite{Neubert:1993um}
\begin{equation}\label{Sdef}
   S(\omega,\mu)
   = \int\frac{dt}{2\pi}\,e^{-i\omega t}\,
   \frac{\langle\bar B(v)| \bar h(tn) S_n(tn) S_n^\dagger(0) h(0)
   |\bar B(v)\rangle }{2M_B} \,.
\end{equation}
Here $v$ denotes the four-velocity of the $B$ meson, and $n$ is a light-like vector pointing along the direction of the final-state hadronic jet. We normalize these vectors such that $v^2=1$, $n^2=0$, $v\cdot n=1$, and $v^0\ge 1$. The soft Wilson line $S_n$ is defined as 
\begin{equation}\label{eq:Sn}
   S_n(x) = {\bf P}\exp\Bigg( ig\int\limits_{-\infty}^0\!du\,
   n\cdot A_s(x+un) \Bigg) \,,
\end{equation}
where the path-ordering symbol {\bf P} means that fields with larger $u$ values stand to the left of those with smaller ones. The conjugate Wilson line $S_n^\dagger$ has the opposite ordering prescription. These definitions imply that $S_n(tn) S_n^\dagger(0)=[tn,0]$ is a straight line segment connecting the points $tn$ and 0, with gauge fields closer to the point $tn$ standing to the left of those closer to 0. Taking the complex conjugate of relation (\ref{Sdef}) and using translational invariance, one finds that the shape function is real. The functions $H_\gamma$, $J$, and $S$ incorporate contributions associated with different scales in the problem. The hard function $H_\gamma$ receives virtual corrections of order the hard scale $\mu_h\sim m_b$, while the shape function $S$ encodes non-perturbative hadronic physics associated with the soft scale $\mu_s\sim p_+\sim\Lambda_{\rm QCD}$. The jet function describes the properties of the final-state hadronic jet, whose invariant mass scales like $\mu_{hc}\sim\sqrt{m_b\,\Lambda_{\rm QCD}}$ in the endpoint region. This intermediate scale is the scale of (anti-)hard-collinear virtualities. Large logarithms arising from ratios of these various scales can be resummed to all orders in perturbation theory by solving renormalization-group equations in the effective theory \cite{Neubert:2004dd,Korchemsky:1994jb,Bauer:2001yt}. 

Beyond the leading power in the heavy-quark expansion, the proper factorization of the various contributions to the decay rate has not yet been discussed systematically in the literature. Such an analysis is the main goal of the present work. In phenomenological discussions of \bsg\ decay one usually starts from expressions for the power-suppressed terms derived in the naive parton model, i.e., by computing the inclusive decay of an on-shell $b$ quark \cite{Ali:1995bi,Chetyrkin:1996vx,Kagan:1998ym}. Including only the phenomenologically relevant contributions from operator products of $Q_1^c$, $Q_{7\gamma}$, and $Q_{8g}$, and setting $V_{ub}=0$ for simplicity, this yields for the first-order power corrections
\begin{equation}\label{Fijpart}
\begin{aligned}
   F_{77}^{\rm part}(E_\gamma,\mu) 
   &= \frac{C_F\alpha_s(\mu)}{4\pi}
    \left( 16\ln\frac{m_b}{p_+} - 15 \right) , \\
   F_{88}^{\rm part}(E_\gamma,\mu) 
   &= \frac{C_F\alpha_s(\mu)}{4\pi}
    \left( \frac29\,\ln\frac{m_b\,p_+}{m_s^2} - \frac13 \right) , \\
   F_{78}^{\rm part}(E_\gamma,\mu) 
   &= \frac{C_F\alpha_s(\mu)}{4\pi}\,\frac{10}{3} \,, \\
   F_{11}^{\rm part}(E_\gamma,\mu) 
   &= F_{18}^{\rm part}(E_\gamma,\mu) 
    = \frac{C_F\alpha_s(\mu)}{4\pi}\,\frac{2}{9} \,, \\
   F_{17}^{\rm part}(E_\gamma,\mu) 
   &= \frac{C_F\alpha_s(\mu)}{4\pi} \left( - \frac{2}{3} \right)
    - \frac{m_b\lambda_2}{9m_c^2}\,\delta(p_+) \,.
\end{aligned}
\end{equation}
In this paper we adopt the scaling $m_c^2={\cal O}(m_b\Lambda_{\rm QCD})$ for the charm-quark mass, meaning that the ratio $m_c^2/m_b$ remains a constant of order $\Lambda_{\rm QCD}$ in the heavy-quark limit. Note that the expression for $F_{17}^{\rm part}$ includes a non-perturbative effect proportional to the HQET parameter $\lambda_2=(M_{B^*}^2-M_B^2)/4\approx 0.12$\,GeV$^2$ \cite{Voloshin:1996gw,Ligeti:1997tc,Grant:1997ec,Buchalla:1997ky}, which is of the same order in power counting as the perturbative contribution. It is related to charm-penguin diagrams with a soft gluon emission. If we were to adopt the alternative counting scheme where $m_c={\cal O}(m_b)$ in the heavy-quark limit, then some of the expressions in (\ref{Fijpart}) would change. In that case
\begin{equation}\label{Fijpart1}
\begin{aligned}
   F_{11}^{\rm part}(E_\gamma,\mu) 
   &= \frac{C_F\alpha_s(\mu)}{4\pi}\,\frac{4}{9} \int_0^1\!dx\,
    (1-x) \left| 1 - F\Big(\frac{z}{x}\Big) \right|^2 , \\
   F_{17}^{\rm part}(E_\gamma,\mu) 
   &= -3 F_{18}^{\rm part}(E_\gamma,\mu)
   = \frac{C_F\alpha_s(\mu)}{4\pi} \left( - \frac{4}{3} \right)
    \int_0^1\!dx\,x\,\mbox{Re}\left[ 1 - F\Big(\frac{z}{x}\Big)
    \right] ,
\end{aligned}
\end{equation}
where $z=(m_c/m_b)^2$, and we have defined the penguin function 
\begin{equation}\label{eq:F(r)}
   F(x) = 4x\arctan^2\left( \frac{1}{\sqrt{4x-1}} \right) .
\end{equation}
The non-perturbative contribution to $F_{17}^{\rm part}$ would be power-suppressed in that case and should be dropped for consistency.

In order to account for non-perturbative effects other than those described by the $\lambda_2$ term, the simplest recipe used in the literature is to replace $p_+\to\omega+p_+$ in the above expressions and convolute them with the leading-order shape function $S(\omega,\mu)$, e.g.\
\begin{equation}\label{F77part1}
\begin{aligned}
   F_{77}(E_\gamma,\mu) 
   &= \frac{C_F\alpha_s(\mu)}{4\pi} 
    \int_{-p_+}^{\bar\Lambda}\!d\omega
    \left( 16\ln\frac{m_b}{\omega+p_+} - 15 \right) S(\omega,\mu) 
    \,, \\
   F_{78}(E_\gamma,\mu) 
   &= \frac{C_F\alpha_s(\mu)}{4\pi}\,\frac{10}{3} 
    \int_{-p_+}^{\bar\Lambda}\!d\omega\,S(\omega,\mu) \,,
\end{aligned}
\end{equation}
and similarly for the other terms \cite{Kagan:1998ym,Lange:2005yw}.

Beyond the leading order in $1/m_b$, the photon spectrum also receives contributions involving more complicated soft functions, usually called subleading shape functions. So far they have been studied only for the direct photon contribution from two insertions of the electromagnetic dipole operator $Q_{7\gamma}$ \cite{Bauer:2001mh,Lee:2004ja,Bosch:2004cb,Beneke:2004in}. In the notation of \cite{Bosch:2004cb}, one obtains at tree level
\begin{equation}\label{F77SSF}
\begin{aligned}
   F_{77}^{\rm SSF}(E_\gamma,\mu) 
   &= p_+\,S(-p_+,\mu) + s(-p_+,\mu) 
    - t(-p_+,\mu) + u(-p_+,\mu) - v(-p_+,\mu) \\
   &\quad\mbox{}- \pi\alpha_s(\mu) \left[ f_u^{(s)}(-p_+,\mu)
    + f_v^{(s)}(-p_+,\mu) \right] 
    + {\cal O}\Big( \frac{\alpha_s(\mu)}{4\pi} \Big) \,.
\end{aligned}
\end{equation}
These same functions also contribute, in other combinations, to the semileptonic \bulv\ decay spectra in the endpoint region. One can think of $t$ and $v$ as non-local generalizations of the $B$-meson matrix element of the subleading HQET chromomagnetic operator, and of $u$ as a non-local generalization of the matrix element of the kinetic operator. The function $s$ arises from an insertion of the subleading HQET Lagrangian into the matrix element for the leading shape function in (\ref{Sdef}). The functions $f_u^{(q)}$ and $f_v^{(q)}$ arise from the matrix elements of non-local four-quark operators. In the present paper we will encounter other four-quark shape functions, which are unique to radiative decays. It is therefore hopeless to try to find weighted distributions of \bsg\ and \bulv\ events, in which the subleading shape functions enter in the same combinations -- a goal pursued in \cite{Lee:2008vs}, working at tree level and neglecting all operators in the effective weak Hamiltonian except $Q_{7\gamma}$. As in all previous analyses of subleading shape-function contributions, it is sufficient for phenomenological purposes to restrict the analysis to the tree level, since so little is known about the functional forms of the subleading shape functions. In the language of the factorization formula (\ref{fact2}), this means that the corresponding hard and jet functions are computed at zeroth order in $\alpha_s/\pi$. We do, however, include jet functions associated with a factor $g^2=4\pi\alpha_s$, which can arise from tree-level hard-collinear gluon exchange. In this work we complete the analysis of subleading shape functions for \bsg\ decay by analyzing the contributions analogous to (\ref{F77SSF}) for the remaining pairs of operators in the effective weak Hamiltonian.

It would be incorrect to simply add the partonic contributions and the contributions from subleading shape functions, such as (\ref{F77part1}) and (\ref{F77SSF}), as this would lead to double counting. In fact, since the partonic expressions (\ref{Fijpart}) and (\ref{Fijpart1}) have not been derived from a proper matching procedure, they secretly contain some soft contributions, which should be subtracted and absorbed into the subleading shape functions. The explicit expressions for $F_{77}^{\rm part}$ and $F_{88}^{\rm part}$ in (\ref{Fijpart}), which contain parametrically large logarithms, already hint at the fact that such a subtraction is required. The dependence of $F_{88}^{\rm part}$ on the strange-quark mass is clearly a sign of an unphysical sensitivity to the IR region, which should not be present in a short-distance coefficient function. For the case of $F_{77}^{\rm part}$ one might think that the large logarithm results from a combination of hard-collinear and hard scales, $\ln(m_b/p_+)=\ln(m_b^2)-\ln(m_b\,p_+)$, in which case it would have a short-distance origin. We will see, however, that it results from a combination of hard-collinear and soft scales, $\ln(m_b/p_+)=\ln(m_b\,p_+)-2\ln(p_+)$. The sensitivity to the soft scale $p_+$ must be subtracted and absorbed into a subleading shape function. 

Our improved expressions for the coefficient functions read
\begin{equation}\label{improved}
\begin{aligned}
   F_{77}(E_\gamma,\mu) 
   &= \frac{C_F\alpha_s(\mu)}{4\pi} 
    \int_{-p_+}^{\bar\Lambda}\!d\omega
    \left( 16\ln\frac{m_b(\omega+p_+)}{\mu^2} + 9 \right) 
    S(\omega,\mu) + F_{77}^{\rm SSF}(E_\gamma,\mu) \,, \\
   F_{88}(E_\gamma,\mu) 
   &= \frac{C_F\alpha_s(\mu)}{4\pi}
    \int_{-p_+}^{\bar\Lambda}\!d\omega
    \left( \frac29\,\ln\frac{m_b(\omega+p_+)}{\mu^2} - \frac13 
    \right) S(\omega,\mu) 
    + 4\pi\alpha_s(\mu)\,f_{88}(-p_+,\mu) \,, \\
   F_{78}(E_\gamma,\mu) 
   &= \frac{C_F\alpha_s(\mu)}{4\pi}\,\frac{10}{3}
    \int_{-p_+}^{\bar\Lambda}\!d\omega\,S(\omega,\mu) 
    + 4\pi\alpha_s(\mu)\,\mbox{Re} \left[
    f_{78}^{\rm (I)}(-p_+,\mu) + f_{78}^{\rm (II)}(-p_+,\mu) 
    \right] , \\
   F_{17}(E_\gamma,\mu) 
   &= \frac{C_F\alpha_s(\mu)}{4\pi} \left( - \frac23 \right)
    \int_{-p_+}^{\bar\Lambda}\!d\omega\,S(\omega,\mu) 
    + \sum_{q=c,u}\,\delta_q\,\mbox{Re}\,f_{17,q}(-p_+,\mu) \,, \\
   F_{11}(E_\gamma,\mu) 
   &= F_{18}(E_\gamma,\mu) 
    = \frac{C_F\alpha_s(\mu)}{4\pi}\,\frac29
    \int_{-p_+}^{\bar\Lambda}\!d\omega\,S(\omega,\mu) \,, 
\end{aligned}
\end{equation}
where we now also include the effects of up-quark loops in $F_{17}$, i.e., we no longer set $V_{ub}=0$. We have defined
\begin{equation}\label{deltaqdef}
   \delta_q = \frac{\mbox{Re} \left[ \lambda_q\,C_1(\mu) 
                    \left(-\lambda_t^*\right) C_{7\gamma}^*(\mu) \right]}%
                   {|\lambda_t|^2\,\mbox{Re}\left[ C_1(\mu)\,C_{7\gamma}^*(\mu) \right]}
    \,, \qquad
   \lambda_q=V_{qb} V_{qs}^* \,,
\end{equation}
where $\delta_c+\delta_u=1$ by unitarity of the CKM matrix. Note that $\delta_u$ is of second order in the Wolfenstein parameter $\lambda\approx 0.22$ and thus numerically very small. The perturbative terms involving convolutions with $S(\omega,\mu)$ are now free of IR-sensitive contributions, and all long-distance physics resides in the shape functions. 

We finish this section with an important remark, which is somewhat orthogonal to the main thrust of this paper but nevertheless relevant. In the analysis of the \bsg\ decay rate and photon spectrum, it is customary to adopt for the charm-quark mass a running mass defined at a hard-collinear scale $\mu_{hc}\sim\sqrt{m_b\Lambda_{\rm QCD}}\sim m_c$ \cite{Gambino:2001ew}. For instance, the default value adopted in \cite{Misiak:2006zs} is $m_c=\overline{m}_c(1.5\,{\rm GeV})$. This scale choice is indeed appropriate for charm-quark mass effects residing in the jet functions entering the factorization formula (\ref{fact2}). We will be concerned with such effects in Section~\ref{sec:Q7Q1}. On the other hand, charm-quark mass effects also enter some of the hard functions in the factorization formula, for instance via the coefficient $H_\gamma(\mu)$ in (\ref{HQexp}) \cite{Neubert:2004dd}, or via phase-space functions such as those shown in (\ref{Fijpart1}). In this case the charm-penguin loops are probed at virtualities of order $m_b$, and it is therefore appropriate to use a running mass $m_c=\overline{m}_c(\mu_h)$ evaluated at a hard scale $\mu_h\sim m_b$. This can have important numerical effects, enhancing the theoretical prediction for the total decay rate by up to 3\%.

\section{Schematic derivation of the factorization formula}
\label{sec:fact}

The endpoint region of the \bsg\ photon spectrum is defined as the kinematical region where the hadronic jet $X_s$ has large energy compared to its invariant mass: $E_X\sim m_b$, but $M_X\sim\sqrt{m_b\Lambda_{\rm QCD}}$. We define two light-like vectors, $n$ and $\bar n$, which are aligned with the directions of the hadronic jet with total momentum $P_X$ and the photon with momentum $q$. These two vectors satisfy $n+\bar n=2v$, where $v$ is the velocity of the $B$ meson. Specifically, we have $P_X=E_X n+(M_X^2/4E_X)\,\bar n$ and $q=E_\gamma\bar n$. In the $B$-meson rest frame, we may choose $n^\mu=(1,0,0,1)$ and $\bar n^\mu=(1,0,0,-1)$. It is convenient to decompose 4-vectors in the light-cone basis spanned by $n$ and $\bar n$,
\begin{equation}
   a^\mu = n\cdot a\,\frac{\bar n^\mu}{2}
    + \bar n\cdot a\,\frac{n^\mu}{2} +  a_\perp^\mu
   \equiv a_+^\mu + a_-^\mu + a_\perp^\mu \,.
\end{equation}
We will often use the short-hand notation $a\sim(n\cdot a,\bar n\cdot a, a_\perp)$. Note that by definition the external momenta $P_X$, $q$, and $M_B v$ have vanishing perpendicular components.

SCET and HQET are the appropriate effective field theories to study the factorization properties of inclusive $B$-meson decay spectra in the endpoint region \cite{Bauer:2001yt,Bosch:2004th,Bosch:2004cb,Lee:2004ja,Beneke:2004in}. We will need several types of SCET modes for our analysis, each one corresponding to a particular physical scale relevant to the process. Indeed, once the relevant modes have been identified, factorization follows from a sequence of simple and by now familiar steps. High-energy scales such as the electro-weak scales $m_t$, $M_W$ or the hard scale set by the heavy-quark mass $m_b$ are integrated out before one enters the low-energy effective theories, and hence there are no fields for such hard quantum fluctuations in SCET or HQET.

The expansion parameter in the factorization analysis is $\lambda=\Lambda_{\rm QCD}/m_b$, where we do not distinguish between $M_B$, $m_b$, $E_X$, and $E_\gamma$, all of which are of the same order in the endpoint region. The partons that make up the final-state hadronic jet $X_s$ carry momenta that generically scale like the total jet momentum $P_X$. In light-cone components this implies $p_{hc}\sim m_b\,(\lambda,1,\sqrt{\lambda})$. We will refer to modes with this scaling behavior as hard-collinear ({\em hc}) fields. The final state as well as the initial $B$ meson also contain soft partons with momenta $p_s\sim m_b\,(\lambda,\lambda,\lambda)$ of order $\Lambda_{\rm QCD}$. The light-like photon momentum $q$ itself does not set a physical scale; however, partons with momenta scaling like $p_{\overline{hc}}\sim m_b\,(1,\lambda,\sqrt{\lambda})$ can convert into a photon accompanied by soft partons, which can be absorbed by the hadronic final state or originate from the initial $B$ meson. We will refer to modes with this scaling behavior as anti-hard-collinear ($\overline{hc}$). The invariant mass of a set of anti-hard-collinear partons scales like $\sqrt{m_b\Lambda_{\rm QCD}}$. Note that processes in which a parton fragments into a photon plus an energetic parton moving along the $\bar n$ direction, such as those considered in \cite{Kapustin:1995fk}, are kinematically not allowed in the endpoint region. The produced energetic parton cannot be absorbed by the low-mass final-state hadronic jet.

In this paper, we denote by $\hc$ and $\ahc$ the effective SCET fields for hard-collinear and anti-hard-collinear quarks, respectively. We define them as
\begin{equation}\label{xidef}
   \hc = W_{\bar n}^\dagger\,\xi_n \sim\sqrt{\lambda} \,, \qquad
   \ahc = W_{n}^\dagger\,\xi_{\bar n} \sim\sqrt{\lambda} \,,
\end{equation}
where $\xi_n$ and $\xi_{\bar n}$ are two-component spinor fields in the effective Lagrangian, which obey $\nslash\,\xi_n=\nbslash\,\xi_{\bar n}=0$. The quantities $W_{\bar n}$ and $W_n$ are the familiar (anti-)hard-collinear Wilson lines of SCET. Similarly, we define the hard-collinear and anti-hard-collinear gluon fields \cite{Hill:2002vw}
\begin{equation}\label{Adef}
   \Ahc^\mu = W_{\bar n}^\dagger\,(iD_{hc}^\mu W_{\bar n})
    \sim (\lambda,0,\sqrt{\lambda}) \,, \qquad
   \Aahc^\mu = W_{n}^\dagger\,(iD_{\overline{hc}}^\mu W_n)
    \sim (0,\lambda,\sqrt{\lambda}) \,.
\end{equation}
Finally, we need the soft heavy- and light-quark fields $h, q\sim\lambda^{3/2}$ and the soft gluon field $A_s^\mu\sim(\lambda,\lambda,\lambda)$. The effective heavy-quark field obeys $\vslash h=h$. For clarity, soft light-quark fields will often be denoted by the flavor of the corresponding particles ($q=u,d,s,\dots$). Finally, note that in general the gluon fields have the same scaling properties as the corresponding momenta (apart from the large components of the (anti-)hard-collinear gluon fields, which have been ``gauged away'' by the introduction of the Wilson lines), and the same is true for derivatives acting on these fields.

The effective fields defined in (\ref{xidef}) and (\ref{Adef}) are invariant under (anti-)hard-collinear gauge transformations, while they transform homogeneously under soft gauge transformations \cite{Beneke:2002ph}. An important technical step in deriving factorization formulae using SCET is the decoupling of the soft gluons from the (anti-)hard-collinear fields. This is accomplished by introducing the soft Wilson lines $S_n(x)$ in (\ref{eq:Sn}) and corresponding Wilson lines $S_{\bar n}$ defined analogously with $n$ replaced by $\bar n$. We then perform the decoupling transformations \cite{Bauer:2001yt,Becher:2003qh}
\begin{equation}\label{eq:redef}
   \hc(x) = S_n(x_-)\,\hc^{(0)}(x) \,, \qquad
   \Ahc^\mu(x) = S_n(x_-)\,\Ahc^{(0)\mu}(x)\,S^\dagger_n(x_-) \,,
\end{equation}
and similarly for the anti-hard-collinear fields. When expressed in terms of the ``sterile" fields with superscript ``(0)", the SCET Lagrangian no longer contains interactions between soft and (anti-)hard-collinear fields at leading order in $\lambda$.

The derivation of the factorization formula (\ref{fact2}) proceeds as follows. In the first step, the QCD matrix element of the product of the two effective weak Hamiltonians in (\ref{THH}) is matched onto operators in the effective theory SCET($hc,\overline{hc},s$), which is the version of SCET containing hard-collinear, anti-hard-collinear, and soft degrees of freedom. In this step the hard functions $H_i^{(n)}$ appear as Wilson coefficients capturing the effects of hard quantum fluctuations. A priori, an operator in the effective weak Hamiltonian can be matched onto any operator in SCET with the right quantum numbers. One finds, however, that up to order $1/m_b$ only a small number of effective-theory operators contribute. Many aspects of this first matching step have been discussed in \cite{Becher:2005fg}, where a factorization theorem was derived for the exclusive decays $B\to K^*\gamma$ and $B\to\rho\gamma$. Note that in the endpoint region anti-hard-collinear partons cannot be part of the final state $X_s$, because this would lead to an invariant hadronic mass $M_X\sim m_b$, in contrast with the required scaling $M_X\sim\sqrt{m_b\Lambda_{\rm QCD}}$. We therefore need insertions from the SCET Lagrangian, which convert all anti-hard-collinear particles into a photon plus a set of soft partons. As will be discussed in Section~\ref{sec:matching}, these insertions are always power suppressed. After this is done, any given contribution to the decay rate is related, in position space, to a matrix element of the generic form
\begin{equation}\label{genericop}
   \mbox{Disc}\,\langle\bar B(v)| 
   \bar h(x) [\phi_s(x_i)\dots] h(0)\,[\phi_{hc}(y_j)\dots]\,
   [\phi_{\overline{hc}}(z_k)\dots\phi'_{\overline{hc}}(z'_l)\dots]
   |\bar B(v)\rangle \,.
\end{equation}
Note that we have introduced two types of anti-hard-collinear fields. The fields $\phi_{\overline{hc}}$ ($\phi'_{\overline{hc}}$) only couple to the photon connected to the initial (final) $B$-meson state via the weak effective Hamiltonian, see Figure~\ref{fig:theorem}. These fields should be thought of as different entities, which do not interact with one another (i.e., there are no interaction terms in the effective Lagrangian coupling $\phi_{\overline{hc}}$ and $\phi'_{\overline{hc}}$). The reason is that the exchange of an anti-hard-collinear particle across the two sides of the forward amplitude is forbidden kinematically, as this would lead to a hadronic final state with mass of order $m_b$. In other words, anti-hard-collinear propagators are never cut.

The soft heavy-quark fields need to be part of any effective-theory operator. The three brackets $[\dots]$ can contain generic products of soft, hard-collinear, and anti-hard-collinear fields. The presence of a hard-collinear jet in the final state requires that the hard-collinear bracket must contain at least two fields, either a pair of strange quarks or of other light partons (gluons or quarks). In the latter case, the strange-quark pair must appear in the soft bracket. The anti-hard-collinear bracket, on the other hand, can be empty. Recall that any field in the effective Lagrangian scales like a positive power of $\sqrt{\lambda}$, so adding an additional field to an operator always leads to further power suppression.

Because of the particular scaling of soft and (anti-)hard-collinear momenta, the soft fields must be multipole expanded when they couple to (anti-)hard-collinear fields \cite{Beneke:2002ni}. The multipole expansion is subtle if there are external momenta in the problem whose components are nearly identical, such as $n\cdot(m_b v)\approx n\cdot q$ in our case. The correct form of the expansion must then be determined on a case-by-case basis for each operator, rather than derived from a simple set of rules. We find that the heavy-quark fields must always be expanded about $x_-$, while other soft fields can depend on either $x_-$ or $x_+$, or both.

At this point the fields belonging to the three types of modes can still interact with each other by the exchange of soft gluons. These interactions are unsuppressed in SCET, but they have an eikonal structure and can be removed by field redefinitions \cite{Bauer:2001yt,Becher:2003qh}. The remaining power-suppressed interactions are treated as Lagrangian insertions and so are included as parts of the operators in (\ref{genericop}). After the decoupling transformations, the forward $B$-meson matrix elements needed for the calculation of inclusive decay spectra can therefore be factorized into a $B$-meson matrix element of soft fields multiplying vacuum expectation values of hard-collinear and anti-hard-collinear fields:
\begin{equation}\label{brackets}
\begin{aligned}
   \langle\bar B(v)| \bar h(x_-) [\phi_s(x_{i\mp})\dots] h(0)
    |\bar B(v)\rangle
   &\times \mbox{Disc}\,\langle 0|[\phi_{hc}^{(0)}(y_j)\dots]
    |0\rangle \\
   &\hspace{-4.5cm}\times
    \langle 0|[\phi_{\overline{hc}}^{(0)}(z_k)\dots]|0\rangle\,
    \langle 0|[\phi_{\overline{hc}}^{\prime(0)}(z_l')\dots]
    |0\rangle \,.
\end{aligned}
\end{equation}
In this process the soft Wilson lines from (\ref{eq:Sn}) arise, which must be included as part of the soft matrix elements. In fact, they render these non-local matrix element gauge invariant.

In the last step, we match SCET onto HQET by integrating out the (anti-)hard-collinear fields. This can be done perturbatively, because the corresponding scales are in the short-distance regime. The Wilson coefficients of this matching are simply the perturbative expressions for the vacuum correlation functions of the (anti-)hard-collinear fields. Their Fourier transforms define the momentum-space jet functions $J_i^{(n)}$ and $\bar J_i^{(n)}$, and the Fourier transforms of the soft matrix elements define the soft functions $S_i^{(n)}$:
\begin{equation}
\begin{aligned}
   J_i^{(n)}
   &\sim \left[ \mbox{Disc}\,\langle 0|[\phi_{hc}^{(0)}(y_j)\dots]
    |0\rangle \right]_{\rm F.T.} \,, \qquad
   \bar J_i^{(n)}\sim \left[ \langle 0|
    [\phi_{\overline{hc}}^{(0)}(z_k)\dots]|0\rangle
    \right]_{\rm F.T.} \,, \\
   &\hspace{2.2cm} S_i^{(n)}\sim \Big[ \langle\bar B|
    \bar h(x_-) [\phi_s(x_{i\mp})\dots] h(0)
    |\bar B\rangle \Big]_{\rm F.T.} \,.
\end{aligned}
\end{equation}
If both anti-hard-collinear brackets in (\ref{brackets}) are empty, the corresponding contribution to the spectrum becomes part of the first term in the factorization formula (\ref{fact2}). If only one of them is empty, the contribution becomes part of the second term. Finally, if both anti-hard-collinear brackets are not empty, the contribution belongs to the last term. Note that the momentum-space functions defined above will, in general, depend on several Fourier variables (one less than the number of space-time variables in the corresponding non-local matrix elements), and as a result the convolutions in the factorization formula (\ref{fact2}) involve multi-dimensional integrals.

This concludes the derivation of the factorization formula. Our main focus in this work is on the resolved photon contributions giving rise to the second and third terms in (\ref{fact2}). We will now present a detailed discussion of the matching procedure for these contributions.

\section{Systematic analysis of resolved photon contributions}
\label{sec:matching}

\subsection{Matching onto SCET}
\label{subsec:SCETmatching}

As outlined in the previous section, the analysis of the \bsg\ photon spectrum in the effective theory consists of two steps. In the first step the effective weak Hamiltonian  summarized in Appendix~\ref{app:Heff} is matched onto operators in SCET consisting of (anti-)hard-collinear and soft fields. The Wilson coefficients arising in this step are the hard functions $H_i^{(n)}$ in (\ref{fact2}). In the second step the (anti-)hard-collinear modes are integrated out, and the theory is matched onto HQET. The Wilson coefficients arising in this step are the jet functions $J_i^{(n)}$ and $\bar J_i^{(n)}$. The remaining hadronic matrix elements define the soft functions $S_i^{(n)}$. For simplicity, we will restrict ourselves to the tree-level approximation for hard quantum corrections, and to the one-loop approximation for (anti-)hard-collinear quantum fluctuations associated with the leading shape function in (\ref{Sdef}). The Wilson coefficients of the new subleading shape functions will be computed at tree level, but including contributions of order $g^2=4\pi\alpha_s$ resulting from tree-level (anti-)hard-collinear gluon exchange. 

In constructing the possible operator basis of SCET for \bsg\ decay, we require that the final state should contain only one anti-hard-collinear particle, which is the photon field with momentum $q^\mu=E_\gamma\bar n$. All the other particles in the final state, including one strange quark, need to be either hard-collinear or soft. At least one of the particles in the final state must be hard-collinear, either the strange quark or a gluon. Note that we can have several anti-hard-collinear and/or soft fields be part of the possible operators, provided that all the anti-hard-collinear particles are converted into the photon plus soft particles via SCET Lagrangian insertions. The number of soft particles in the final state is restricted only in the sense that adding soft fields to an operator always leads to power suppression. From these simple requirements, it is straightforward to find the possible SCET operator basis systematically.

\begin{figure}
\begin{center}
\epsfig{file=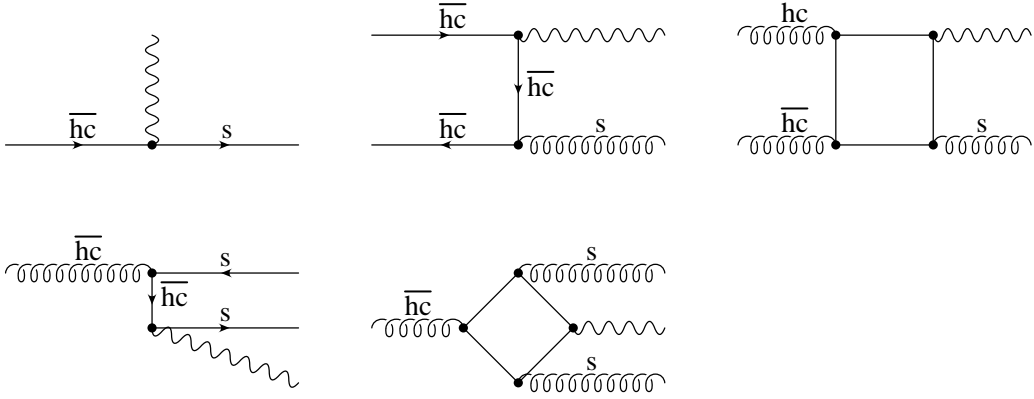,width=14cm}
\parbox{15.5cm}{\caption{\label{fig:conv} 
${\cal O}(\lambda^{1/2})$ (top row) and ${\cal O}(\lambda)$ (bottom row) conversions of anti-hard-collinear particles into a photon accompanied by soft particles. Only some representative diagrams are shown.}}
\end{center}
\end{figure}

Since the SCET expansion parameter is $\sqrt\lambda$ with $\lambda\sim\Lambda_{\rm QCD}/m_b$, we need to consider operators of leading, next-to-leading, and next-to-next-to-leading order in SCET power counting in order to systematically analyze $\Lambda_{\rm QCD}/m_b$-suppressed contributions to the photon spectrum \cite{Beneke:2002ph,Beneke:2003pa,Becher:2005fg}. The possible effective weak-interaction operators can then be divided into two classes, depending on whether they contain a photon field or not. The first class of operators leads to direct photon contributions, whereas the second class gives rise to resolved photon contributions. In the latter case, the operators must be supplemented with Lagrangian insertions that convert the anti-hard-collinear particles into a photon. 

As our main interest is in the resolved photon contributions, we begin by systematically analyzing how anti-hard-collinear partons can be converted into a photon. These conversions can be derived from the SCET($hc,\overline{hc},s$) Lagrangian. For power counting the photon field scales like an anti-hard-collinear field, and hence the conversion of any number of anti-hard-collinear particles into a photon is unsuppressed as long as it is allowed by the rules of the leading-order SCET Lagrangian. However, each conversion involving a soft parton costs a certain power of $\sqrt\lambda$. At ${\cal O}(\sqrt{\lambda})$ we find the possibilities (here and below one could add any number of anti-hard-collinear gluons on the left-hand side of the relations)
\begin{equation}\label{conversions1}
   \ahc\to A_\perp^{\rm em} + q \,, \qquad
   \ahc + \ahcbar\to A_\perp^{\rm em} + A_s \,, \qquad
   \Aahc + \Aahc\to A_\perp^{\rm em} + A_s \,.
\end{equation}
Similarly, at ${\cal O}(\lambda)$ we have
\begin{equation}\label{conversions2}
   \Aahc\to A_\perp^{\rm em} + q + \bar q \,, \qquad
   \Aahc\to A_\perp^{\rm em} + A_s + A_s \,.
\end{equation}
These five possibilities are illustrated in Figure~\ref{fig:conv}. In each case the last conversion, which involves three gluon fields, is not needed for our tree-level analysis.

\begin{figure}
\centering
\begin{tabular}{ccc}
$Q_{7\gamma}$:~~~ & ${\cal O}(\lambda^{5/2})$ & \\
 & \epsfig{file=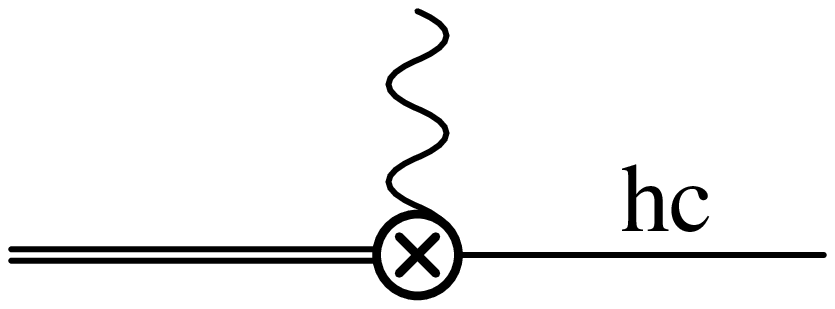,width=3.5cm} & \\[5mm]
$Q_{8g}$:~~~ & ${\cal O}(\lambda^{5/2})
 +{\cal O}(\lambda^{1/2})~\mbox{conversion}$ &\hspace{4mm} 
 ${\cal O}(\lambda^{5/2})+{\cal O}(\lambda)~\mbox{conversion}$ \\
 & \epsfig{file=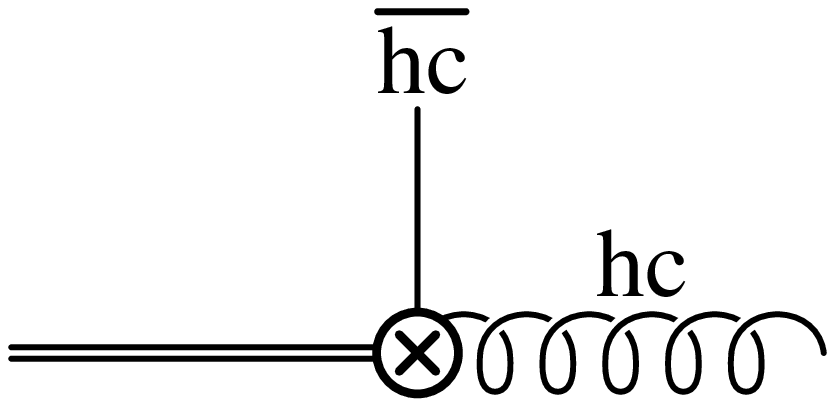,width=3.5cm} & \hspace{4mm} 
 \epsfig{file=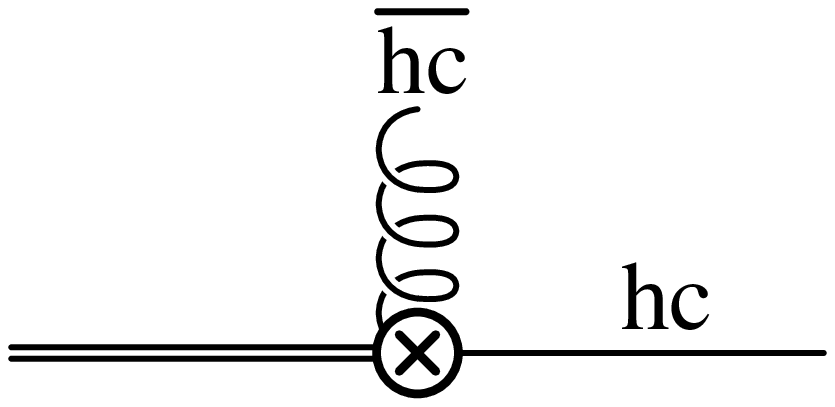,width=3.5cm} \\[5mm]
$Q_1^{c,u}$:~~~ & \hspace{-6mm} 
 ${\cal O}(\lambda^3)+{\cal O}(\lambda^{1/2})~\mbox{conversion}$ & \\
 & \hspace{2mm} \epsfig{file=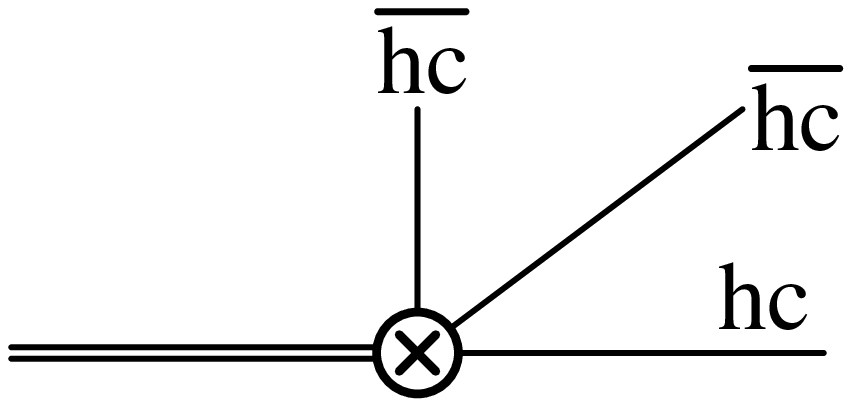,width=3.9cm} &
\end{tabular}
\parbox{15.5cm}{\caption{\label{fig:graphs1} 
Relevant operators arising in the matching of the effective weak Hamiltonian onto SCET. While many other operators exist, only those shown here contribute to the resolved photon contributions at tree-level in perturbative matching.}}
\end{figure}

In the matching of the effective weak Hamiltonian onto SCET we focus on the contributions of the operators $Q_{7\gamma}$, $Q_{8g}$, and $Q_1^{c,u}$. Other four-quark operators, which could be treated in a way analogous to $Q_1^{c,u}$, give rise to negligible effects. There is a large number of SCET operators appearing in the matching relations, all of which are listed in Appendix~\ref{app:SCETops}. Here we will explicitly present only those operators that are needed for our tree-level analysis of resolved photon contributions. The unique leading-order operator arising in the matching relation for $Q_{7\gamma}$ is
\begin{equation}\label{eq:Q7}
   Q_{7\gamma}(x)\to \frac{-em_b}{4\pi^2}\,e^{-im_bv\cdot x}\,
   \hcbar(x)\,\frac{\nbslash}{2}\,
   [in\cdot\partial\Aslash_\perp^{\rm em}(x)]\,(1+\gamma_5)\,
   h(x_-) \,,
\end{equation}
where the derivative acting on the photon field produces a factor $-n\cdot q=-2E_\gamma$. This operator arises at ${\cal O}(\lambda^{5/2})$ in SCET power counting. It is shown in the first row in Figure~\ref{fig:graphs1}. The power suppression of other operators must be evaluated in comparison with this scaling. While at tree level only $Q_{7\gamma}$ matches onto the operator in (\ref{eq:Q7}), at ${\cal O}(\alpha_s)$ the other operators contribute as well, with Wilson coefficients that can be found in \cite{Neubert:2004dd}. From the many operators arising at subleading power in SCET, we only need those that can give rise to $1/m_b$-suppressed resolved photon contributions. They must contain at least one anti-hard-collinear field. From the matching relation for $Q_{8g}$, we need the leading-order operators
\begin{equation}\label{eq:Q8_3p}
\begin{aligned}
   Q_{8g}(x) 
   &\to \frac{-g m_b}{4\pi^2}\,e^{-im_bv\cdot x}\,\bigg[ \,
    \hcbar(x)\,\frac{\nbslash}{2}\,
    [in\cdot\partial\calAslash_{\overline{hc}\perp}(x)]\,
    (1+\gamma_5)\,h(x_-) \\
   &\hspace{3.08cm}\mbox{}+ 
    \ahcbar(x)\,\frac{\nslash}{2}\,
    [i\bar n\cdot\partial\calAslash_{hc\perp}(x)]\,
    (1+\gamma_5)\,h(x_-) \bigg] \,.
\end{aligned}
\end{equation}
They are shown in the second row in Figure~\ref{fig:graphs1}. The conversions of the anti-hard-collinear fields into the photon plus soft fields follow from (\ref{conversions1}) and (\ref{conversions2}) and give rise to power suppression, as indicated in the figure. Finally, from the matching relation for the current-current operators $Q_1^q$, we need the ${\cal O}(\lambda^3)$ operator
\begin{equation}\label{Q1q}
   Q_1^q(x)\to e^{-im_bv\cdot x}\,
   \hcbar\gamma^\mu(1-\gamma_5) h(x_-)\,
   \ahcbar(x)\gamma_\mu(1-\gamma_5) \ahc(x)
\end{equation}
illustrated in the third row in Figure~\ref{fig:graphs1}. Here $\hc$ is the strange-quark field, while $\ahc$ are fields for the quark with flavor $q=c,u$ in the four-quark operator. The anti-hard-collinear quark pair must be converted into a photon plus a hard-collinear or soft gluon via a penguin loop, as indicated by the second relation in (\ref{conversions1}) and the second graph in the first row of Figure~\ref{fig:conv}. 

\begin{figure}
\begin{center}
\vspace{0.5cm}
\epsfig{file=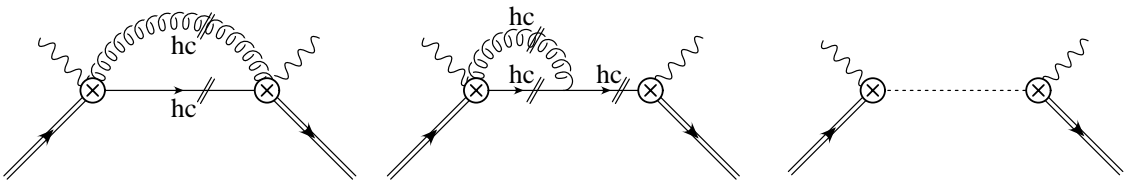,width=15cm} \hspace{2mm}
\parbox{15.5cm}{\caption{\label{fig:direct} 
Diagrams representing the ${\cal O}(1/m_b)$ direct photon contributions arising from hard-collinear loops. The two graphs on the left represent products of SCET operators (see Appendix~\ref{app:SCETops} for a complete list), while the graph on the right represents non-local operators built out of the soft fields remaining after the matching onto HQET.}}
\end{center}
\end{figure}

In our analysis we will also include power corrections of ${\cal O}(\alpha_s)$ to the direct photon contributions in the factorization formula (\ref{fact2}). They involve a hard-collinear loop and give rise to a convolution of subleading jet functions with the leading shape function. The operators $Q_{7\gamma}$, $Q_{8g}$, and $Q_1^{c,u}$ can all be matched onto ${\cal O}(\lambda^3)$ and ${\cal O}(\lambda^{7/2})$ SCET operators containing a photon field accompanied by hard-collinear quark and gluon fields. A complete list of the corresponding operators can be found in Appendix~\ref{app:SCETops}. A specific example, which we will need in the analysis in Section~\ref{subsec:Q8Q8ssf}, is the unique ${\cal O}(\lambda^3)$ four-particle operator containing the photon field in the matching relation for the dipole operator $Q_{8g}$, which reads
\begin{equation}\label{eq:Q8_4p}
   Q_{8g}(x)\to \frac{-g m_b}{4\pi^2}\,e^{-im_bv\cdot x}\,
   \hcbar(x)\,\frac{1}{i\bar n\cdot\overleftarrow{\partial}}\,
   e_d\,e\Aslash_\perp^{\rm em}(x)\,
   [i\bar n\cdot\partial\calAslash_{hc\perp}(x)]\,(1+\gamma_5)\,
   h(x_-) \,,
\end{equation}
where $e_d=-1/3$ is the electric charge of a down-type quark in units of $e$. We refrain from listing the relevant operators descending from $Q_{7\gamma}$ and $Q_1^{c,u}$ here. The resulting ${\cal O}(1/m_b)$ direct photon contributions are of the form shown in Figure~\ref{fig:direct}, where the two diagrams on the left show the possible operator products in SCET. In the second graph, the operator on the right vertex represents the leading-order matching contribution given in (\ref{eq:Q7}). This diagram therefore only arises in pairings of the form $Q_i-Q_{7\gamma}$. The graph on the right illustrates the structure of soft fields remaining after the hard-collinear fields have been integrated out in the second matching step. Here the dashed horizontal line represents a Wilson line along the $n$ direction. In this case the soft matrix element in HQET is the leading shape function $S(\omega,\mu)$.

\subsection{Analysis of the $\bm{Q_{7\gamma}-Q_{7\gamma}}$ contribution}
\label{subsec:42}

We have discussed in Section~\ref{sec:review} that at order $1/m_b$ in the heavy-quark expansion the $Q_{7\gamma}-Q_{7\gamma}$ terms in the \bsg\ photon spectrum contain a parton-model contribution of order $\alpha_s$ as well as a tree-level contribution from subleading shape functions. They are given in (\ref{Fijpart}) and (\ref{F77SSF}), respectively. We have argued that $F_{77}^{\rm\,part}(E_\gamma,\mu)$ contains a soft contribution, which needs to be extracted and absorbed into the subleading shape-function contribution to avoid double counting. We will now discuss this subtraction in more detail.

Evaluating the diagrams in Figure~\ref{fig:direct} with two insertions of $Q_{7\gamma}$, we find that they give rise to a convolution of the leading shape function in (\ref{Sdef}) with a subleading jet function $J_{\rm subl}(p^2)$ resulting from the cuts of hard-collinear loop diagrams. This jet function is divergent, and using dimensional regularization with $d=4-2\epsilon$ space-time dimensions we obtain the bare expression \cite{Paz:2009ut}
\begin{equation}\label{sjf_partonic}
   J_{\rm subl}^{\rm bare}(p^2) = \theta(p^2)\,\frac{d-2}{2}\,
   \frac{C_F\alpha_s(\mu)}{4\pi} \left( - \frac{16}{\bar\epsilon} 
   - 16\ln\frac{\mu^2}{p^2} + 9 \right) ,
\end{equation}
where $1/\bar\epsilon\equiv 1/\epsilon-\gamma_E+\ln 4\pi$, and the factor of $(d-2)$ arises from the Dirac algebra in $d$ dimensions. Subtracting the pole term, we obtain
\begin{equation}\label{F77_sjf}
   F_{77}^{(a)}(E_\gamma,\mu)
   = \frac{C_F\alpha_s(\mu)}{4\pi}
   \int_{-p_+}^{\bar\Lambda}\!d\omega
   \left( 16\ln\frac{m_b\,(\omega+p_+)}{\mu^2} + 25 
   - 16 c_{\rm RS} \right) S(\omega,\mu) \,,
\end{equation}
where the scheme-dependent constant $c_{\rm RS}$ vanishes in the $\overline{\rm MS}$ scheme, $c_{\overline{\rm MS}}=0$. The $\ln(m_b)$ term in the above result agrees with that in the parton-model inspired expression for $F_{77}$ given in (\ref{F77part1}). However, the $p_+$-dependent terms and the constant accompanying the logarithm are different in the two expressions. This difference is accounted for by the subleading shape-function contribution $F_{77}^{(b)}(E_\gamma,\mu)\equiv F_{77}^{\rm SSF}(E_\gamma,\mu)$ given in (\ref{F77SSF}).

In order to show that this is indeed the case, we derive the partonic expressions resulting from a naive, perturbative calculation of the subleading shape functions. To this end, we need to calculate the one-loop corrections to the matrix elements of the soft operators corresponding to the subleading shape functions $\omega\,S(\omega,\mu)$, $s(\omega,\mu)$, $t(\omega,\mu)$, $u(\omega,\mu)$, $v(\omega,\mu)$, as well as the tree-level matrix elements of the soft operators corresponding to $f_u(\omega,\mu)$ and $f_v(\omega,\mu)$ between on-shell heavy-quark states with velocity $v$. Without loss of generality we take $v_\perp=0$, and as a result the matrix elements of the soft operators corresponding to $t$ and $v$ vanish at one-loop order, since they only contain gluons with transverse polarization. The tree-level matrix elements of the soft operators corresponding to $f_u$ and $f_v$ also vanish, since they contain scaleless integrals over the $\bar n$-components of the light-quark momenta. Finally, the matrix element of the soft operator corresponding to $s$ is non-zero only when the heavy quarks are off shell with a non-zero $n$-component of the residual momentum. As a result it does not contribute when we set the residual momentum to zero in the partonic calculation. We are therefore left with only $\omega\,S$ and $u$. The former can be extracted from the calculations performed in \cite{Bosch:2004th}. After setting $v\cdot k$ to zero, we find
\begin{equation}\label{eq:wS}
   \omega\,S_{\rm bare}(\omega)
   = \theta(-\omega)\,\frac{C_F\alpha_s(\mu)}{4\pi}
    \left( - \frac{4}{\bar\epsilon} - 4\ln\frac{\mu^2}{(-\omega)^2} + 4 \right) .
\end{equation}
For the latter, one finds by explicit calculation that
\cite{Paz:2009ut}
\begin{equation}\label{eq:u}
   u_{\rm bare}(\omega)
   = \theta(-\omega)\,\frac{C_F\alpha_s(\mu)}{4\pi}
    \left( \frac{12}{\bar\epsilon} + 12\ln\frac{\mu^2}{(-\omega)^2} - 20 \right) .
\end{equation}
Subtracting the UV poles in the $\overline{\rm MS}$ scheme, we obtain from (\ref{F77SSF})
\begin{equation}\label{F77b_part}
   F_{77}^{\rm SSF}(E_\gamma,\mu) \big|_{\rm pert}
   = \frac{C_F\alpha_s(\mu)}{4\pi} \left( 16\ln\frac{\mu^2}{p_+^2} - 24 \right) ,
\end{equation}
where the subscript ``pert'' indicates that these are naive perturbative expressions for non-pertur\-bative hadronic functions. Adding to this result the naive perturbative expression for $F_{77}^{(a)}(E_\gamma,\mu)$ in (\ref{F77_sjf}) obtained by replacing $S(\omega,\mu)\to\delta(\omega)$ yields
\begin{equation}\label{F77xx}
   F_{77}^{(a)}(E_\gamma,\mu) \big|_{\rm pert}
    + F_{77}^{\rm SSF}(E_\gamma,\mu) \big|_{\rm pert}
   = \frac{C_F\alpha_s(\mu)}{4\pi}
    \left( 16\ln\frac{m_b}{p_+} + 1 - 16 c_{\rm RS} \right) , 
\end{equation}
which coincides with expression for $F_{77}^{\rm part}(E_\gamma,\mu)$ in (\ref{Fijpart}) apart from the constant term.

The remaining difference has its origin in the $(d-2)$ prefactor in (\ref{sjf_partonic}), which results from the $d$-dimensional Dirac algebra. The reason is that relation (\ref{F77SSF}) was derived in \cite{Bosch:2004cb} by working in $d=4$ dimensions. Indeed, it is convenient to renormalize the subleading shape functions before evaluating the traces arising from the Dirac structures specific to a given process. Only in that case the definitions of the subleading shape functions are the same for different processes, such as $\bar B\to X_s\gamma$ and $\bar B\to X_u l\,\bar\nu$ decays. However, in this case it is crucial that the same subtraction scheme is adopted in the calculation of the subleading jet-function contribution in (\ref{F77_sjf}). In other words, we should work in the $\overline{\rm DR}$ subtraction scheme \cite{Siegel:1979wq,Capper:1979ns}, where the Dirac algebra is performed in $d=4$ dimensions, while loop integrals are evaluated with $d=4-2\epsilon$. In this scheme $c_{\rm RS}=1$ in (\ref{F77_sjf}) and (\ref{F77xx}), and the latter expression then agrees with that in (\ref{Fijpart}). Our final result is therefore
\begin{equation}\label{F77final}
   F_{77}(E_\gamma,\mu)
   = \frac{C_F\alpha_s(\mu)}{4\pi}
    \int_{-p_+}^{\bar\Lambda}\!d\omega
    \left( 16\ln\frac{m_b(\omega+p_+)}{\mu^2} + 9 \right)
    S(\omega,\mu) + F_{77}^{\rm SSF}(E_\gamma,\mu).
\end{equation}
This is the first equation in (\ref{improved}), which replaces the ``incorrect'' (in the sense of improper factorization) result in (\ref{Fijpart}). 

Throughout this paper we set the strange-quark mass to zero, which turns out to be an excellent approximation numerically. Let us nevertheless briefly comment on finite strange-quark mass effects. Taking $m_s$ to be non-zero gives rise to a subleading jet function proportional to $m_s^2/p_{hc}^2$, where $p_{hc}$ is the momentum of the hard-collinear jet containing the strange quark \cite{Chay:2005ck}. If we adopt the scaling $m_s\sim{\cal O}(\Lambda_{\rm QCD})$, this function scales as $\Lambda_{\rm QCD}/m_b$ in the endpoint 
region and contributes to $F_{77}$. From \cite{Chay:2005ck}, we find that the extra contribution is
\begin{equation}\label{Fms}
   F_{77}^{m_s}(E_\gamma,\mu) 
   = - \tilde H_{77} \int_{-p_+}^{\bar\Lambda}\!d\omega\,m_b\,
    j_m(\omega+p_+,\mu)\,S(\omega,\mu) \,,
\end{equation}
where we have defined $j_m=\frac{1}{\pi}\,{\rm Im}\, J^m_{m_b}$. Both $\tilde H_{77}=1+{\cal O}(\alpha_s)$ and $J^m_{m_b}$ can be found in \cite{Chay:2005ck}. At the lowest order in $\alpha_s$,
\begin{equation}\label{jsubm}
   j_m(\omega+p_+,\mu) = \frac{m_s^2}{m_b}\,\delta^\prime(\omega+p_+)
    + {\cal O}(\alpha_s) \,,
\end{equation}
and indeed $j_m$ is suppressed by $m_s^2/(m_b\Lambda_{\rm QCD})$ compared to the leading-order jet function $J(p^2,\mu)$. One could argue that for $m_s\approx 100$\,MeV and $\Lambda_{\rm QCD}\sim 500$\,MeV the parameter $m_s$ should scale as a higher power of the SCET expansion parameter $\lambda$, e.g.~$m_s\sim\lambda^2$. This would imply that $j_m$ can be neglected at order $\Lambda_{\rm QCD}/m_b$. Contributions of $j_m$ to coefficient functions $F_{ij}$ other than $F_{77}$ are further suppressed by hard functions $\tilde H_{ij}={\cal O}(\alpha_s)$. We will not consider them in the following, since they are bound to be tiny.

\subsection{Analysis of the $\bm{Q_1^q-Q_{7\gamma}}$ contributions}
\label{sec:Q7Q1}

Evaluating the diagrams in Figure~\ref{fig:direct} for this pairing of operators, we obtain the direct photon contribution
\begin{equation}\label{F17a}
   F_{17}^{(a)}(E_\gamma,\mu) 
   = \frac{C_F\alpha_s(\mu)}{4\pi} \left( -\frac23 \right)
   \int_{-p_+}^{\bar\Lambda}\!d\omega\,S(\omega,\mu) \,,
\end{equation}
which involves a convolution of the leading shape function (\ref{Sdef}) with a jet function consisting of the cut of a hard-collinear loop. This is the expected extension of the first term in the expression for $F_{17}^{\rm part}(E_\gamma,\mu)$ given in (\ref{Fijpart}).

\begin{figure}
\begin{center}
\begin{tabular}{cc}
 \epsfig{file=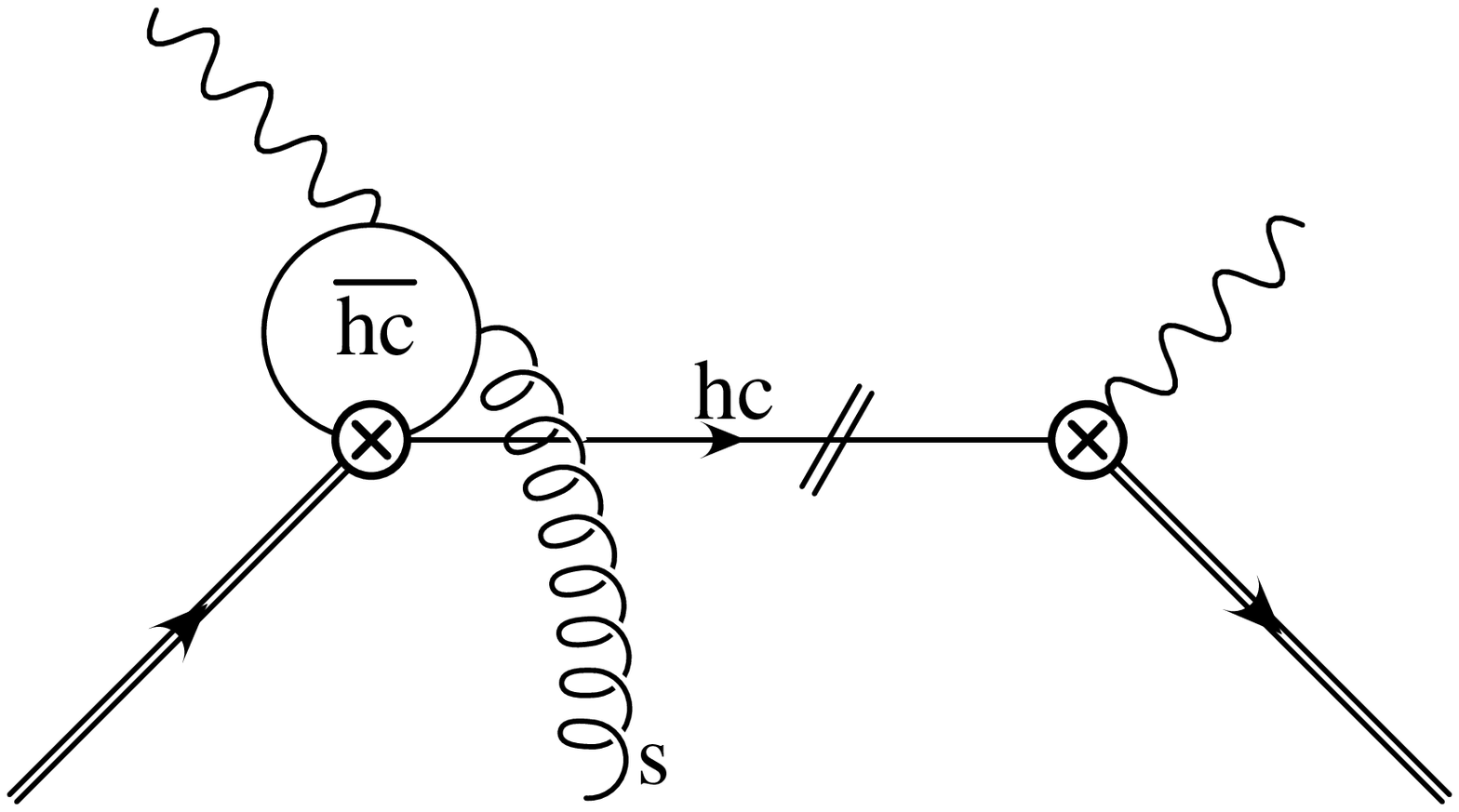,width=6cm} & \hspace{1cm}
 \epsfig{file=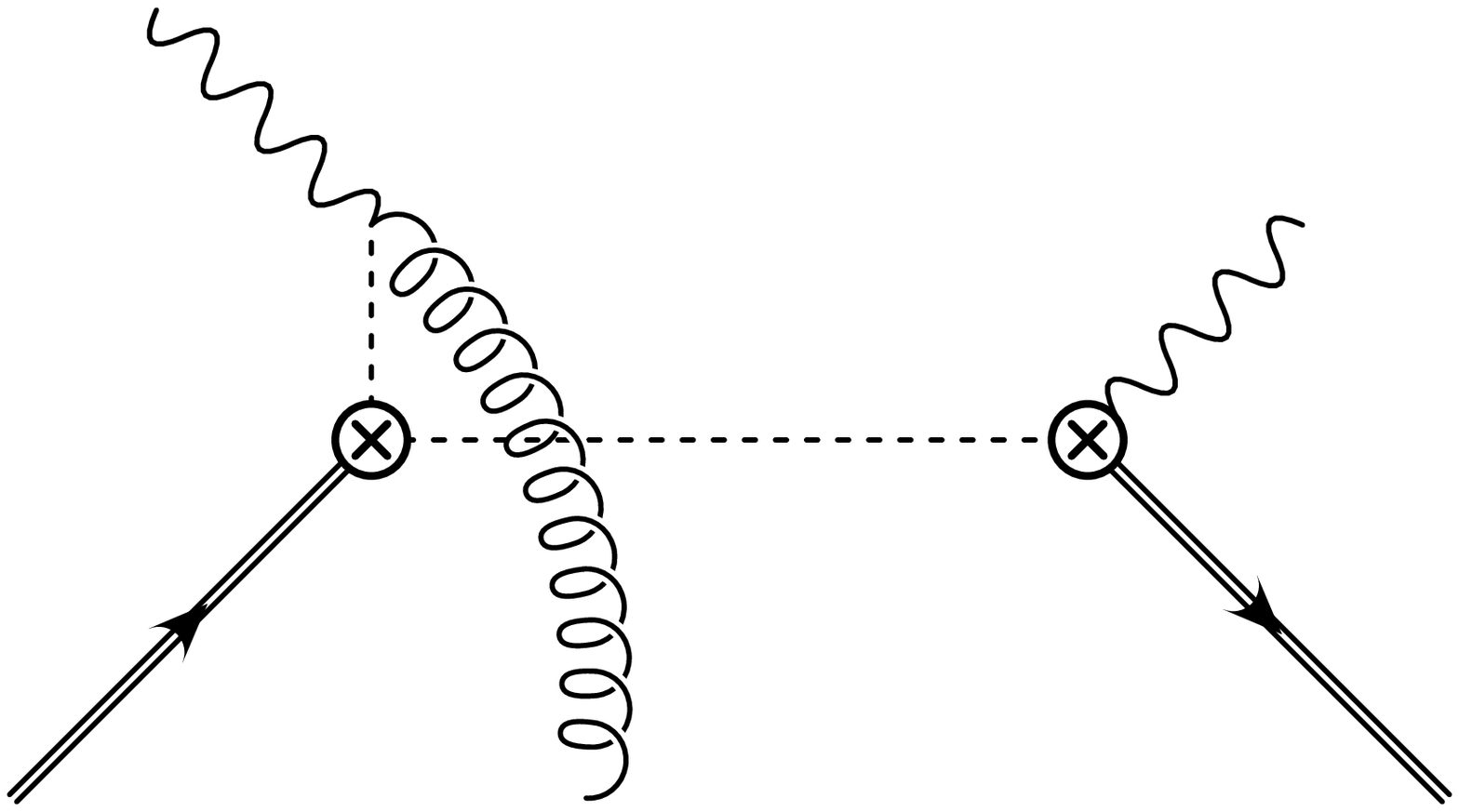,width=6cm}
\end{tabular}
\parbox{15.5cm}{\caption{\label{fig:graphs17} 
Diagrams arising from the matching of the $Q_1^q-Q_{7\gamma}$ contribution onto SCET (left) and HQET (right). Horizontal (vertical) dashed lines denote non-localities obtained after (anti-)hard-collinear fields have been integrated out.}}
\end{center}
\end{figure}

Much more interesting is the single resolved photon contribution arising from this operator pair. As shown in the left graph of Figure~\ref{fig:graphs17}, it is obtained by combining the ${\cal O}(\lambda^3)$ SCET four-quark operator in (\ref{Q1q}), which contains two anti-hard-collinear quark fields in addition to a hard-collinear strange quark and a heavy quark, with the leading-power contribution (\ref{eq:Q7}) descending from $Q_{7\gamma}$. According to the rules (\ref{conversions1}), the conversion of the two anti-hard-collinear fields into a photon and a soft gluon costs a factor of $\lambda^{1/2}$, giving a total suppression with respect to the leading term of $\lambda\sim\Lambda_{\rm QCD}/m_b$. When the (anti-)hard-collinear fields are integrated out in the second matching step, we obtain the HQET diagram shown in the right graph of Figure~\ref{fig:graphs17}. Here and below, horizontal (vertical) dotted lines represent Wilson lines along the $n$ ($\bar n$) direction. The HQET matrix element corresponding to the graph on the right in Figure~\ref{fig:graphs17} contains a soft gluon field in addition to the two heavy quarks. In the language of the factorization formula (\ref{fact2}) this corresponds to a single resolved contribution with a subleading shape function. For the contribution from the operator $Q_1^c$ we obtain
\begin{equation}\label{F17b}
\begin{split}
   F_{17,c}^{(b)}(E_\gamma,\mu) 
   &= \frac23\,(1-\delta_u) 
    \int_{-\infty}^{\bar\Lambda}\!d\omega\,\delta(\omega+p_+) \\
   &\quad\times \mbox{Re} \int_{-\infty}^\infty 
   \frac{d\omega_1}{\omega_1+i\varepsilon} \left[ 1 
   - F\!\left( \frac{m_c^2-i\varepsilon}{2E_\gamma\,\omega_1} \right)
   \right] g_{17}(\omega,\omega_1,\mu) \,,
\end{split}
\end{equation}
where the CKM-suppressed parameter $\delta_u$ has been introduced in (\ref{deltaqdef}), and we have defined the subleading shape function
\begin{eqnarray}\label{g17def}
   g_{17}(\omega,\omega_1,\mu) 
   &=& \int\frac{dr}{2\pi}\,e^{-i\omega_1 r}\!
    \int\frac{dt}{2\pi}\,e^{-i\omega t} \\
   &&\times \frac{\langle\bar B| \big(\bar h S_n\big)(tn)\,
    \nbslash (1+\gamma_5) \big(S_n^\dagger S_{\bar n}\big)(0)\,
    i\gamma_\alpha^\perp\bar n_\beta\,
    \big(S_{\bar n}^\dagger\,g G_s^{\alpha\beta} S_{\bar n} 
    \big)(r\bar n)\,
    \big(S_{\bar n}^\dagger h\big)(0) |\bar B\rangle}{2M_B} \,.
    \nonumber
\end{eqnarray}
The penguin function $F(x)$ has been defined in (\ref{eq:F(r)}). For $0<x<1/4$ this function develops an imaginary part. Note that we adopt a power counting for the charm-quark mass such that $m_c^2={\cal O}(m_b\Lambda_{\rm QCD})$. The argument of the penguin function in the convolution (\ref{F17b}) then counts as ${\cal O}(1)$.

The structure of the soft Wilson lines in (\ref{g17def}), which are directed either along $n$ or $\bar n$, follows when the decoupling transformation is applied to the (anti-)hard-collinear fields in SCET to remove their soft interactions from the effective Lagrangian and absorb them into eikonal factors. The Wilson lines reflect the space-time topology of the HQET diagrams shown on the right-hand side in Figure~\ref{fig:graphs17}. Let us label the two weak vertices by coordinates 0 (left) and $x=tn+x_+ +x_\perp$ (right), and the vertex of the soft gluon by $y=r\bar n+y_- +y_\perp$. The multipole expansion of the effective-theory fields implies that $x_{+,\perp}$ and $y_{-,\perp}$ can be set to zero at this order. Gauge invariance then requires that the fields $\bar h(tn)$ and $G_s(r\bar n)$ are joined by a Wilson line, and the rules of SCET determine that this line consists of two segments: a straight line $[tn,0]$ along the light-like direction $n$ followed by a straight line $[0,r\bar n]$ along the light-like direction $\bar n$. The fields $G_s(r\bar n)$ and $h(0)$ are joined by a straight Wilson line $[r\bar n,0]$ along the light-like direction $\bar n$. Using that $[tn,0]=S_n(tn) S_n^\dagger(0)$ etc., we recover the structure of the Wilson lines in the non-local operator in (\ref{g17def}). We note for completeness that soft functions closely related to our functions $g_{17}$ in (\ref{g17def}) and $g_{11}$ in (\ref{g11def}) were introduced, in a context not related to \bsg\ decay, in \cite{Chay:2007ej}.

There is more to the space-time structure of the soft operator that is worth pointing out. Since hard-collinear fields in SCET carry large momentum components, the particles created by these fields always move forward in time. As a result, after convolution with the jet functions, the quantum fields in the definition of the subleading shape functions are ordered in the same way as they appear in Feynman graphs \cite{Bosch:2004cb}. The operators considered in the present paper contain fields that propagate along the two light-like directions $n$ and $\bar n$, as indicated by the dotted lines in the right graph in Figure~\ref{fig:graphs17}. If we assign coordinate 0 to the first of the weak vertices in the figure, the gluon is emitted at space-time point $r\bar n$ with $r>0$. This is ensured by the $i\varepsilon$ prescription in the jet function in (\ref{F17b}), since
\begin{equation}\label{thetar}
   \int\!d\omega_1\,e^{-i\omega_1 r}\,\frac{i}{\omega_1+i\varepsilon}
   = 2\pi\,\theta(r) \,.
\end{equation}
The gluon thus lives at a later time than the field $h(0)$ (recall that $n^0=\bar n^0=+1$), and indeed it appears to the left of that field.

Another comment is in order concerning the structure of the result (\ref{F17b}). From the diagrams shown in Figure~\ref{fig:graphs17} we derive one half times the expression (\ref{F17b}) without the real-part prescription in front of the integral and in expressions (\ref{HQexp}) and (\ref{deltaqdef}). The mirror diagrams not shown in the figure, in which the two weak vertices are interchanged, give an analogous contribution with the complex conjugate Wilson coefficients and CKM matrix elements, the complex conjugate penguin function $F^*(r)$,\footnote{This is because the fields in the charm-quark loop of the mirror graphs are anti-time-ordered.} 
the propagator factor $1/(\omega_1-i\varepsilon)$, and the soft function
\begin{eqnarray}
   g_{17}'(\omega,\omega_1,\mu) 
   &=& \int\frac{dr}{2\pi}\,e^{i\omega_1 r}\!
   \int\frac{dt}{2\pi}\,e^{-i\omega t} \\
   &&\hspace{-15mm}
    \times \frac{\langle\bar B|\big(\bar h S_{\bar n}\big)(tn)\,
    (-i\gamma_\alpha^\perp\bar n_\beta)\,
    \big(S_{\bar n}^\dagger\,g G_s^{\alpha\beta} S_{\bar n} 
    \big)(tn+r\bar n)\,
   \big(S_{\bar n}^\dagger S_n\big)(tn)\,
   \nbslash (1+\gamma_5)
   \big(S_n^\dagger h\big)(0) |\bar B\rangle}{2M_B} \,, \nonumber
\end{eqnarray}
which is related to the original one by complex conjugation: $g_{17}'(\omega,\omega_1,\mu) =[g_{17}(\omega,\omega_1,\mu)]^*$. To show this, one uses translational invariance to shift all position arguments by $-tn$ and then changes the sign of the integration variable $t$. The sum of the diagrams in Figure~\ref{fig:graphs17} plus their mirror graphs thus gives a real result, and after averaging over CP-conjugate decay modes we obtain (\ref{F17b}).

The real-part symbols in (\ref{deltaqdef}) and (\ref{F17b}) refer to different kinds of complex parameters. The various products of Wilson coefficients and CKM factors carry, in general, CP-violating weak phases. The convolution of the jet and soft functions, on the other hand, can carry CP-even, strong-rescattering phases, which in principle can result either from anti-hard-collinear loops (i.e., the jet functions $\bar J_i$) or from the soft matrix elements themselves. (However, in Section~\ref{sec:PT} we will argue that the soft functions are real.) When both types of phases are present, a non-zero direct CP asymmetry arises. The subleading power corrections investigated in this paper provide new mechanisms for generating such an asymmetry. A more detailed exploration of their phenomenological relevance is left for further study.

The range of support of the soft functions in HQET can be derived by noting that the light-cone projections $n\cdot p_i$ and $\bar n\cdot p_i$ of all parton momenta in the $B$ meson must be non-negative, and that the total momentum of all partons in the $B$ meson is $M_B v$. Since in HQET the momentum of a heavy quark is decomposed as $p_b=m_b v+k$, where $k$ is the residual momentum, it follows that
\begin{equation}
   \sum_{i\ne b} n\cdot p_i + n\cdot k = \bar\Lambda \,, \qquad
   \sum_{i\ne b} \bar n\cdot p_i + \bar n\cdot k = \bar\Lambda \,,
\end{equation}
where $n\cdot k>-m_b$ and $\bar n\cdot k>-m_b$. In the heavy-quark limit $m_b\to\infty$ it follows that $-\infty<n\cdot k\le\bar\Lambda$ and $0\le n\cdot p_i<\infty$ (for $i\ne b$), and similarly for $\bar n\cdot k$ and $\bar n\cdot p_i$. In the special case of the soft function $g_{17}$ in (\ref{g17def}), the variable $\omega$ corresponds to the residual-momentum component $n\cdot k$ of the initial-state heavy quark, while $\omega_1$ can either correspond to the component $\bar n\cdot p_g$ of a gluon in the final-state $B$ meson or to the component $-\bar n\cdot p_g$ of a gluon in the initial-state $B$ meson. It thus follows that $-\infty<\omega\le\bar\Lambda$ and $-\infty<\omega_1<\infty$. This implies that the penguin-loop function $F(x)$ is sampled over both positive and negative values of its argument.

In principle, each of the six QCD penguin four-quark operators in the effective weak Hamiltonian can give rise to a similar contribution, either via loops of massless quarks ($q=u,d,s$) or via a charm-quark loop. ($b$-quark loops lead to further power suppression.) Of these options only $Q_1^c$ has both a large Wilson coefficient $C_1\sim 1$ and a large CKM factor $V_{cb} V_{cs}^*={\cal O}(\lambda^2)$, so it will give rise to the dominant effects. However, for academic reasons we will also consider the case of the CKM-suppressed operator $Q_1^u$. Using that $F(0)=0$, it follows that the resolved photon contribution resulting from this operator is given by 
\begin{equation}\label{F17bu}
   F_{17,u}^{(b)}(E_\gamma,\mu) 
   = \frac23\,\delta_u\,
    \mbox{Re} \int_{-\infty}^{\bar\Lambda}\!d\omega\,\delta(\omega+p_+) 
    \int_{-\infty}^\infty \frac{d\omega_1}{\omega_1+i\varepsilon}\,
    g_{17}(\omega,\omega_1,\mu) \,.
\end{equation}
Note that the soft function is the same as in (\ref{F17b}).

Let us now investigate the convergence properties of the convolution integrals in (\ref{F17b}) and (\ref{F17bu}). In the UV region, for $\omega_1\gg\Lambda_{\rm QCD}$, the first integral approaches the form of the second one, since mass effects become negligible. It follows that the convolution over $\omega_1$ converges as long as the soft function $g_{17}$ vanishes for $\omega_1\to\pm\infty$. In general, the asymptotic behavior of the soft functions for large values of the $\omega_i$ variables can be analyzed using short-distance methods \cite{Bosch:2004th}. This shows, for instance, that the leading shape function behaves as $S(\omega,\mu)\sim 1/\omega$ modulo logarithms for $\omega\to-\infty$. For the present case, naive dimensional analysis suggests the behavior $g_{17}(\omega,\omega_1,\mu)\propto\omega_1$ for large $\omega_1$ but fixed $\omega$, in which case the convolution integral would diverge linearly. To obtain such a contribution, however, would require a non-zero matrix element of the soft operator between two on-shell $b$ quarks. But this matrix element vanishes by Lorentz invariance. A non-zero contribution is only obtained if, in addition to the heavy quarks, one adds a soft external gluon. This costs two orders in power counting, so that the asymptotic fall-off is at least as strong as $g_{17}(\omega,\omega_1,\mu)\propto 1/\omega_1$ for $\omega_1\to\pm\infty$. It follows that the convolution integrals (\ref{F17b}) and (\ref{F17bu}) are UV convergent. 

The behavior of the soft functions in the IR region  cannot be derived from a perturbative analysis. In the present case, however, it suffices to make the reasonable assumption that $g_{17}(\omega,\omega_1,\mu)$ is non-singular at $\omega_1=0$. Using the expansion
\begin{equation}\label{Fexpand}
   1 - F(x) = - \frac{1}{12x} - \frac{1}{90 x^2} - \frac{1}{560 x^3} - \dots
\end{equation}
valid for large $x$, we find that for small $\omega_1$ the convolution integral (\ref{F17b}) arising from the charm-quark loop behaves as
\begin{equation}
   \int\limits_{\omega_1\approx 0}
    \frac{d\omega_1}{\omega_1+i\varepsilon} \left[ 1 
    - F\!\left( \frac{m_c^2-i\varepsilon}{2E_\gamma\,\omega_1} \right)
    \right] g_{17}(\omega,\omega_1,\mu) 
   \approx - \frac{E_\gamma}{6m_c^2}\, 
    \int\limits_{\omega_1\approx 0}\!d\omega_1\,
    g_{17}(\omega,\omega_1,\mu) \,.
\end{equation}
For the convolution integral (\ref{F17bu}) arising from the up-quark loop, we find instead
\begin{equation}\label{PVreg}
   \int\frac{d\omega_1}{\omega_1+i\varepsilon}\,
    g_{17}(\omega,\omega_1,\mu) 
   = \mbox{P}\!\!\int\frac{d\omega_1}{\omega_1}\,
    g_{17}(\omega,\omega_1,\mu) - i\pi\,g_{17}(\omega,0,\mu) \,,
\end{equation}
where the symbol P denotes the Cauchy principal value of the integral. We conclude that the convolution integrals indeed exist as long as the subleading shape function is non-singular at $\omega_1=0$. Note that for the case of the up-quark loop it is important that the integral over $\omega_1$ runs over both positive and negative values. Previous authors have already pointed out that the up-quark loop contribution to the \bsg\ decay rate, while strongly CKM suppressed, is described by an uncalculable long-distance contribution \cite{Voloshin:1996gw,Grant:1997ec,Buchalla:1997ky}. Our relation (\ref{PVreg}) provides a rigorous field-theoretic definition of this contribution in terms of a well-defined, non-local soft matrix element.

For phenomenological purposes it is useful to define a new function
\begin{equation}\label{f17def}
   f_{17,q}(\omega,\mu)
   = \frac23 \int_{-\infty}^\infty 
   \frac{d\omega_1}{\omega_1+i\varepsilon} \left[ 1 
   - F\!\left( \frac{m_q^2-i\varepsilon}{(m_b+\omega)\,\omega_1} 
   \right) \right] g_{17}(\omega,\omega_1,\mu) \,.
\end{equation}
Our final expression for the $Q_1^q-Q_{7\gamma}$ contribution can then be written as
\begin{equation}\label{F17final}
   F_{17}(E_\gamma,\mu) 
   = \frac{C_F\alpha_s(\mu)}{4\pi} \left( -\frac23 \right)
   \int_{-p_+}^{\bar\Lambda}\!d\omega\,S(\omega,\mu) 
   + \sum_{q=c,u}\,\delta_q\,\mbox{Re}\,f_{17,q}(-p_+,\mu) \,.
\end{equation}
Note that the argument of the penguin function entering $f_{17,q}(-p_+,\mu)$ is $m_q^2/[(m_b-p_+)\,\omega_1]=m_q^2/(2E_\gamma\,\omega_1)$, as it should be.

Our result differs from the parton-model expression given in (\ref{Fijpart}) by the now familiar integral over the leading shape function in the first term, and by the fact that a non-trivial function $f_{17,c}(-p_+,\mu)$ replaces $(-\frac{m_b\lambda_2}{6m_c^2})\,\delta(p_+)$. The latter expression would be obtained if one made the unjustified approximation of neglecting the dependence of the soft fields in (\ref{g17def}) on $t$ and $r$, i.e., of evaluating all fields and Wilson lines at the origin. This would replace
\begin{equation}\label{g17part}
   g_{17}(\omega,\omega_1,\mu)
   \to 2\lambda_2\,\delta(\omega)\,\delta(\omega_1) \,,
\end{equation}
and from (\ref{f17def}) we would then recover the parton-model expression given above. However, there is no reason why (\ref{g17part}) should provide a decent model for the soft function. All we know is that the definition of the soft function $g_{17}$ in (\ref{g17def}) implies the normalization condition
\begin{equation}\label{g17norm}
   \int_{-\infty}^{\bar\Lambda}\!d\omega 
   \int_{-\infty}^\infty\!d\omega_1\,g_{17}(\omega,\omega_1,\mu) 
   = \frac{\langle\bar B|\,\bar h\,\nbslash\,i\gamma_\alpha^\perp
           \bar n_\beta\,g G_s^{\alpha\beta}\,h\,|\bar B\rangle}{2M_B} 
   = 2\lambda_2 \,,
\end{equation}
where we have used a general relation derived in \cite{Falk:1992wt} to evaluate the matrix element of the local quark-gluon operator in terms of the hadronic parameter $\lambda_2$. Moreover, the trace formalism of HQET \cite{Neubert:1993mb,Falk:1992wt} implies that the soft function can be written as
\begin{equation}\label{traceformalism}
   g_{17}(\omega,\omega_1,\mu)
   = \mbox{\rm Tr} \left[ \frac{1+\vslash}{2}\,\nbslash (1+\gamma_5)\,
    i\gamma_\alpha^\perp\,\frac{1+\vslash}{2}\,
    \Xi^{\alpha_\perp}(v,\bar n,\omega,\omega_1,\mu) \right] 
   = 4\Xi_2(\omega,\omega_1,\mu) \,,
\end{equation}
where we have used that the most general decomposition of the quantity $\Xi^{\alpha_\perp}$ is of the form $\Xi^{\alpha_\perp}(v,\bar n,\omega,\omega_1,\mu)=i\gamma^{\alpha_\perp} (\Xi_1+\nbslash\,\Xi_2)$ with scalar functions $\Xi_i\equiv\Xi_i(\omega,\omega_1,\mu)$. It follows from this argument that the factor $(1+\gamma_5)$ in (\ref{g17def}) can be replaced by 1, since the part of the trace involving $\gamma_5$ vanishes. It is then easy to see that
\begin{equation}\label{g17int}
   \int_{-\infty}^{\bar\Lambda}\!d\omega\,
    g_{17}(\omega,\omega_1,\mu) 
   = \int_{-\infty}^{\bar\Lambda}\!d\omega\, 
    \Big[ g_{17}(\omega,-\omega_1,\mu) \Big]^* .
\end{equation}

One can constrain the function $g_{17}(\omega,\omega_1,\mu)$ further by looking at its first moments with respect to $\omega$ and $\omega_1$. These can be related to linear combinations of HQET matrix elements with three covariant derivatives. Such matrix elements can be expressed in terms of two hadronic parameters, $\rho_1$ and $\rho_2$, via \cite{Mannel:1994kv}
\begin{equation}
   \frac{\langle\bar B|\,\bar h\,\Gamma_{\alpha\delta\beta}\,
         iD^\alpha iD^\delta iD^\beta h\,|\bar B\rangle}{2M_B} 
   = \frac12\,\mbox{Tr} \left( \Gamma_{\alpha\delta\beta}\,
    \frac{1+\vslash}{2} \left[ (g^{\alpha\beta}-v^\alpha v^\beta)\,
    v^\delta\,\frac{\rho_1}{3} + i\sigma^{\alpha\beta} v^\delta\,
    \frac{\rho_2}{2} \right] \frac{1+\vslash}{2} \right) .
\end{equation}
Thus, we find
\begin{equation}\label{g17mom1}
\begin{aligned}
   \int_{-\infty}^{\bar\Lambda}\!d\omega\,\omega 
   \int_{-\infty}^\infty\!d\omega_1\,g_{17}(\omega,\omega_1,\mu)
   &= \frac{\langle\bar B|\,\bar h\,\nbslash\gamma_{\perp\alpha}\, 
    in\cdot D\,[iD_\perp^\alpha,i\bar n\cdot D]\,h\,|\bar B\rangle}{2M_B}
   = - \rho_2 \,, \\
   \int_{-\infty}^{\bar\Lambda}\!d\omega 
   \int_{-\infty}^\infty\!d\omega_1\,\omega_1\,g_{17}(\omega,\omega_1,\mu)
   &= \frac{\langle\bar B|\,\bar h\,\nbslash\gamma_{\perp\alpha}
    \left[ [iD_\perp^\alpha,i\bar n\cdot D], i\bar n\cdot D\right]
    h\,|\bar B\rangle}{2M_B} = 0 \,,
\end{aligned} 
\end{equation}
where the HQET parameter $\rho_2$ is related to the parameter $\rho_{LS}^3$ introduced in \cite{Bigi:1994ga} via $\rho_{LS}^3=3\rho_2$. The vanishing of the first moment with respect to $\omega_1$ of $g_{17}$ is not a coincidence. As we will see in Section~\ref{sec:PT}, $g_{17}$ is in fact a real function. Relation (\ref{g17int}) then implies that all the odd moments in $\omega_1$ vanish. 

As a final comment, let us add that even in the limit where the charm quark is treated as a heavy quark, $m_c={\cal O}(m_b)$, the penguin contribution to the photon spectrum must still be described by a subleading shape function. In this limit the argument of the penguin function in (\ref{F17b}) is of order $m_b/\Lambda_{\rm QCD}$. Expanding then the function $[1-F(x)]$ to first order in $1/x$ leads to\footnote{Only the first term in this expansion gives rise to a UV-convergent convolution integral.} 
\begin{equation}
   f_{17,c}(\omega,\mu)
   \to - \frac{m_b+\omega}{18 m_c^2}
   \int\frac{dt}{2\pi}\,e^{-i\omega t}\,
   \frac{\langle\bar B| \big(\bar h S_n\big)(tn)\,
   \nbslash (1+\gamma_5)\,i\gamma_\alpha^\perp\bar n_\beta\,
   \big(S_n^\dagger\,g G_s^{\alpha\beta}\,h\big)(0) 
   |\bar B\rangle}{2M_B} \,.
\end{equation}
Integrating this expression over $\omega$, and dropping higher power corrections, we obtain a contribution to the total decay rate proportional to
\begin{equation}
   \int_{-\bar\Lambda}^{m_b}\!dp_+\,f_{17,c}(-p_+,\mu)
   \to - \frac{m_b\lambda_2}{9m_c^2} \,.
\end{equation}
This agrees with the result found in \cite{Voloshin:1996gw,Ligeti:1997tc,Grant:1997ec,Buchalla:1997ky} (the correct sign was obtained in the last reference).

\subsection{Analysis of the $\bm{Q_1^q-Q_1^q}$ and $\bm{Q_1^q-Q_{8g}}$ contributions}
\label{sec:Q8Q1}

\begin{figure}
\begin{center}
\begin{tabular}{cc}
\epsfig{file=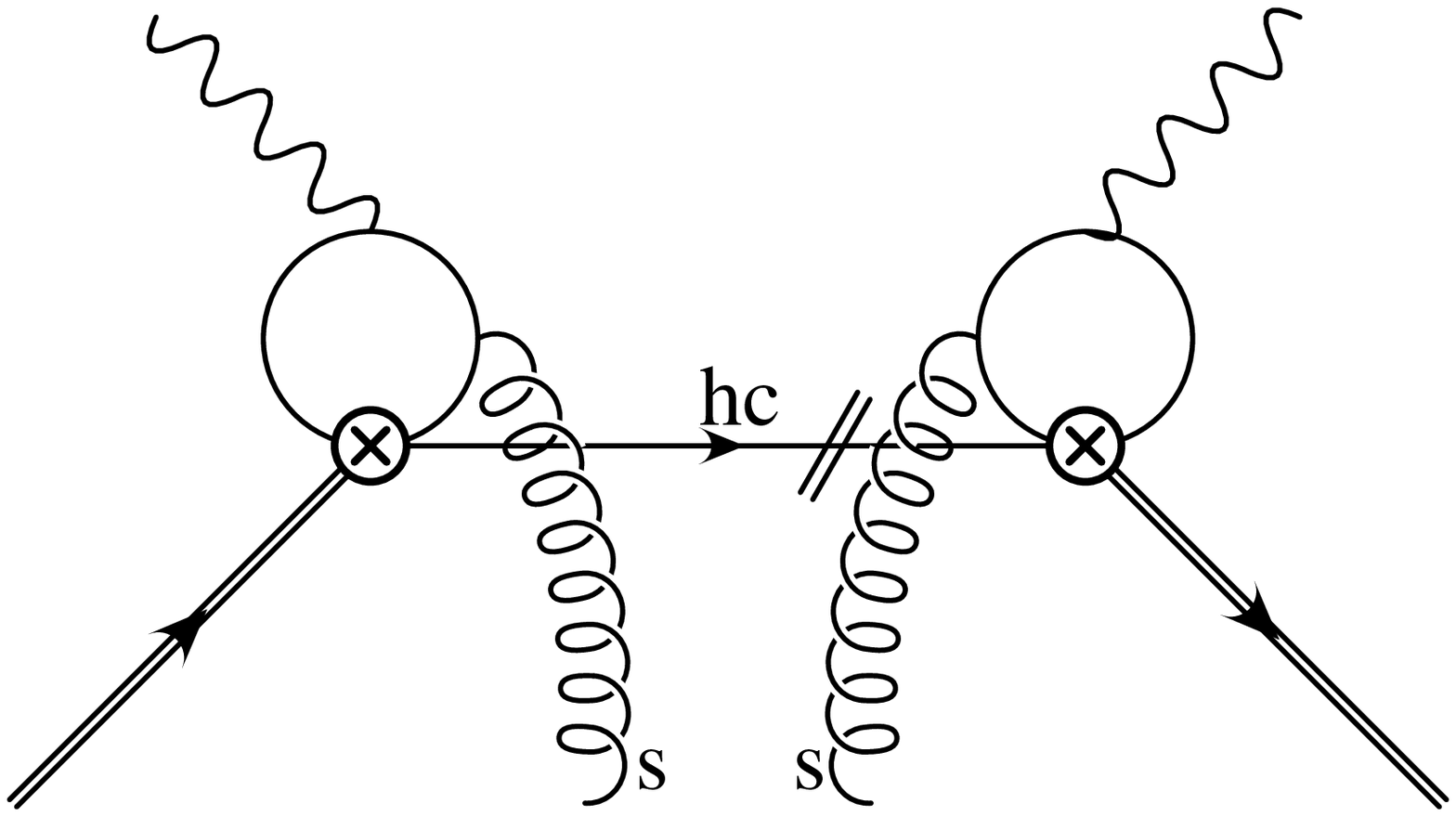,width=6cm} & \hspace{1cm}
\epsfig{file=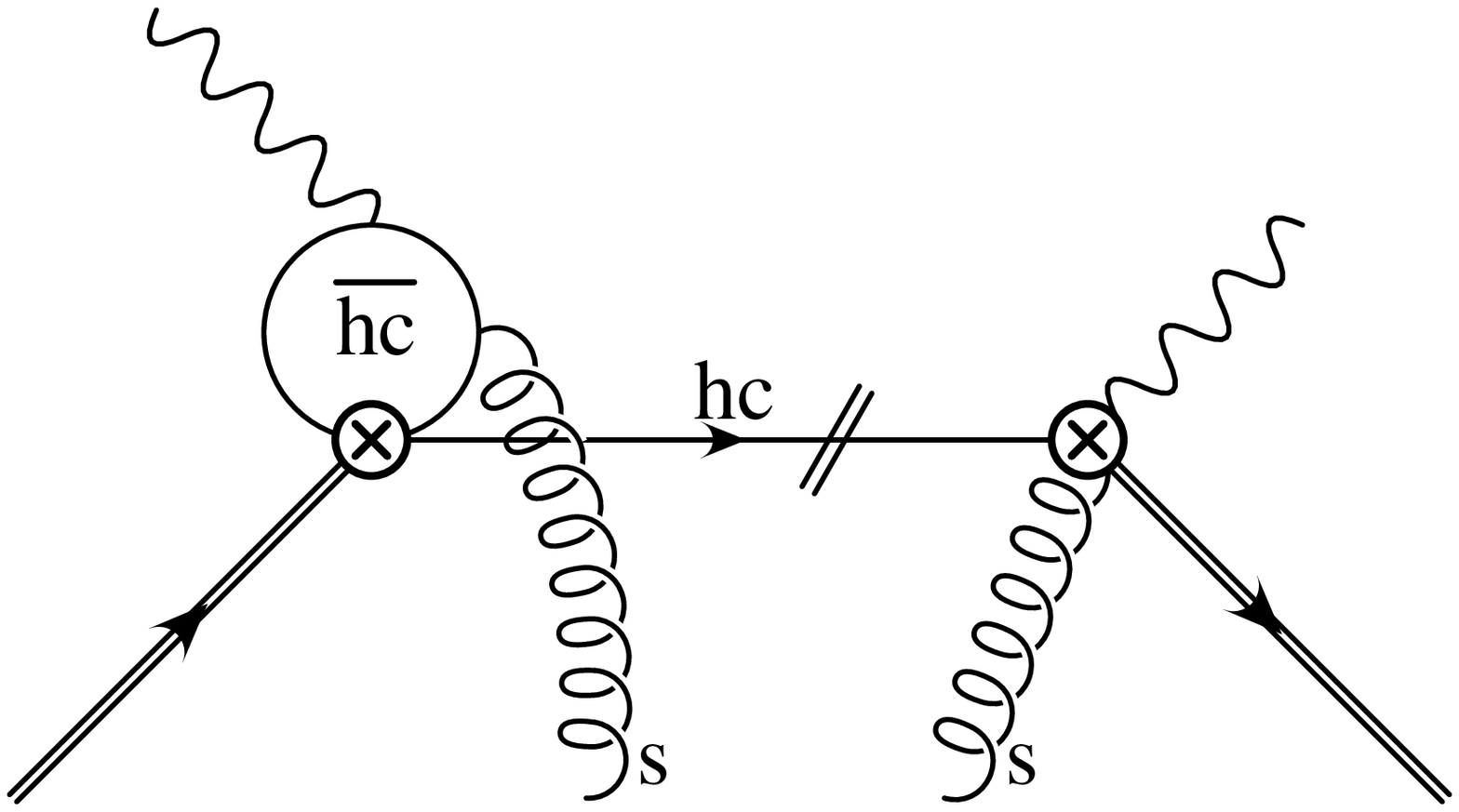,width=6cm} 
\end{tabular}
\parbox{15.5cm}{\caption{\label{fig:Q1Q1}
Examples of SCET diagrams giving rise to resolved photon contributions suppressed by at least two powers of of $1/m_b$. The left graph arises from the pairing $Q_1^q-Q_1^q$, while the right one contributes to the $Q_1^q-Q_{8g}$ term.}}
\end{center}
\end{figure}

The power-counting rules described in Appendix~\ref{app:SCETops} show that for these two cases there do not exist operators arising at order $1/m_b$ in the heavy-quark expansion that contain soft fields other than the two heavy quarks. In particular, the diagrams shown in Figure~\ref{fig:Q1Q1} contribute to the \bsg\ photon spectrum only at order $1/m_b^2$. This is an important finding. Since for the case of two charm-quark loops the first diagram in the figure is proportional to the large Wilson coefficient $|C_1|^2\sim 1$, it has sometimes been mentioned as a potentially dangerous source of power corrections. It follows from our analysis that this contribution scales as $(\Lambda_{\rm QCD}/m_b)^2$ relative to the leading term. It is therefore expected to be a small correction (see below).

It thus remains to calculate the leading power corrections to the direct photon term in the factorization formula, by analyzing graphs of the type shown in Figure~\ref{fig:direct}. Note that here only the first diagram on the left in this figure can contribute. After a straightforward calculation (summing over $q=c,u$), we find once again contributions involving a convolution of the leading shape function with a jet function consisting of the cut of a hard-collinear loop. The results are the same in the two cases and given by
\begin{equation}\label{F11}
   F_{11}^{(a)}(E_\gamma,\mu) 
   = \frac{2E_\gamma}{m_b}\,F_{18}^{(a)}(E_\gamma,\mu) 
   = \frac{C_F\alpha_s(\mu)}{4\pi}\,\frac29
   \int_{-p_+}^{\bar\Lambda}\!d\omega\,S(\omega,\mu) \,. 
\end{equation}
This is the obvious generalization of the parton-model results in (\ref{Fijpart}). Note that the prefactor $m_b/2E_\gamma$ in the result for $F_{18}$ follows from the SCET calculation, and we have therefore presented it here. In the endpoint region this factor equals 1 up to power corrections, and it could therefore be omitted, as we have done in (\ref{improved}).

In order to substantiate the statement about the smallness of the ${\cal O}(1/m_b^2)$ double resolved photon contribution represented by the first diagram in Figure~\ref{fig:Q1Q1}, we have evaluated this graph explicitly. The result is
\begin{eqnarray}\label{F11b}
   F^{(b)}_{11}(E_\gamma,\mu)
   &=& -\frac1{m_b} \left|\frac{\lambda_c}{\lambda_t}\right|^2
    \frac29 \int_{-\infty}^{\bar\Lambda}\!d\omega\,\delta(\omega+p_+) 
    \int_{-\infty}^\infty\!d\omega_1\,\int_{-\infty}^\infty\!d\omega_2 \\
   &&\times \frac{1}{\omega_1+i\varepsilon} \left[ 1 
    - F\!\left( \frac{m_c^2-i\varepsilon}{2E_\gamma\,\omega_1} \right)
    \right] \frac{1}{\omega_2-i\varepsilon} \left[ 1 
    - F^*\!\left( \frac{m_c^2-i\varepsilon}{2E_\gamma\,\omega_2} \right)
    \right] g_{11}(\omega,\omega_1,\omega_2,\mu) \,, \nonumber
\end{eqnarray}
where the $1/m_b$ prefactor indicates the additional power suppression. We have defined the soft function
\begin{eqnarray}\label{g11def}
   g_{11}(\omega,\omega_1,\omega_2,\mu) 
   &=& \int\frac{dt}{2\pi}\,e^{-i\omega t}
    \int\frac{dr}{2\pi}\,e^{-i\omega_1 r} 
    \int\frac{du}{2\pi}\,e^{i\omega_2 u}\,g_{\mu\nu}\,
    \bar n^\alpha\bar n^\beta \\
   &&\hspace{-34mm}\times 
    \frac{\langle\bar B| \big(\bar h S_{\bar n}\big)(tn)\,
    \big(S_{\bar n}^\dagger\,g G_s^{\nu\beta} S_{\bar n} 
    \big)(tn+u\bar n)
    \,\Gamma \big(S^\dagger_{\bar n}S_n\big)(tn)\,
    \big(S^\dagger_{n}S_{\bar n}\big)(0)\,
    \big(S_{\bar n}^\dagger\,g G_s^{\mu\alpha} S_{\bar n} 
    \big)(r\bar n)\,
    \big(S_{\bar n}^\dagger h\big)(0) |\bar B\rangle}{2M_B} \,,
    \nonumber
\end{eqnarray}
where $\Gamma=\nbslash(1-\gamma_5)$. This function satisfies $g_{11}(\omega,\omega_1,\omega_2,\mu)=\left[g_{11}(\omega,\omega_2,\omega_1,\mu)\right]^*$, which implies that $F^{(b)}_{11}$ is real. In order to obtain an estimate of the magnitude of this contribution, we expand the penguin functions to first order using (\ref{Fexpand}). This yields
\begin{equation}
   F^{(b)}_{11}(E_\gamma,\mu)
   \approx - \frac{1}{648} \left( \frac{2E_\gamma}{m_b} \right)^2 
   \left|\frac{\lambda_c}{\lambda_t}\right|^2
   \frac{m_b}{m_c^4}\,
   \int_{-\infty}^\infty\!d\omega_1\,\int_{-\infty}^\infty\!d\omega_2\,\,
   g_{11}(-p_+,\omega_1,\omega_2,\mu) \,,
\end{equation}
where the remaining double integral over the soft function scales like $\Lambda_{\rm QCD}^3$. For any reasonable value of this quantity, the prefactor $1/648$ and the additional $1/m_b$ suppression render this contribution negligible. For instance, if we model the double integral by $\Lambda_{11}^4\,S(-p_+,\mu)$ with some hadronic scale $\Lambda_{11}\sim\Lambda_{\rm QCD}$, the contribution of this term relative to the leading direct photon contribution in (\ref{HQexp}) is approximately given by
\begin{equation}
   - \frac{1}{648}\,\left| \frac{C_1(\mu)}{C_{7\gamma}(\mu)} \right|^2 
    \left( \frac{\Lambda_{11}}{m_c} \right)^4 
   \approx - 2\cdot 10^{-4} \left( \frac{\Lambda_{11}}{0.5\,{\rm GeV}} \right)^4 .
\end{equation}

\subsection{Analysis of the $\bm{Q_{8g}-Q_{8g}}$ contribution}\label{subsec:Q8Q8ssf}

\begin{figure}
\begin{center}
\begin{tabular}{cc}
\epsfig{file=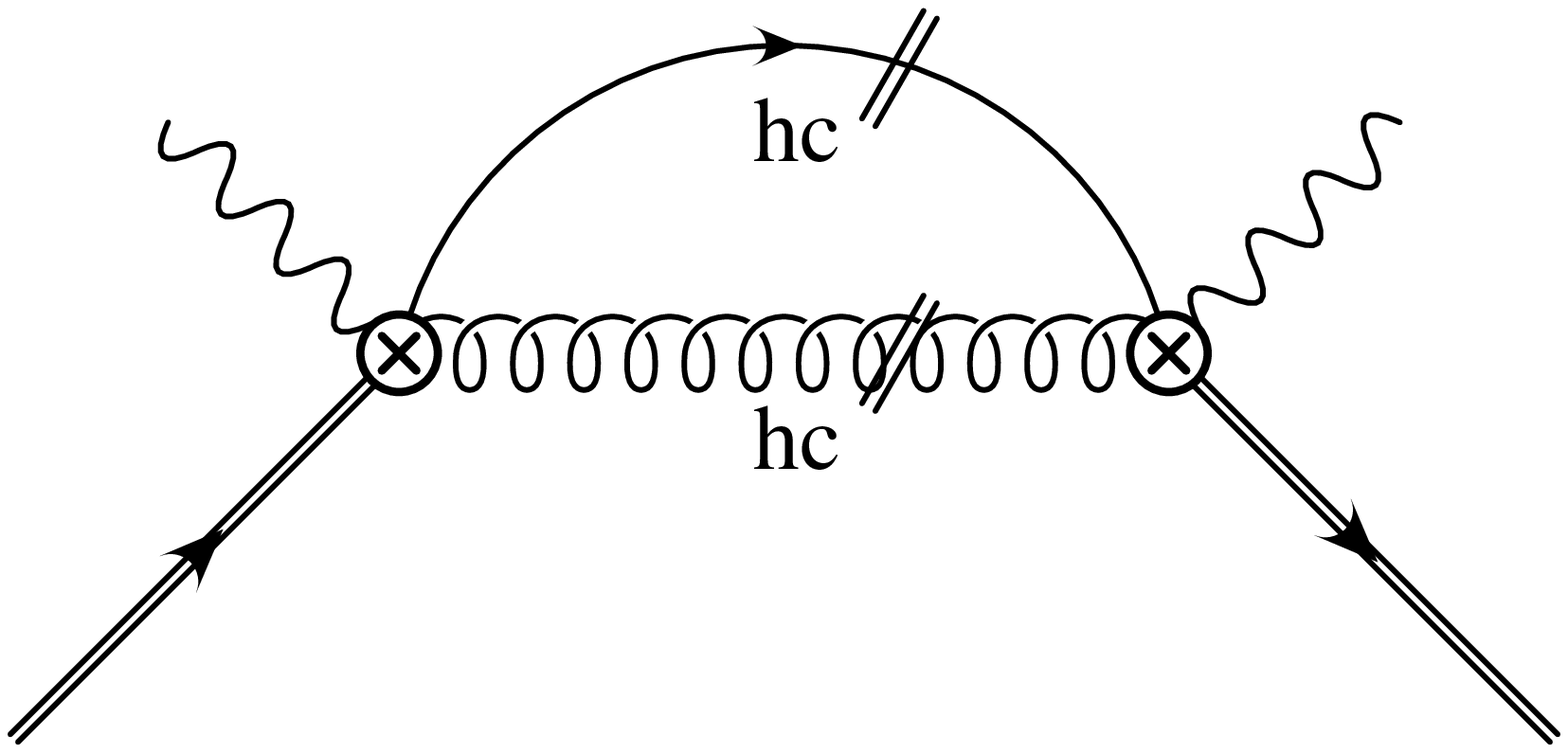,width=6cm} & \hspace{1cm}
\epsfig{file=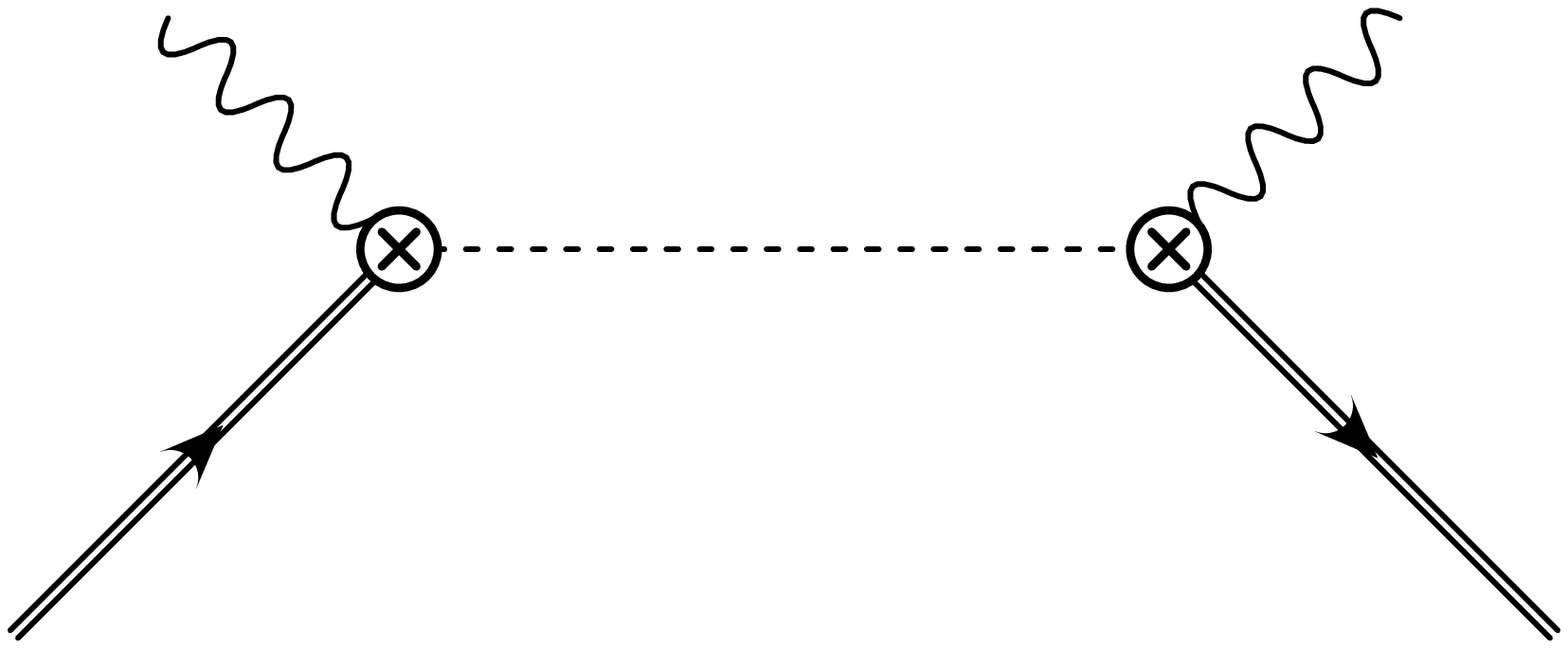,width=6cm} \\[0.3cm]
\epsfig{file=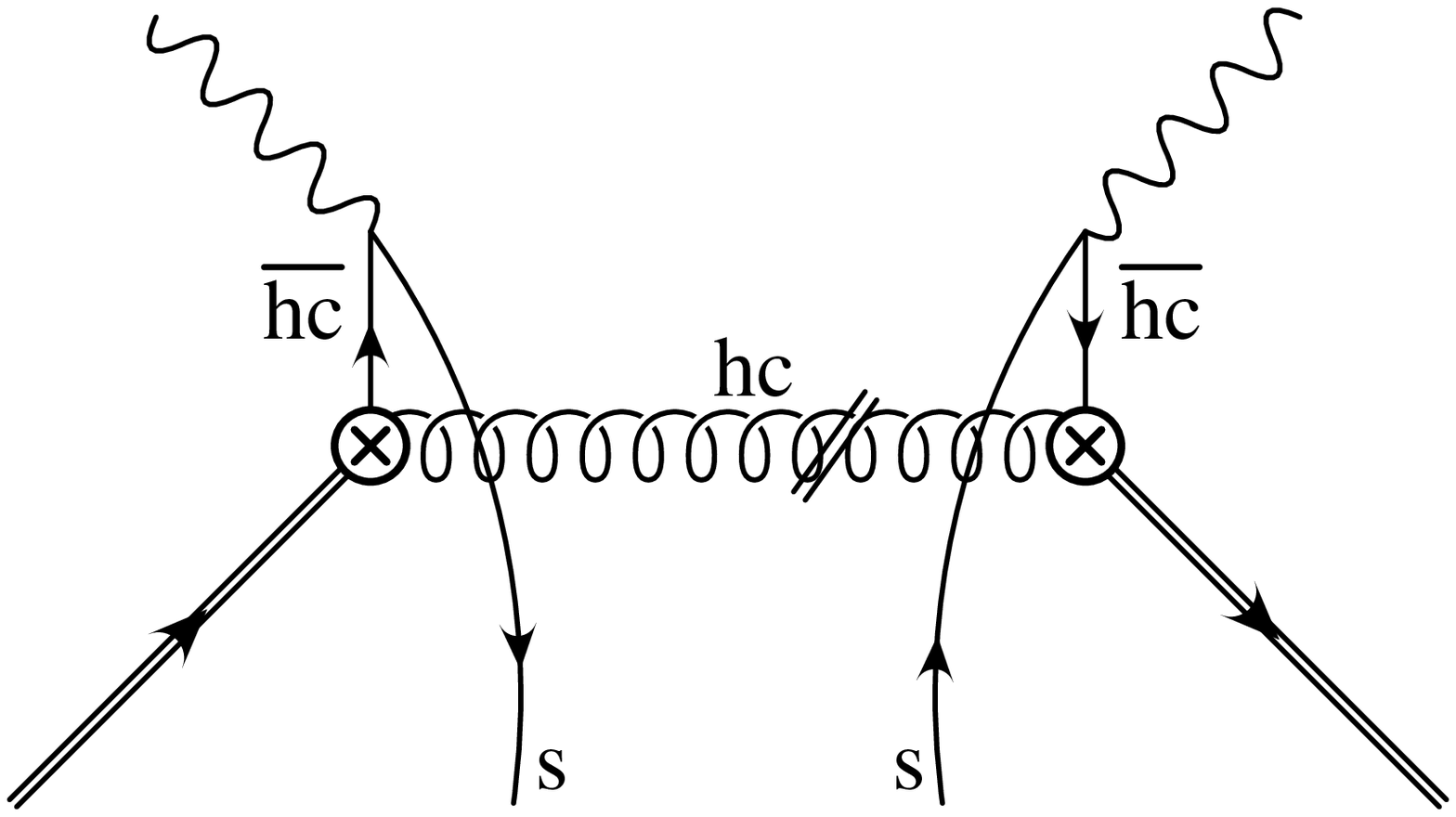,width=6cm} & \hspace{1cm}
\epsfig{file=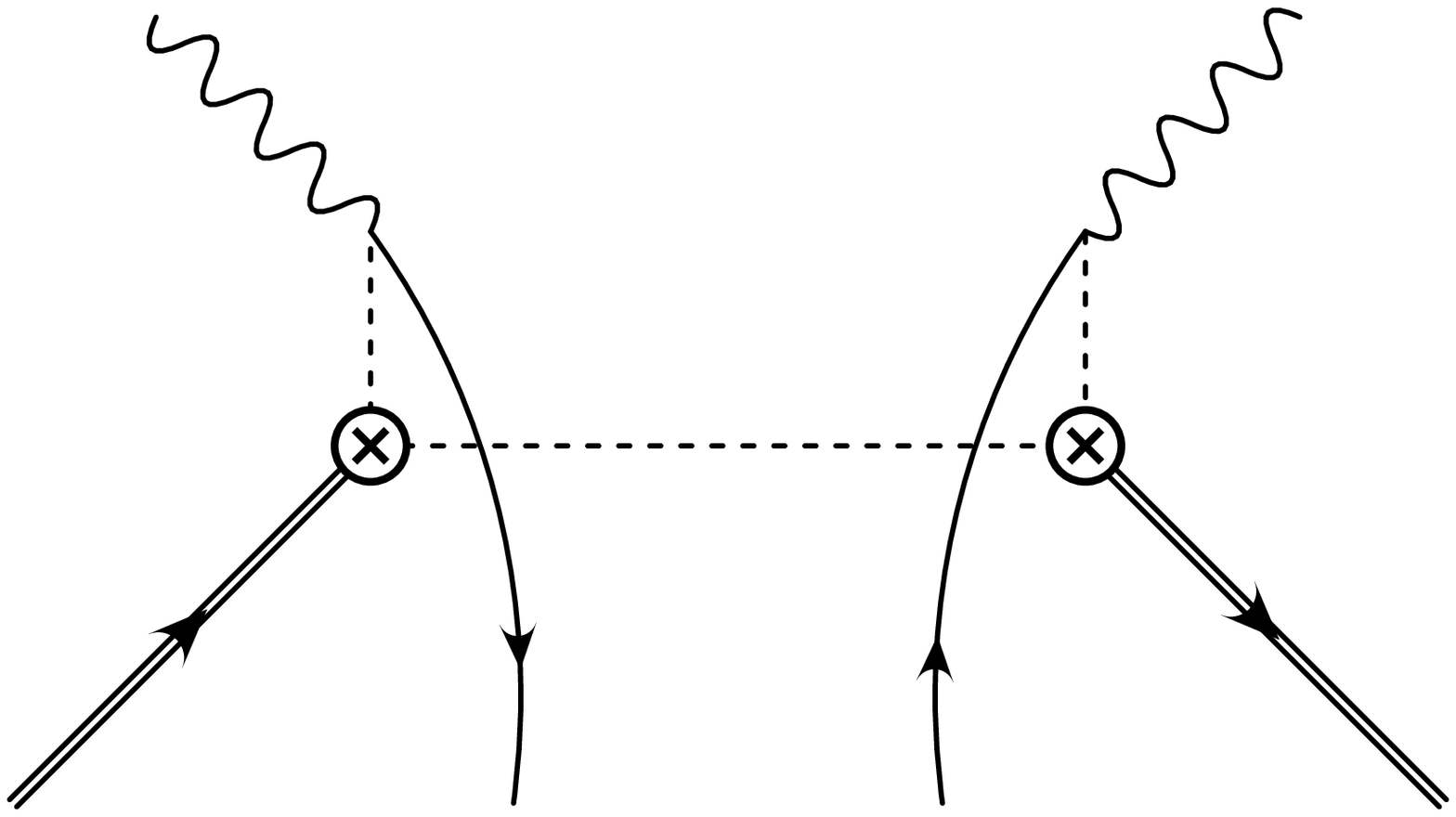,width=6cm}
\end{tabular}
\parbox{15.5cm}{\caption{\label{fig:graphs88} 
Diagrams arising from the matching of the $Q_{8g}-Q_{8g}$ contribution onto SCET (left) and HQET (right). Dashed lines denote non-localities obtained after (anti-)hard-collinear fields have been integrated out.}}
\end{center}
\end{figure}

As we shall see, this contribution is more subtle than the remaining ones, so we will present its calculations in more detail. As mentioned in Section~\ref{subsec:SCETmatching}, the relevant matching relations for the dipole operator $Q_{8g}$ contains three SCET operators: the ${\cal O}(\lambda^3)$ four-particle operator in (\ref{eq:Q8_4p}) containing the photon along with a hard-collinear strange quark and a hard-collinear gluon, and the two leading-order operators in (\ref{eq:Q8_3p}) containing either a hard-collinear strange quark and an anti-hard-collinear gluon, or an anti-hard-collinear strange quark and a hard-collinear gluon (see the second row in Figure~\ref{fig:graphs1}). After the conversion of the anti-hard-collinear partons into a photon plus soft particles only the second operator contributes at ${\cal O}(\lambda^3)$, while the third one receives a stronger power suppression. Note, in particular, that matching $Q_{8g}$ onto an operator containing any soft fields does not give rise to a contribution at order $1/m_b$ in the heavy-quark expansion. It follows that the $Q_{8g}-Q_{8g}$ term receives two contributions: a direct photon contribution from a pair of two SCET operators of the form shown in (\ref{eq:Q8_4p}), and a double resolved photon contribution from a pair of two SCET operators of the form shown in the second line in (\ref{eq:Q8_3p}), followed by the ${\cal O}(\lambda^{1/2})$ conversions of the anti-hard-collinear quark fields into photons. The resulting SCET diagrams are shown in the left panels in Figure~\ref{fig:graphs88}. When the (anti-)hard-collinear fields are integrated out in the second matching step, we obtain the HQET diagrams shown in the right panels. 

The direct photon contribution involves a convolution of the leading shape function in (\ref{Sdef}) with a subleading jet function consisting of the cut of a hard-collinear loop. In the present case the jet function is divergent and needs to be regularized. Using dimensional regularization and subtracting its $1/\bar\epsilon$ pole in the $\overline{\rm MS}$ scheme, we obtain
\begin{equation}\label{F88a}
   F_{88}^{(a)}(E_\gamma,\mu) 
   = \frac{C_F\alpha_s(\mu)}{4\pi}
   \left( \frac{m_b}{2E_\gamma} \right)^2
   \int_{-p_+}^{\bar\Lambda}\!d\omega
   \left( \frac29 \ln\frac{m_b(\omega+p_+)}{\mu^2} + \frac19 
   - \frac49\,c_{\rm RS} \right) S(\omega,\mu) \,. 
\end{equation}
The scheme-dependent constant $c_{\rm RS}$ vanishes in the $\overline{\rm MS}$ scheme, $c_{\overline{\rm MS}}=0$. If, one the other hand, we adopt the dimensional reduction scheme, in which the Dirac algebra is performed in 4 rather than $d=4-2\epsilon$ dimensions, then $c_{\overline{\rm DR}}=1$. We will see later that the final answer for $F_{88}$ is scheme independent. 

The double resolved photon contribution gives rise to a more complicated structure, as the resulting soft matrix element contains four quark fields located at different space-time points. We find
\begin{equation}\label{F88b}
   F_{88}^{(b)}(E_\gamma,\mu) 
   = \frac89\,\pi\alpha_s(\mu) \left( \frac{m_b}{2E_\gamma} \right)^2 
   \int_{-\infty}^{\bar\Lambda}\!d\omega\,\delta(\omega+p_+)
   \int_{-\infty}^\infty \frac{d\omega_1}{\omega_1+i\varepsilon} 
   \int_{-\infty}^\infty \frac{d\omega_2}{\omega_2-i\varepsilon}\,
   g_{88}^{\rm cut}(\omega,\omega_1,\omega_2,\mu) \,,
\end{equation}
where we have defined the subleading shape function
{\small
\begin{eqnarray}\label{g88def}
   g_{88}^{\rm cut}(\omega,\omega_1,\omega_2,\mu) 
   &=& \int\frac{dr}{2\pi}\,e^{-i\omega_1 r}\! 
    \int\frac{du}{2\pi}\,e^{i\omega_2 u}\!
    \int\frac{dt}{2\pi}\,e^{-i\omega t} 
    \sum \hspace{-6.7mm} \int\limits_{~~{\cal X}_s} \nonumber\\
   &&\hspace{-3.3cm} \times 
    \frac{\langle\bar B| \big(\bar h S_n\big)(tn)\,T^A
          \big(S_n^\dagger S_{\bar n}\big)(tn)\,
          \overline{\Gamma}_{\bar n}
          \big(S_{\bar n}^\dagger s\big)(tn+u\bar n)
          |{\cal X}_s\rangle\,\langle {\cal X}_s|
          \big(\bar s S_{\bar n}\big)(r\bar n)
          \Gamma_{\bar n}
          \big(S_{\bar n}^\dagger S_{n}\big)(0)\,T^A 
          \big(S_n^\dagger h\big)(0)
          |\bar B\rangle}{2M_B} \nonumber
\end{eqnarray}
}
\begin{eqnarray}
   &=& \int\frac{dr}{2\pi}\,e^{-i\omega_1 r}\!\int\frac{du}{2\pi}\,e^{i\omega_2 u}\!
    \int\frac{dt}{2\pi}\,e^{-i\omega t} \\
   &&\mbox{}\times 
    \frac{\langle\bar B| \big(\bar h S_n\big)(tn)\,T^A
          \big(S_n^\dagger S_{\bar n}\big)(tn)\,
          \overline{\Gamma}_{\bar n}
          \big(S_{\bar n}^\dagger s\big)(tn+u\bar n)
          \big(\bar s S_{\bar n}\big)(r\bar n)
          \Gamma_{\bar n}
          \big(S_{\bar n}^\dagger S_{n}\big)(0)\,T^A 
          \big(S_n^\dagger h\big)(0)
          |\bar B\rangle}{2M_B} \,. \nonumber
\end{eqnarray}
As before, the Wilson lines render the soft matrix element gauge invariant. The sum over soft intermediate states ${\cal X}_s$ with strangeness $S=-1$ in the first equation arises since in this particular case the hard-collinear jet does not contain the strange quark. Note that only color-octet partonic states contribute to the sum, not physical hadronic ones. Performing the complete sum over states gives rise to the second equation, in which the two strange-quark fields are {\em not\/} time ordered but appear in the order shown in the formula. In the definition above 
\begin{equation}\label{projectors}
   \Gamma_{\bar n} 
   = \frac{\nbslash\nslash}{4}\,(1+\gamma_5) \,, \qquad
   \overline{\Gamma}_{\bar n} 
   = \frac{\nslash\nbslash}{4}\,(1-\gamma_5)
\end{equation}
are projectors onto two-component light-quark spinors. In deriving the result (\ref{F88b}) we have simplified the Dirac structure using the identity \cite{Lange:2003pk}
\begin{equation}\label{Dirac88}
   \gamma_\perp^\alpha\,\frac{\nslash\nbslash}{4}\,\gamma^\mu_\perp
    \otimes \gamma_{\perp\mu}\,\frac{\nbslash\nslash}{4}\,
    \gamma_{\perp\alpha}
   = (d-2)^2\,\frac{\nslash\nbslash}{4}
    \otimes\frac{\nbslash\nslash}{4} \,,
\end{equation}
where $d$ is the number of space-time dimensions. 

According to the discussion of the previous section, it follows that $g_{88}^{\rm cut}$ has support for $-\infty<\omega\le\bar\Lambda$ and $-\infty<\omega_{1,2}<\infty$. Note the difference in the sign of the $i\varepsilon$ terms in the two anti-hard-collinear propagators in (\ref{F88b}), which is due to the fact that the anti-hard-collinear fields connected to the right weak vertex in the diagrams are anti-timed-ordered. Consequently, after convolution with the jet functions the position variables $r$ and $u$ in (\ref{g88def}) are restricted to positive values, such that the fields in the soft $\langle{\cal X}_s|\dots|\bar B\rangle$ matrix elements are time ordered, while those in the $\langle\bar B|\dots|{\cal X}_s\rangle$ matrix elements are anti-time-ordered, as it should be. Finally, we observe that the second equality in (\ref{g88def}) implies the relation
\begin{equation}\label{g88sym}
   \Big[ g_{88}^{\rm cut}(\omega,\omega_1,\omega_2,\mu) \Big]^* 
   = g_{88}^{\rm cut}(\omega,\omega_2,\omega_1,\mu) \,,
\end{equation}
and since the convolution in (\ref{F88b}) is symmetric in $\omega_1$ and $\omega_2$ up to complex conjugation, it follows that the final result is real. Contrary to (\ref{g17norm}) there is no useful normalization condition for the soft function $g_{88}^{\rm cut}$.

For phenomenological purposes, it will be convenient to define a new, real function
\begin{equation}\label{f88def}
   f_{88}(\omega,\mu)
   = \frac29 
   \int_{-\infty}^\infty \frac{d\omega_1}{\omega_1+i\varepsilon} 
   \int_{-\infty}^\infty \frac{d\omega_2}{\omega_2-i\varepsilon}\,
   g_{88}^{\rm cut}(\omega,\omega_1,\omega_2,\mu) \,,
\end{equation}
in terms of which the second contribution to the photon spectrum is simply
\begin{equation}\label{F88bres}
   F_{88}^{(b)}(E_\gamma,\mu) 
   = 4\pi\alpha_s(\mu) 
   \left( \frac{m_b}{2E_\gamma} \right)^2 f_{88}(-p_+,\mu) \,.
\end{equation}
Note that the poles at $\omega_1=0$ and $\omega_2=0$ are regularized by the $i\varepsilon$ prescriptions, in analogy with (\ref{PVreg}).

Let us briefly comment on the structure of our result in light of the general factorization formula (\ref{fact2}). The present case is our only example of a double resolved photon contribution at order $1/m_b$. The jet function for the hard-collinear gluon is given by the cut of the gluon propagator, which up to trivial prefactors yields $J(p^2)=\delta(p^2)$ with $p^2=m_b(\omega+p_+)$, in analogy with the tree-level expression for the quark jet function in (\ref{HQexp}). The jet functions for the two anti-hard-collinear quark propagators are, up to a trivial numerator factor, given by $\bar J(p^2)=1/(p^2+i\varepsilon)$, where $p^2=2E_\gamma\,\omega_{1,2}$ in the present case. Hence, the triple convolution can be recast in the form (omitting scale dependences for brevity)
\begin{eqnarray}
   &&\int d\omega\,\delta(p_+ +\omega)
    \int\frac{d\omega_1}{\omega_1+i\varepsilon} 
    \int\frac{d\omega_2}{\omega_2-i\varepsilon}\,
    g_{88}^{\rm cut}(\omega,\omega_1,\omega_2) \\
   &=& H \int m_b\,d\omega\,J\big(m_b(p_+ +\omega)\big)
    \int 2E_\gamma\,d\omega_1\,\bar J(2E_\gamma\,\omega_1)
    \int 2E_\gamma\,d\omega_2\,\big[\bar J(2E_\gamma\,\omega_2)
    \big]^*\,g_{88}^{\rm cut}(\omega,\omega_1,\omega_2) \,, \nonumber
\end{eqnarray}
in agreement with the factorization formula (\ref{fact2}). The hard matching coefficient $H=1$ at tree level. 

Our presentation above has hidden an important subtlety. The scale dependence should cancel (up to terms of order $\alpha_s^2$) in the sum of the two contributions (\ref{F88a}) and (\ref{F88b}), but in order for this to happen the convolution (\ref{f88def}) must contain a $\mu$-dependent term at zeroth order in the strong coupling. This fact is incompatible with a multiplicative renormalization of SCET operators in the usual (convolution) sense. The resolution of this puzzle is that the convolution integrals over the soft function themselves are not convergent. In order to demonstrate this, we calculate the asymptotic behavior of the soft function for large values $\omega_{1,2}\gg\Lambda_{\rm QCD}$, corresponding to highly energetic light quarks. This behavior can be extracted using short-distance methods \cite{Bosch:2004th}. At leading order in perturbation theory, we simply need to replace the light-quark fields in the definition (\ref{g88def}) by a cut propagator and perform some phase-space integrations. Working in $d=4-2\epsilon$ dimensions, we obtain
\begin{equation}\label{g88pert}
   g_{88}^{\rm cut}(\omega,\omega_1,\omega_2,\mu)
    \Big|_{\omega_{1,2}\gg\Lambda_{\rm QCD}}
   = \frac{C_F}{(4\pi)^{2-\epsilon}}\,
   \frac{\theta(\omega_1)\,\omega_1^{1-\epsilon}}{\Gamma(1-\epsilon)}
   \,\delta(\omega_1-\omega_2)
   \int_\omega^{\bar\Lambda}\!d\omega'\,S(\omega',\mu)
   \left( \omega'-\omega \right)^{-\epsilon} + \dots \,.
\end{equation}
Corrections to this result are suppressed by powers of $\alpha_s$ or $\Lambda_{\rm QCD}/\omega_{1,2}$. The limit $\epsilon\to 0$ is smooth and gives rise to a dependence $g_{88}^{\rm cut}\propto\omega_1\,\delta(\omega_1-\omega_2)$. It is then obvious that the double convolution integral in (\ref{f88def}) is logarithmically divergent in the UV region.\footnote{Note that according to (\ref{g88pert}) there are no UV divergences from the region of large negative values of $\omega_{1,2}$. In this region the convolution integrals are cut off by non-perturbative dynamics.} 
When the convolution is understood in the usual sense as an integral over renormalized functions, then this divergence is not regularized.

On the other hand, our explicit expression (\ref{g88pert}) shows that the convolution integral would be regularized by the dimensional regulator if the limit $\epsilon\to 0$ was taken after the convolutions have been evaluated. In that case we obtain a $1/\epsilon$ pole from the UV-divergent convolution integral, which needs to be subtracted in the $\overline{\rm MS}$ scheme. We thus proceed as follows: we introduce a hard cutoff $\Lambda_{\rm UV}$ and split up the convolution integral in a low-momentum region defined by $\omega_1,\omega_2<\Lambda_{\rm UV}$ and a high-momentum region defined by the complement. In the high-momentum region we can replace the soft function by the perturbative expression (\ref{g88pert}) up to higher-order terms in $\alpha_s$ and power-suppressed contributions. We then evaluate the high-momentum contribution to the double convolution integral before taking the limit $\epsilon\to 0$. In doing so, we must remember to reinstate a factor $(1-\epsilon)^2$ from the Dirac algebra, see (\ref{Dirac88}), and a factor $\mu^{2\epsilon}$ from the conversion of the bare coupling constant $g^2$ into the renormalized coupling $4\pi\alpha_s(\mu)$. In this way, we obtain
\begin{equation}\label{f88reg}
\begin{aligned}
   f_{88}(\omega,\mu)
   &= \frac29 \left[ (1-\epsilon)^2\,\mu^{2\epsilon}\!
    \int_{-\infty}^\infty \frac{d\omega_1}{\omega_1+i\varepsilon} 
    \int_{-\infty}^\infty \frac{d\omega_2}{\omega_2-i\varepsilon}\,
    g_{88}^{\rm cut,bare}(\omega,\omega_1,\omega_2) 
    \right]_{\overline{\rm MS}~{\rm subtracted}} \\
   &= \frac29 \int_{-\infty}^{\Lambda_{\rm UV}}\!
    \frac{d\omega_1}{\omega_1+i\varepsilon} 
    \int_{-\infty}^{\Lambda_{\rm UV}}\!
    \frac{d\omega_2}{\omega_2-i\varepsilon}\,
    g_{88}^{\rm cut}(\omega,\omega_1,\omega_2,\mu) \\
   &\quad\mbox{}- \frac{C_F}{72\pi^2}
    \int_\omega^{\bar\Lambda}\!d\omega'\,S(\omega',\mu) 
    \left( \ln\frac{\Lambda_{\rm UV}(\omega'-\omega)}{\mu^2} 
    + 2 - 2 c_{\rm RS} \right) ,
\end{aligned}
\end{equation}
where $c_{\rm RS}$ is the same scheme-dependent constant as in (\ref{F88a}). This expression is independent of the auxiliary scale $\Lambda_{\rm UV}$, which for consistency should be taken to be several times $\Lambda_{\rm QCD}$, so that perturbation theory can be trusted. In the above result the dependence on the factorization scale of dimensional regularization is explicit, and it is now evident that the sum of the two contributions (\ref{F88a}) and (\ref{F88b}) is both scale and scheme independent. Indeed, we find
\begin{equation}\label{F88fin}
\begin{aligned}
   F_{88}(E_\gamma,\mu) 
   &= \frac{C_F\alpha_s(\mu)}{4\pi} 
    \left( \frac{m_b}{2E_\gamma} \right)^2
    \left( \frac29\,\ln\frac{m_b}{\Lambda_{\rm UV}} - \frac13 \right) 
    \int_{-p_+}^{\bar\Lambda}\!d\omega\,S(\omega,\mu) \\
   &\quad\mbox{}+ \frac89\,\pi\alpha_s(\mu)
    \left( \frac{m_b}{2E_\gamma} \right)^2
    \int_{-\infty}^{\Lambda_{\rm UV}}\!
    \frac{d\omega_1}{\omega_1+i\varepsilon} 
    \int_{-\infty}^{\Lambda_{\rm UV}}\!
    \frac{d\omega_2}{\omega_2-i\varepsilon}\,
    g_{88}^{\rm cut}(-p_+,\omega_1,\omega_2,\mu) \,.
\end{aligned}
\end{equation}
The large logarithm $\ln(m_b/\Lambda_{\rm UV})$ in the first term results from the ratio of the hard-collinear scale $m_b(\omega+p_+)$ in (\ref{F88a}) and the soft (yet perturbative) scale $\Lambda_{\rm UV}(\omega+p_+)$ contained in the function $f_{88}(-p_+,\mu)$ in (\ref{F88bres}). Resumming these large logarithms would require solving evolution equations in the effective theory. In the case of UV-divergent convolution integrals, the derivation of such equations is an open problem. 

Note that our result (\ref{F88fin}) is insensitive to the mass of the strange quark, as it should be. This is in contrast with the parton-model expression derived in \cite{Ali:1995bi} and shown in (\ref{Fijpart}). The IR regulator $m_s$ introduced in the parton-model calculation is replaced in real QCD by a subleading shape function, i.e., by a hadronic matrix element of a non-local operator. We have checked that one recovers the parton-model expression for $F_{88}$ if one calculates the soft matrix element in perturbation theory, i.e., if one assumes the validity of (\ref{g88pert}) also at small values of $\omega_{1,2}$ and introduces $m_s$ as an IR regulator, which replaces $\omega_1^{-\epsilon}(\omega'-\omega)^{-\epsilon}\to[\omega_1(\omega'-\omega)-m_s^2]^{-\epsilon}$ in this formula. Of course, such a treatment cannot be justified due to the non-perturbative nature of QCD at low energies.

It was argued in \cite{Kapustin:1995fk} that the IR-sensitive terms in the $Q_{8g}-Q_{8g}$ contribution to the \bsg\ photon spectrum can be absorbed into photon fragmentation functions of a strange quark or gluon. While no formal proof of this assertion was given in that paper, it is likely to be true in the kinematic region away from the endpoint, where the splitting processes $s\to\gamma+s$ and $g\to\gamma+g$ can be treated using the collinear approximation. In the language of SCET this means that the partons after the splitting are still anti-hard-collinear fields, and hence the photon energy cannot be near the endpoint. In the endpoint region, on the other hand, these partons are soft, and they do not factorize from the remaining soft matrix element. Hence, in this region the non-perturbative physics is encoded in a complicated subleading four-quark shape function rather than a simpler fragmentation function. 

\subsection{Analysis of the $\bm{Q_{7\gamma}-Q_{8g}}$ contribution}
\label{sec:Q7Q8}

Evaluating the diagrams in Figure~\ref{fig:direct} for this operator pair, we find the direct photon contribution
\begin{equation}
   F_{78}^{(a)}(E_\gamma,\mu) 
   = \frac{C_F\alpha_s(\mu)}{4\pi}\,\frac{m_b}{2E_\gamma}\,
    \frac{10}{3} \int_{-p_+}^{\bar\Lambda}\!d\omega\,S(\omega,\mu) \,,
\end{equation}
which generalizes the parton-model result in (\ref{Fijpart}). The case of the $Q_{7\gamma}-Q_{8g}$ interference term is special in that, even though the parton-model expression for $F_{78}^{\rm part}(E_\gamma,\mu)$ does not indicate any problematic feature that would call for non-trivial soft contributions, there actually do exist some ${\cal O}(1/m_b)$ effects that are described by subleading shape functions. Moreover, these effects remain non-local even for the total decay rate \cite{Lee:2006wn}. 

\begin{figure}
\begin{center}
\epsfig{file=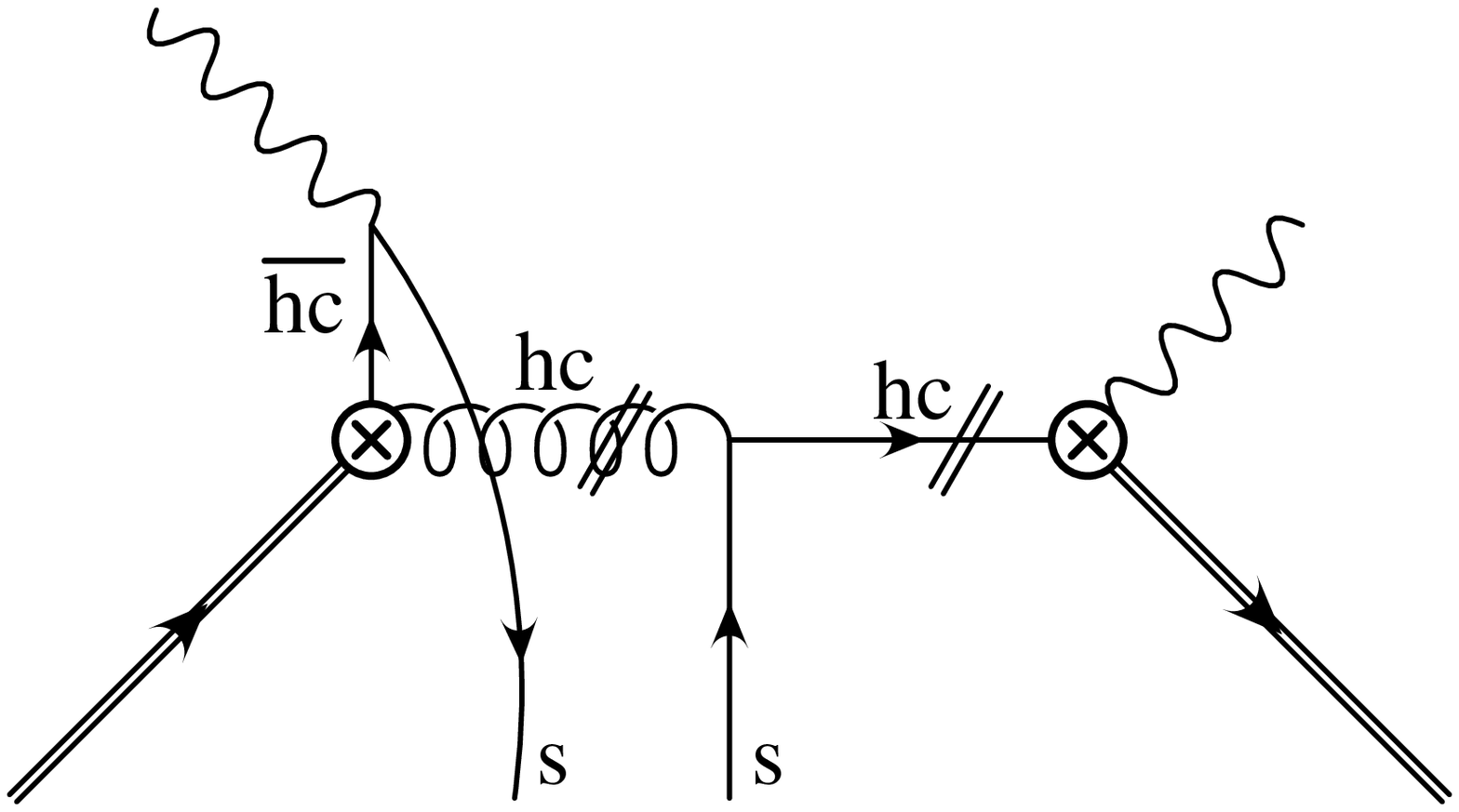,width=6cm} \hspace{2cm}
\epsfig{file=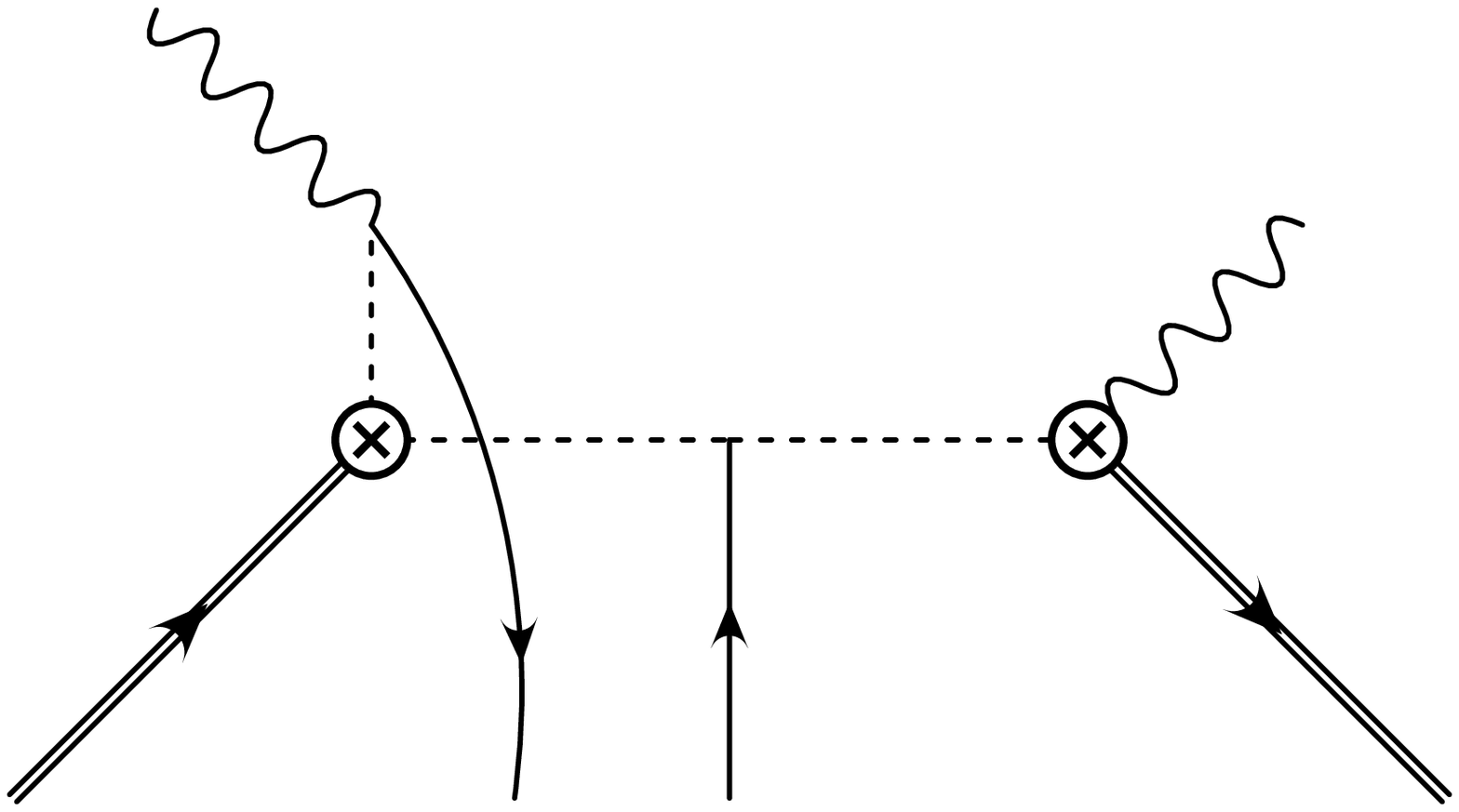,width=6cm} 
\epsfig{file=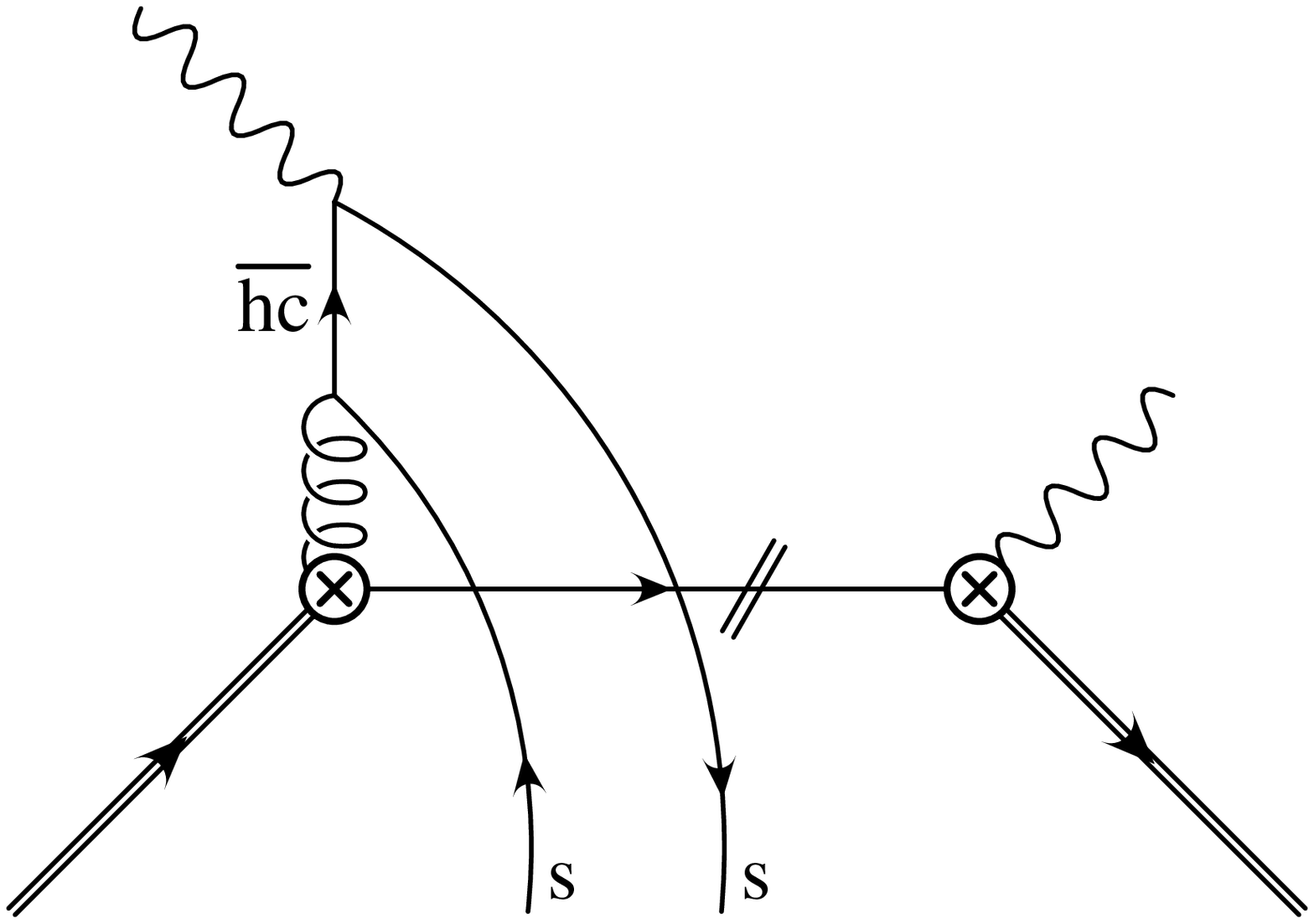,width=6cm} \hspace{2cm}
\epsfig{file=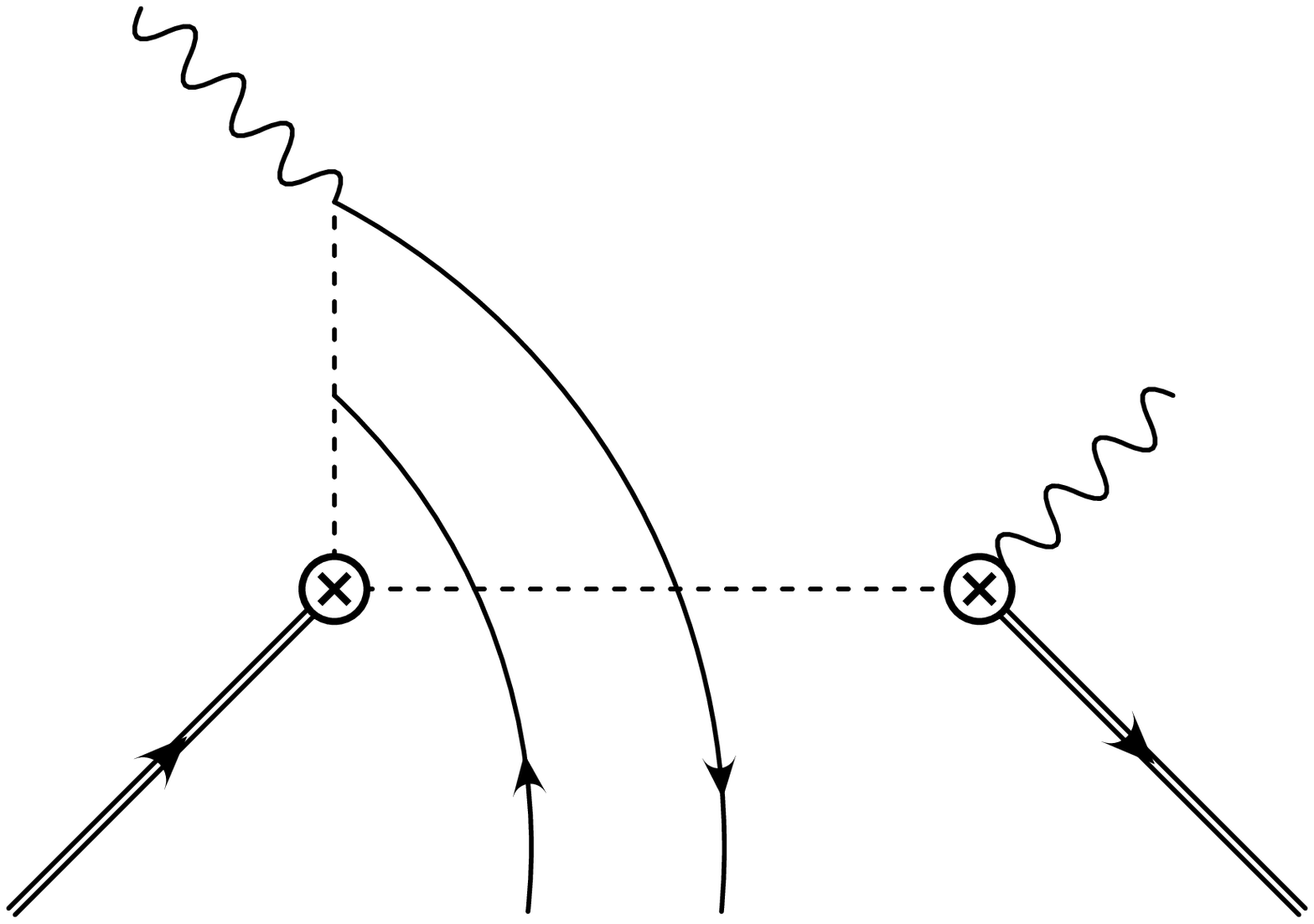,width=6cm} 
\parbox{15.5cm}{\caption{\label{fig:Q7Q8b}
Diagrams arising from the matching of the $Q_{7\gamma}-Q_{8g}$ contribution onto SCET (left) and HQET (right). Dashed lines denote non-localities obtained after (anti-)hard-collinear fields have been integrated out.}}
\end{center}
\end{figure}

In order to study the resolved photon contributions, we must combine either one of the two SCET operators in (\ref{eq:Q8_3p}) arising from the matching relation for $Q_{8g}$ with the leading-order operator in (\ref{eq:Q7}) descending from $Q_{7\gamma}$ (see also Figure~\ref{fig:graphs1}). In both cases, the conversion of the anti-hard-collinear fields gives rise to one or more soft quark fields. The relevant SCET diagrams are depicted in the left panels in Figure~\ref{fig:Q7Q8b}, while the corresponding soft graphs resulting after the second matching step are shown in the right panels. In the first case, the second soft quark is generated by an insertion of a subleading term in the SCET Lagrangian.

Evaluating the first contribution in detail, we find 
\begin{equation}\label{F78c}
\begin{aligned}
   F_{78}^{(b)}(E_\gamma,\mu) 
   &= \frac{16}{3}\,\pi\alpha_s(\mu)\,\frac{m_b}{2E_\gamma}\,
    \mbox{Re} \int_{-\infty}^{\bar\Lambda}\!d\omega\,
    \delta(\omega+p_+) 
    \int_{-\infty}^{\infty} \frac{d\omega_1}{\omega_1+i\varepsilon} 
    \int_{-\infty}^{\infty} \frac{d\omega_2}{\omega_2-i\varepsilon} \\
   &\quad\times \left[ \bar g_{78}(\omega,\omega_1,\omega_2,\mu)
    - \bar g_{78}^{\rm cut}(\omega,\omega_1,\omega_2,\mu) \right] .
\end{aligned}
\end{equation}
The soft function $\bar g_{78}$ arises when the hard-collinear strange-quark line in the left diagram in the first row of Figure~\ref{fig:Q7Q8b} is cut, while the function $\bar g_{78}^{\rm cut}$ originates from the cut through the hard-collinear gluon line. In this latter case the soft strange-quark line must also be cut. Specifically, we define the corresponding subleading shape functions as
\begin{equation}\label{g78bardef}
\begin{aligned}
   \bar g_{78}(\omega,\omega_1,\omega_2,\mu) 
   &= \int\frac{dr}{2\pi}\,e^{-i\omega_1 r}\!\int\frac{du}{2\pi}\,e^{i\omega_2 u}\!
    \int\frac{dt}{2\pi}\,e^{-i\omega t} \\
   &\hspace{-2.0cm}\times 
    \frac{\langle\bar B| \big(\bar h S_n\big)(tn)\,T^A\,
          \overline{\Gamma}_n\,\big(S_n^\dagger s\big)(un)
          \big(\bar s S_{\bar n}\big)(r\bar n)\,\Gamma_{\bar n}\,
          \big(S_{\bar n}^\dagger S_{n}\big)(0)\,T^A          
          \big(S_n^\dagger h\big)(0) |\bar B\rangle}{2M_B} \,, 
\end{aligned}
\end{equation}
and
\begin{eqnarray}\label{g78barcutdef}
   &&\hspace{-0.5cm} 
   \bar g_{78}^{\rm cut}(\omega,\omega_1,\omega_2,\mu) 
   = \int\frac{dr}{2\pi}\,e^{-i\omega_1 r}\!\int\frac{du}{2\pi}\,e^{i\omega_2 u}\!
    \int\frac{dt}{2\pi}\,e^{-i\omega t} 
    \sum \hspace{-6.7mm} \int\limits_{~~{\cal X}_s} \nonumber\\
   &&\times 
    \frac{\langle\bar B| \big(\bar h S_n\big)(tn)\,T^A\,
          \overline{\Gamma}_n\,\big(S_n^\dagger s\big)((t+u)n)
          |{\cal X}_s\rangle\,\langle {\cal X}_s|
          \big(\bar s S_{\bar n}\big)(r\bar n)\,\Gamma_{\bar n}\,
          \big(S_{\bar n}^\dagger S_{n}\big)(0)\,T^A          
          \big(S_n^\dagger h\big)(0) |\bar B\rangle}{2M_B} 
    \nonumber\\
   && \hspace{2.5cm} = \int\frac{dr}{2\pi}\,e^{-i\omega_1 r}\!
    \int\frac{du}{2\pi}\,e^{i\omega_2 u}\!
    \int\frac{dt}{2\pi}\,e^{-i\omega t} \\
   && \times 
    \frac{\langle\bar B| \big(\bar h S_n\big)(tn)\,T^A\,
          \overline{\Gamma}_n\,\big(S_n^\dagger s\big)((t+u)n)
          \big(\bar s S_{\bar n}\big)(r\bar n)\,\Gamma_{\bar n}\,
          \big(S_{\bar n}^\dagger S_{n}\big)(0)\,T^A          
          \big(S_n^\dagger h\big)(0) |\bar B\rangle}{2M_B} \,,
    \nonumber 
\end{eqnarray}
where $\Gamma_{\bar n}$ was introduced in (\ref{projectors}), and $\Gamma_n$ is defined in the same way as $\Gamma_{\bar n}$ but with $n$ and $\bar n$ interchanged. One half of the contribution shown in (\ref{F78c}), but without the real part prescription, arises from the original diagrams, while the mirror diagrams not shown in the figure give the complex conjugate of the above expressions. The two results combined give a real result, as indicated in (\ref{F78c}). Note that there is no need to insert a time-ordering symbol in front of the light-quark fields in the definition of $\bar g_{78}$ in (\ref{g78bardef}), since after convolution with the jet functions the integration variables $r$ and $u$ are restricted to take positive values, and hence the light-quark fields have a space-like separation. In the second equation in (\ref{g78barcutdef}), on the other hand, the fields are not time ordered because the non-local four-fermion operator arises upon performing a sum over intermediate states, as shown in the first equation.

Consider next the contribution shown in the second row of Figure~\ref{fig:Q7Q8b}. In this case the soft light-quark pair can carry any flavor. We obtain
\begin{eqnarray}\label{F78b}
   F_{78}^{(c)}(E_\gamma,\mu) 
   &=& 4\pi\alpha_s(\mu)
    \,\frac{m_b}{2E_\gamma}\,\mbox{Re} 
    \int_{-\infty}^{\bar\Lambda}\!d\omega\,\delta(\omega+p_+)
    \int_{-\infty}^\infty\!d\omega_1 
    \int_{-\infty}^\infty\!d\omega_2\,
    \frac{1}{\omega_1-\omega_2+i\varepsilon} \\
   &&\hspace{-1.2cm}
    \times \left[ \left( \frac{1}{\omega_1+i\varepsilon}
    + \frac{1}{\omega_2-i\varepsilon} \right) 
    g_{78}^{(1)}(\omega,\omega_1,\omega_2,\mu)
    - \left( \frac{1}{\omega_1+i\varepsilon}
    - \frac{1}{\omega_2-i\varepsilon} \right) 
    g_{78}^{(5)}(\omega,\omega_1,\omega_2,\mu) \right] , \nonumber
\end{eqnarray}
where we have defined the subleading shape functions
\begin{eqnarray}\label{g78def}
   &&\hspace{-0.3cm} g_{78}^{(1)}(\omega,\omega_1,\omega_2,\mu) 
   = \int\frac{dr}{2\pi}\,e^{-i\omega_1 r}\!
    \int\frac{du}{2\pi}\,e^{i\omega_2 u}\!
    \int\frac{dt}{2\pi}\,e^{-i\omega t} \nonumber\\
   &&\times 
    \frac{\langle\bar B| \big(\bar h S_n\big)(tn) 
          \big(S_n^\dagger S_{\bar n}\big)(0)\,T^A\,
          \nbslash (1+\gamma_5)\,
          \big(S_{\bar n}^\dagger h\big)(0)\,{\bf T} 
          \sum{}_q\,e_q\,\big(\bar q S_{\bar n}\big)(r\bar n)\,
          \nbslash\,T^A
          \big(S_{\bar n}^\dagger q\big)(u\bar n)
          |\bar B\rangle}{2M_B} \,, \nonumber\\
   &&\hspace{-0.3cm} g_{78}^{(5)}(\omega,\omega_1,\omega_2,\mu) 
   = \int\frac{dr}{2\pi}\,e^{-i\omega_1 r}\!
    \int\frac{du}{2\pi}\,e^{i\omega_2 u}\!
    \int\frac{dt}{2\pi}\,e^{-i\omega t} \\
   &&\times 
    \frac{\langle\bar B| \big(\bar h S_n\big)(tn) 
          \big(S_n^\dagger S_{\bar n}\big)(0)\,T^A\,
          \nbslash (1+\gamma_5)\,
          \big(S_{\bar n}^\dagger h\big)(0)\,{\bf T}
          \sum{}_q\,e_q\,\big(\bar q S_{\bar n}\big)(r\bar n)\,
          \nbslash\gamma_5\,T^A
          \big(S_{\bar n}^\dagger q\big)(u\bar n)
          |\bar B\rangle}{2M_B} \,, \nonumber
\end{eqnarray}
where the sum extends over light quark flavors ($q=u,d,s$), and $e_q$ denote the quark electric charges in units of $e$. One half of the contribution shown in (\ref{F78b}), but without the real part prescription, arises from the original diagrams, while the mirror diagrams not shown in the figure give the complex conjugate of the above expressions. 

In these definitions the light-quark fields are time-ordered, as indicated by the {\bf T} symbols. That this is the appropriate ordering can be seen as follows. After convolution with the jet functions, for the terms containing the propagator $1/(\omega_1+i\varepsilon)$ in the second line of (\ref{F78b}) the integration variables $r$ and $u$ are restricted to the range $r>u>0$. These terms correspond to the Feynman graph shown on the left in the second row of Figure~\ref{fig:Q7Q8b}, in which $\bar q(r\bar n)$ should appear to the left of $q(u\bar n)$. For the terms containing the propagator $1/(\omega_2-i\varepsilon)$ the integration variables are restricted to the range $u>r>0$. These terms correspond to the analogous Feynman graph with the opposite direction of the fermion arrow on the light-quark line, for which $\bar q(r\bar n)$ should appear to the right of $q(u\bar n)$. Hence, the proper ordering is indeed the ordering according to (light-cone) time. On the other hand, arguments along the lines discussed in \cite{Jaffe:1983hp} suggest that the time-ordering prescription is, in fact, not required for forward matrix elements and fields at light-like separation. We assume in what follows that the {\bf T} symbol can be dropped in (\ref{g78def}).

Very little is known about the complicated four-quark shape-functions defined in (\ref{g78bardef}), (\ref{g78barcutdef}), and (\ref{g78def}). Following the general arguments presented in Section~\ref{subsec:42}, we conclude that the soft functions $\bar g_{78}$ and $g_{78}^{(1,5)}$ have support for $-\infty<\omega\le\bar\Lambda$ and $-\infty<\omega_{1,2}<\infty$. However, in the case of $\bar g_{78}^{\rm cut}$ we must require that $\omega_{1,2}>0$. Note also the symmetry property
\begin{equation}\label{g78sym}
   \int_{-\infty}^{\bar\Lambda}\!d\omega 
    \left[ g_{78}^{(1,5)}(\omega,\omega_1,\omega_2,\mu) \right]^*
   = \int_{-\infty}^{\bar\Lambda}\!d\omega\,
    g_{78}^{(1,5)}(\omega,\omega_2,\omega_1,\mu) \,, 
\end{equation}
which follows from the definitions of the soft functions in (\ref{g78def}). 

The fact that in the case of $g_{78}^{(1,5)}$ the operators involve light quarks of all flavors offers a strategy for modeling their matrix elements between $B$-meson states. Unlike the case of the four-quark operators encountered for $g_{88}^{\rm cut}$, $\bar g_{78}$, and $\bar g_{78}^{\rm cut}$, here it is possible to (very roughly) estimate the matrix element by inserting the vacuum intermediate state between the two light-quark fields. The ``vacuum-insertion approximation'' (VIA) is used extensively in the study of local four-quark operator matrix elements, and we see no reason why it should work less accurately for non-local operators. Following \cite{Lee:2006wn}, we then obtain 
\begin{equation}\label{eq:78}
   \int_{-\infty}^{\bar\Lambda}\!d\omega\,
    g_{78}^{(1,5)}(\omega,\omega_1,\omega_2,\mu) \Big|_{\rm VIA}
   = -e_{\rm spec}\,\frac{F^2(\mu)}{8} \left( 1 - \frac{1}{N_c^2} \right)
    \phi_+^B(-\omega_1,\mu)\,\phi_+^B(-\omega_2,\mu) \,,
\end{equation}
where $e_{\rm spec}$ denotes the charge of the spectator quark inside the $B$ meson, i.e., $e_{\rm spec}=2/3$ for $B^\pm$, and $e_{\rm spec}=-1/3$ for $B^0$ and $\bar B^0$. The quantity $F(\mu)$ is the HQET matrix element corresponding to the asymptotic value of the product $f_B\sqrt{M_B}$ in the heavy-quark limit \cite{Neubert:1991sp}. Finally, $\phi_+^B(\omega,\mu)$ is the leading light-cone distribution amplitude of the $B$ meson \cite{Grozin:1996pq}. It is a real function with support for $\omega>0$, which vanishes at $\omega=0$ and asymptotically falls off like $1/\omega$ modulo logarithms \cite{Lee:2005gza}. Useful forms for this function have been derived based on QCD sum rules \cite{Grozin:1996pq,Ball:2003fq,Braun:2003wx}, the relativistic quark model \cite{Le Yaouanc:2007mz}, and model-independent moment relations obtained using the operator-product expansion \cite{Lee:2005gza,Kawamura:2010tj}. The support of $\phi_+^B(\omega,\mu)$ implies that only negative values of $\omega_1$ and $\omega_2$ give rise to non-zero contributions in (\ref{eq:78}), which is in accordance with the fact that $\bar q$ ($q$) describes an anti-quark in the initial (final) state.

To conclude this analysis, we define the phenomenological functions 
\begin{eqnarray}
   f_{78}^{\rm (I)}(\omega,\mu) 
   &=& \frac43 
    \int_{-\infty}^{\infty} \frac{d\omega_1}{\omega_1+i\varepsilon} 
    \int_{-\infty}^{\infty} \frac{d\omega_2}{\omega_2-i\varepsilon} 
    \left[ \bar g_{78}(\omega,\omega_1,\omega_2,\mu)
    - \bar g_{78}^{\rm cut}(\omega,\omega_1,\omega_2,\mu) \right] , 
    \nonumber\\
   f_{78}^{\rm (II)}(\omega,\mu) 
   &=& \int_{-\infty}^\infty\!d\omega_1 
    \int_{-\infty}^\infty\!d\omega_2\,
    \frac{1}{\omega_1-\omega_2+i\varepsilon} \\
   &&\hspace{-1.0cm}
    \times \left[ \left( \frac{1}{\omega_1+i\varepsilon}
    + \frac{1}{\omega_2-i\varepsilon} \right) 
    g_{78}^{(1)}(\omega,\omega_1,\omega_2,\mu)
    - \left( \frac{1}{\omega_1+i\varepsilon}
    - \frac{1}{\omega_2-i\varepsilon} \right) 
    g_{78}^{(5)}(\omega,\omega_1,\omega_2,\mu) \right] . \nonumber
\end{eqnarray}
In the VIA, we obtain
\begin{eqnarray}\label{VIA4ever}
   \int_{-\infty}^{\bar\Lambda}\!d\omega\,
    f_{78}^{\rm (II)}(\omega,\mu) \Big|_{\rm VIA}
   &=& -e_{\rm spec}\,\frac{F^2(\mu)}{8} \left( 1 - \frac{1}{N_c^2} \right) 
    \left\{ \frac{1}{\lambda_B^2(\mu)}  
    + 2\pi i \int_0^\infty\!d\omega\,
    \frac{\left[ \phi_+^B(\omega,\mu) \right]^2}{\omega} \right\} , \nonumber\\
   \int_{-\infty}^{\bar\Lambda}\!d\omega\,\mbox{Re}\,
    f_{78}^{\rm (II)}(\omega,\mu) \Big|_{\rm VIA}
   &=& -e_{\rm spec}\,\frac{F^2(\mu)}{8} \left( 1 - \frac{1}{N_c^2} \right)
    \frac{1}{\lambda_B^2(\mu)} \,,
\end{eqnarray}
where $\lambda_B=\int_0^\infty\!d\omega\,\phi_+^B(\omega,\mu)/\omega$ denotes the first inverse moment of the $B$-meson light-cone distribution amplitude \cite{Beneke:2001ev}. In terms of the functions $f_{78}^{\rm (I,II)}$, the direct and resolved photon contributions to the \bsg\ photon spectrum can be summarized as
\begin{equation}\label{F78final}
\begin{split}
   F_{78}(E_\gamma,\mu) 
   &= \frac{C_F\alpha_s(\mu)}{4\pi}\,\frac{m_b}{2E_\gamma}\,
    \frac{10}{3} \int_{-p_+}^{\bar\Lambda}\!d\omega\,S(\omega,\mu) \\
   &\quad\mbox{}+ 4\pi\alpha_s(\mu)\,\frac{m_b}{2E_\gamma}\,
    \mbox{Re} \left[ f_{78}^{\rm (I)}(-p_+,\mu) 
    + f_{78}^{\rm (II)}(-p_+,\mu) \right] .
\end{split}
\end{equation}

\section{Constraints from $\bm{P}\bm{T}$ invariance}
\label{sec:PT}

In the expressions presented in the previous section there are four potential sources of complex phases: weak (CP-violating) phases from the CKM matrix elements and the Wilson coefficients, and strong (CP-conserving) phases from the new jet functions $\bar J_i$ and the various subleading shape functions. The CKM phases are non-zero only for the $Q_1^q-Q_{7\gamma}$ contributions (with $q=c,u$), where they are suppressed by two powers of the Cabbibo angle. The Wilson coefficients are real in the Standard Model, although they can be complex in many of its extensions (see e.g.\ \cite{Kagan:1998bh}).

A unique property of the resolved photon contribution is that the new jet functions $\bar J_i$ are given in terms of full propagators (dressed by Wilson lines) and not by cut propagators. As a result, these functions are in general complex and give rise to strong phases. Since the relevant scale of the jet functions is $\sqrt{2E_\gamma\,\Lambda_{\rm QCD}}$, which is perturbative in the endpoint region, these strong phases are calculable in perturbation theory. The other potential source of strong phases are the soft functions, whose phases are of a non-perturbative nature. In studying the effects of the resolved photon contributions on the rate and $CP$ asymmetry in $B\to X_s\gamma$ decay, it is important to obtain some handle on these non-perturbative phases. In order to do so, we employ the invariance of strong-interaction matrix elements under parity ($P$) and time reversal ($T$).

Under the combined transformation $PT$, a spinor field $\psi(x)$ transforms as $PT\,\psi(x)\,PT=\Lambda_{PT}\,\psi(-x)$, where in the Weyl representation of the Dirac matrices $\Lambda_{PT}=-\gamma^0\gamma^1\gamma^3$ up to an irrelevant phase factor. The soft Wilson line $S_n(x)$ in (\ref{Sdef}) transforms into $S_n(-x)$.\footnote{Strictly speaking the lower limit of integration is also changed from $-\infty$ to $+\infty$, but this provides an equally valid definition of the same object.} 
Finite-length Wilson lines, as they appear in the definitions of the soft functions, transform as $PT\,[tn,0]\,PT=PT\,S_n(tn)\,S_n^\dagger(0)\,PT=S_n(-tn)\,S_n^\dagger(0)=[-tn,0]$. Finally, the external $B$-meson states transform as $PT\,|\bar B(v)\rangle=-|\bar B(v)\rangle$. Also, because the time-reversal transformation is anti-linear, matrix elements get complex conjugated under application of $PT$. Consider now the definition of the soft function $g_{17}$ in (\ref{g17def}). Using the fact that the position-space strong-interaction matrix element is $PT$ invariant, we find that 
\begin{eqnarray}\label{g17PT}
   g_{17}(\omega,\omega_1,\mu) 
   &=& \int\frac{dr}{2\pi}\,e^{-i\omega_1 r}\!
    \int\frac{dt}{2\pi}\,e^{-i\omega t} \\
   &&\times \frac{\langle\bar B| \big(\bar h S_n\big)(-tn)\,
    \nbslash (1-\gamma_5) \big(S_n^\dagger S_{\bar n}\big)(0)\,
    i\gamma_\alpha^\perp\bar n_\beta\,
    \big(S_{\bar n}^\dagger\,g G_s^{\alpha\beta} S_{\bar n} 
    \big)(-r\bar n)\,
    \big(S_{\bar n}^\dagger h\big)(0) |\bar B\rangle^*}{2M_B} \,,
    \nonumber
\end{eqnarray}
where we have used that $\Lambda_{PT}^\dagger\,\nbslash\,i\gamma_\alpha^\perp\,\Lambda_{PT}=\nbslash\,i\gamma_\alpha^\perp$ and $\Lambda_{PT}^\dagger\,\nbslash\gamma_5\,i\gamma_\alpha^\perp\,\Lambda_{PT}=-\nbslash\gamma_5\,i\gamma_\alpha^\perp$. However, we have already argued after (\ref{traceformalism}) that the term containing $\gamma_5$ vanishes. Hence, $PT$ transforms the position-space matrix element into the complex conjugate of the same matrix element with all position arguments $x_i$ replaced by $-x_i$. By taking the complex conjugate of relation (\ref{g17PT}) and reversing the sign of the integration variables $r$ and $t$, it then follows that $g_{17}(\omega,\omega_1,\mu)$ is a real function.

An analogous argument can be presented for the soft functions $\bar g_{78}$ in (\ref{g78bardef}) and $g_{78}^{(1,5)}$ in (\ref{g78def}), where it is important however that we can avoid the time-ordering prescription for the soft light-quark fields. The HQET trace formalism can be used to show that in the definitions of these matrix elements only even numbers of $\gamma_5$ matrices can give rise to non-vanishing contributions, and then the Dirac structures are in all cases even under $PT$. More specifically, in analogy to (\ref{traceformalism}) we can write
\begin{equation}\label{tracegg}
\begin{split}
   \bar g_{78}(\omega,\omega_1,\omega_2,\mu)
   &= \mbox{\rm Tr} \left[ \frac{1+\vslash}{2}\,\Gamma_A\,
    \Xi_1(v,\bar n)\,\Gamma_B\,\frac{1+\vslash}{2}\,
    \Xi_2(v,\bar n) \right] \\
   &\quad\mbox{}+ 
    \mbox{\rm Tr} \left[ \gamma_5\,\frac{1+\vslash}{2}\,
    \Gamma_A\,\Xi_3(v,\bar n) \right]\,
    \mbox{\rm Tr} \left[ \Xi_4(v,\bar n)\,\Gamma_B\,
    \frac{1+\vslash}{2}\,\gamma_5 \right] ,
\end{split}
\end{equation}
where $\Gamma_A=\overline{\Gamma}_n$ and $\Gamma_B=\Gamma_{\bar n}$, and for brevity we have suppressed the dependence of the coefficient functions $\Xi_i$ on $\omega$, $\omega_1$, $\omega_2$, and $\mu$. A similar expression, but with different matrices $\Gamma_A$ and $\Gamma_B$, holds for $g_{78}^{(1,5)}$ after a Fierz transformation. The most general Lorentz-invariant decompositions of the functions $\Xi_i$ involve products of up to four $\vslash$, $\nbslash$, and $\gamma_\perp^\alpha$ matrices, where all transverse indices must be contracted. No $\gamma_5$ matrices appear in this decomposition. Note also that the relation $n+\bar n=2v$ allows us to eliminate $\nslash$ in favor of $\vslash$ and $\nbslash$. With only two independent external vectors $v$ and $\bar n$, however, it is impossible to saturate the four indices of an $\epsilon_{\alpha\beta\gamma\delta}$ symbol, and hence only even numbers of $\gamma_5$ matrices in the product structure $\Gamma_A\otimes\Gamma_B$ can give rise to non-zero traces. For the case of $\bar g_{78}$ considered above, it follows that we can replace $16\,\overline{\Gamma}_n\otimes\Gamma_{\bar n}\to\nbslash\nslash\otimes\nbslash\nslash-\nbslash\nslash\gamma_5\otimes\nbslash\nslash\gamma_5$. Both of these product structures are even under $PT$. In the case of $g_{78}^{(1)}$ and $g_{78}^{(5)}$, we find similarly that $\nbslash(1+\gamma_5)\otimes\nbslash\to\nbslash\otimes\nbslash$ and $\nbslash(1+\gamma_5)\otimes\nbslash\gamma_5\to\nbslash\gamma_5\otimes\nbslash\gamma_5$. Once again, these product structures are even under $PT$. It follows that the functions $\bar g_{78}$, $g_{78}^{(1)}$, and $g_{78}^{(5)}$ are all real.

Let us finally consider the functions $\bar g_{88}^{\rm cut}$ in (\ref{g88def}) and $\bar g_{78}^{\rm cut}$ in (\ref{g78barcutdef}), which are defined in terms of sums over intermediate states $|{\cal X}_s\rangle$, which without loss of generality can be chosen to be eigenstates of $PT$ with eigenvalues $\pm 1$. After summing over the polarizations of the intermediate states and integrating over their momenta, we find that each term in the sum over states can be written as a product of two traces, in analogy to the second term in (\ref{tracegg}). The same arguments as above then show that $\bar g_{88}^{\rm cut}$ and $\bar g_{78}^{\rm cut}$ are real.

In conclusion, we find that all of the subleading shape functions are real. The strong phases mentioned in the introduction to this section thus arise only from the new jet functions $\bar J_i$. Given that the soft functions are real, it now follows from (\ref{g17int}), (\ref{g88sym}), and (\ref{g78sym}) that $\int d\omega\,g_{17}(\omega,\omega_1,\mu)$ is an even function of $\omega_1$, and that $g_{88}^{\rm cut}(\omega,\omega_1,\omega_2,\mu)$ and $\int d\omega\,g_{78}^{(1,5)}(\omega,\omega_1,\omega_2,\mu)$ are symmetric under the exchange of $\omega_1$ and $\omega_2$.

\section{Partially integrated decay rate}
\label{sec:Gtot}

In the various parts of Section~\ref{sec:matching}, we have derived explicit expressions for the direct and resolved photon contributions to the coefficient functions $F_{ij}(E_\gamma,\mu)$ entering the master formula for the \bsg\ photon spectrum in (\ref{HQexp}). The results are given in (\ref{F77final}), (\ref{F17final}), (\ref{F11}), (\ref{F88fin}), and (\ref{F78final}). For phenomenological purposes, it is most interesting to study the partial \bsg\ decay rate 
\begin{equation}\label{eqn:partialDecayRate}
   \Gamma(E_0) \equiv \int_{E_0}^{M_B/2}\!dE_\gamma\,\frac{d\Gamma}{dE_\gamma} \,, 
\end{equation}
obtained by integrating the photon spectrum over a region $E_0<E_\gamma<M_B/2$. Provided that $\Delta\equiv m_b-2E_0$ is much larger than $\Lambda_{\rm QCD}$, the direct photon contributions to this integrated rate can be calculated in terms of local operator matrix elements \cite{Neubert:2004dd} using a combined expansion in powers of $\Delta/m_b$ and $\Lambda_{\rm QCD}/\Delta$. In the limit $E_0\to 0$ one obtains the total decay rate, and $\Delta=m_b$; however, the rates measured experimentally are obtained with values of $E_0$ larger than 1.7\,GeV, so that $\Delta<1.25$\,GeV.

An important feature of the resolved photon contributions studied in this work is that they do not reduce to local operator matrix element in the limit $\Delta\gg\Lambda_{\rm QCD}$. Rather, the corresponding contributions to the integrated decay rate must still be described in terms of matrix elements of non-local operators. This implies that the corresponding theoretical uncertainties do not reduce significantly as the cutoff $E_0$ is taken out of the endpoint region. We will now illustrate this by deriving expressions for the first-order power corrections to the integrated decay rate
\begin{equation}\label{Gtot}
\begin{split}
   \Gamma(E_0) &= \frac{G_F^2\alpha|V_{tb} V_{ts}^*|^2}{32\pi^4}\,
    \overline{m}_b^2(\mu)\,m_b^3\,\Bigg[\,
    |H_\gamma(\mu)|^2 \left[ 1 + {\cal O}(\alpha_s) \right] \\
   &\hspace{4.9cm}\mbox{}+ \frac{1}{m_b}\,\sum_{i\le j}\,
    \mbox{Re}\big[C_i^*(\mu)\,C_j(\mu)\big]\,\bar F_{ij}(\Delta,\mu)
    + \dots \Bigg] \,,
\end{split}
\end{equation}
valid for $\Delta\gg\Lambda_{\rm QCD}$. Here $m_b$ denotes the pole mass of the $b$ quark. The dots represent terms of order $1/m_b^2$ and higher, which we ignore. The integrated coefficient functions are obtained as
\begin{equation}\label{Fijbar}
   \bar F_{ij}(\Delta,\mu) = \int_{-\bar\Lambda}^\Delta dp_+\,
   F_{ij}(E_\gamma,\mu) \,,
\end{equation}
where $p_+=m_b-2E_\gamma$. As will be explained below, with the exception of $g_{88}^{\rm cut}$ the non-perturbative soft functions have support for values $\omega={\cal O}(\Lambda_{\rm QCD})$.\footnote{We ignore radiative tails of these functions, which can exhibit power behavior and extend to larger $\omega$ values. These effects only contribute at higher orders in $\alpha_s$.} 
In the limit $\Delta\gg\Lambda_{\rm QCD}$, the $\omega$ integrals in the definitions of the subleading shape function can then be performed over the entire range from $-\infty$ to $\bar\Lambda$, and this leads to simplifications. However, the integrals over the remaining $\omega_i$ variables cannot be simplified. 

For the direct photon contributions, we need the integrals
\begin{equation}
\begin{aligned}
   \int_{-\bar\Lambda}^\Delta dp_+ \int_{-p_+}^{\bar\Lambda}\!d\omega\,
    S(\omega,\mu)
   &\approx \Delta \,, \\
   \int_{-\bar\Lambda}^\Delta dp_+ \int_{-p_+}^{\bar\Lambda}\!d\omega\,
    \ln\frac{m_b(\omega+p_+)}{\mu^2}\,S(\omega,\mu)
   &\approx \Delta \left( \ln\frac{m_b\Delta}{\mu^2} - 1 \right) ,
\end{aligned}
\end{equation}
where the approximate expressions on the right are valid up to corrections of order $\alpha_s(\Delta)$ and $(\Lambda_{\rm QCD}/\Delta)^2$, both of which are known \cite{Neubert:2004dd}. It follows that the direct photon terms contribute to (\ref{Gtot}) at order $\Delta/m_b$ in power counting and can be computed using a local operator-product expansion. In the formal limit $E_0\to 0$ these terms are promoted to ${\cal O}(1)$ contributions. 

Let us now discuss what happens to the subleading shape-function contributions to the integrated decay rate. For the operator pair $Q_{7\gamma}-Q_{7\gamma}$, the subleading shape-function contributions also reduce to matrix elements of local operators, as discussed in detail in \cite{Bauer:2001mh,Lee:2004ja,Bosch:2004cb,Beneke:2004in}. From (\ref{F77b_part})--(\ref{jsubm}) it follows that
\begin{equation}\label{F77barres}
   \bar F_{77}(\Delta,\mu) = \frac{C_F\alpha_s(\mu)}{4\pi}\,\Delta
   \left( 16 \ln\frac{m_b}{\Delta} + 1 \right) .
\end{equation}
Note that, at order $1/m_b$, the non-zero strange-quark mass effect discussed in Section~\ref{subsec:42} integrates to zero in the partially integrated decay rate (at tree level in $\alpha_s$), as long as $\Delta\gg\Lambda_{\rm QCD}$. Similarly, for the operator pairs $Q_1^q-Q_1^q$ and $Q_1^q-Q_{8g}$, only direct photon contributions contribute at order $1/m_b$, and we obtain
\begin{equation}
   \bar F_{11}(\Delta,\mu) = \bar F_{18}(\Delta,\mu)
   = \frac{C_F\alpha_s(\mu)}{4\pi}\,\frac29\,\Delta \,.
\end{equation}

The remaining contributions, all of which contain resolved photon terms, are more interesting. For the operator pairs $Q_1^q-Q_{7\gamma}$, we obtain
\begin{equation}
   \bar F_{17}(\Delta,\mu) 
   = \frac{C_F\alpha_s(\mu)}{4\pi}\,\left( - \frac23 \right) \Delta
    + \frac23\,(1-\delta_u)\,\mbox{Re} \int_{-\infty}^\infty 
    \frac{d\omega_1}{\omega_1+i\varepsilon} \left[ 1 
    - F\!\left( \frac{m_c^2-i\varepsilon}{m_b\,\omega_1} \right)
    \right] h_{17}(\omega_1,\mu) \,,
\end{equation}
where 
\begin{equation}\label{eqn:h17}
\begin{aligned}
   h_{17}(\omega_1,\mu) 
   &= \int_{-\Delta}^{\bar\Lambda}\!d\omega\,
    g_{17}(\omega,\omega_1,\mu)
   \approx \int_{-\infty}^{\bar\Lambda}\!d\omega\,
    g_{17}(\omega,\omega_1,\mu) \\
   &= \int\frac{dr}{2\pi}\,e^{-i\omega_1 r}\,
    \frac{\langle\bar B| \big(\bar h S_{\bar n}\big)(0)\,
    \nbslash\,i\gamma_\alpha^\perp\bar n_\beta\,
    \big(S_{\bar n}^\dagger\,g G_s^{\alpha\beta} S_{\bar n} 
    \big)(r\bar n)\,
    \big(S_{\bar n}^\dagger h\big)(0) |\bar B\rangle}{2M_B} \,. 
\end{aligned}
\end{equation}
The integral over $p_+$ in (\ref{Fijbar}) eliminates the $\delta(\omega+p_+)$ distribution in (\ref{F17b}), and integrating the soft function $g_{17}(\omega,\omega_1,\mu)$ in (\ref{g17def}) over $\omega$ then eliminates the $t$-integral and sets $t=0$, so that part of the non-localities of the operator are eliminated. However, the gluon field is still smeared out on the $\bar n$ light-cone. Note that there is no contribution from the up-quark penguin loop to the integrated rate. As noted in Section~\ref{sec:PT}, the integral over $\omega$ of $g_{17}(\omega,\omega_1,\mu)$ is symmetric in $\omega_1$, so that the integral over $F_{17,u}^{(b)}$ in (\ref{F17bu}) vanishes.\footnote{It has been pointed out in the past that up-quark penguin loops might give rise to an ${\cal O}(\Lambda_{\rm QCD}/m_b)$ uncertainty in the integrated rate for $\bar B\to X_d\gamma$ decay \cite{Buchalla:1997ky}, where unlike in \bsg\ they are not CKM suppressed. Applying our analysis to $\bar B\to X_d\gamma$ shows that this contribution actually vanishes, removing that source of uncertainty in the integrated decay rate. Note that the same is not true for the CP asymmetry in $\bar B\to X_d\gamma$ decay, where the corresponding contribution is proportional to $h_{17}(0)$, which is non-zero in general. We further comment on $\bar B\to X_d\gamma$ decay in the conclusions.} 
In the approximation where the penguin function is expanded to first order using (\ref{Fexpand}), one would obtain
\begin{equation}
   \bar F_{17}(\Delta,\mu) 
   \approx \frac{C_F\alpha_s(\mu)}{4\pi}\,\left( - \frac23 \right) \Delta
    - (1-\delta_u)\,\frac{m_b\lambda_2}{9m_c^2} \,,
\end{equation}
which equals the integral over the partonic expression in (\ref{Fijpart}), where we had neglected the small correction proportional to $\delta_u$.

For the case of the pair $Q_{7\gamma}-Q_{8g}$, we obtain
\begin{equation}
   \bar F_{78}(\Delta,\mu) 
   = \frac{C_F\alpha_s(\mu)}{4\pi}\,\frac{10}{3}\,\Delta 
    + 4\pi\alpha_s(\mu)\,\mbox{Re}  
    \int_{-\infty}^{\infty} \frac{d\omega_1}{\omega_1+i\varepsilon} 
    \int_{-\infty}^{\infty} \frac{d\omega_2}{\omega_2-i\varepsilon}\,
    h_{78}^{(5)}(\omega_1,\omega_2,\mu) \,, 
\end{equation}
where in analogy with (\ref{eqn:h17}) we have introduced
\begin{eqnarray}\label{h785def}
   h_{78}^{(5)}(\omega_1,\omega_2,\mu) 
   &=& \int\frac{dr}{2\pi}\,e^{-i\omega_1 r}\!
    \int\frac{du}{2\pi}\,e^{i\omega_2 u} \\
   &&\times 
    \frac{\langle\bar B| \big(\bar h S_{\bar n}\big)(0)\,T^A\,
          \nbslash\gamma_5\,
          \big(S_{\bar n}^\dagger h\big)(0)\,
          \sum{}_q\,e_q\,\big(\bar q S_{\bar n}\big)(r\bar n)\,
          \nbslash\gamma_5\,T^A
          \big(S_{\bar n}^\dagger q\big)(u\bar n)
          |\bar B\rangle}{2M_B} \,. \nonumber
\end{eqnarray}
Note that the contribution from $g_{78}^{(1)}$ vanishes, since the integral over $\omega$ of this function is symmetric under the exchange of $\omega_1$ and $\omega_2$. Likewise, the contributions from the functions $\bar g_{78}$ and $\bar g_{78}^{\rm cut}$ to the integrated decay rate cancel each other. This follows from the fact that the two non-local operators in (\ref{g78bardef}) and (\ref{g78barcutdef}) coincide for $t=0$. In the VIA we obtain
\begin{equation}
   \bar F_{78}(\Delta,\mu) \Big|_{\rm VIA} 
   = \frac{C_F\alpha_s(\mu)}{4\pi}\,\frac{10}{3}\,\Delta 
    - \frac{\pi\alpha_s(\mu)}{2}\,e_{\rm spec} \left( 1 - \frac{1}{N_c^2} \right) 
    \frac{F^2(\mu)}{\lambda_B^2(\mu)} \,.
\end{equation}
The second term coincides with the result derived first in \cite{Lee:2006wn}.

Finally, for the case of the operator pair $Q_{8g}-Q_{8g}$, we find from (\ref{F88fin})
\begin{equation}\label{barF88}
\begin{aligned}
   \bar F_{88}(\Delta,\mu) 
   &= \frac{C_F\alpha_s(\mu)}{4\pi} 
    \left( \frac29\,\ln\frac{m_b}{\Lambda_{\rm UV}} - \frac13 \right) \Delta \\
   &\quad\mbox{}+ \frac89\,\pi\alpha_s(\mu)
    \int_{-\infty}^{\Lambda_{\rm UV}}\!\frac{d\omega_1}{\omega_1+i\varepsilon} 
    \int_{-\infty}^{\Lambda_{\rm UV}}\!\frac{d\omega_2}{\omega_2-i\varepsilon}\,
    h_{88}^{\rm cut}(\Delta,\omega_1,\omega_2,\mu) \,,
\end{aligned}
\end{equation}
where
\begin{equation}\label{h88cut}
   h_{88}^{\rm cut}(\Delta,\omega_1,\omega_2,\mu) 
   = \int_{-\Delta}^{\bar\Lambda}\!d\omega\,
   g_{88}^{\rm cut}(\omega,\omega_1,\omega_2,\mu) \,.
\end{equation}
Naively, we would expect that for $\Delta\gg\Lambda_{\rm QCD}$ this function becomes independent of $\Delta$ and reduces to the expression
\begin{eqnarray}\label{h88naive}
   h_{88}^{\rm cut}(\omega_1,\omega_2,\mu) 
   &\stackrel{?}{=}& \int\frac{dr}{2\pi}\,e^{-i\omega_1 r}\! 
    \int\frac{du}{2\pi}\,e^{i\omega_2 u} \\
   &&\hspace{-2.5cm} \times 
    \frac{\langle\bar B| \big(\bar h S_n\big)(0)\,T^A
          \big(S_n^\dagger S_{\bar n}\big)(0)\,
          \overline{\Gamma}_{\bar n}
          \big(S_{\bar n}^\dagger s\big)(u\bar n)
          \big(\bar s S_{\bar n}\big)(r\bar n)
          \Gamma_{\bar n}
          \big(S_{\bar n}^\dagger S_{n}\big)(0)\,T^A 
          \big(S_n^\dagger h\big)(0)
          |\bar B\rangle}{2M_B} \,, \nonumber
\end{eqnarray}
in which case the second term in (\ref{barF88}) would be strictly positive. However, in the present case the limit $t\to 0$ in (\ref{g88def}) is singular, since then the separation between the two light-quark fields $s$ and $\bar s$ becomes light-like. As a result, the integral over $\omega$ in (\ref{h88cut}) diverges linearly as $\Delta$ is raised to infinity, and hence it must be evaluated at large but finite $\Delta$. In other words, unlike for the other soft functions, the support of the function $g_{88}^{\rm cut}$ is not restricted to values $\omega={\cal O}(\Lambda_{\rm QCD})$ but extends to large negative values of $\omega$. This is in accordance with the asymptotic behavior derived in (\ref{g88pert}).

The convolutions of the soft functions with anti-hard-collinear jet functions in the results given above cannot be expressed in terms of local operator matrix elements, but rather define unknown hadronic parameters of order $\Lambda_{\rm QCD}$. These are the sources of genuine, first-order power corrections to the integrated decay rate, which are not reduced by lowering the cutoff $E_0$ on the photon energy.

\section{Phenomenological implications}
\label{sec:conc}

The results of the previous sections can be used to quantify the effect of the resolved photon terms on the \bsg\ photon spectrum, as well as on the decay rate and CP asymmetry.  In this paper we will restrict our attention to the decay rate integrated over a sufficiently wide energy range. The photon spectrum and CP asymmetry will be studied in a future publication.
 
In order to estimate the irreducible theoretical uncertainty from these new non-local effects on the integrated decay rate, we define the function
\begin{equation}\label{calFdef}
   {\cal F}_E(\Delta) 
   = \frac{\Gamma(E_0) - \Gamma(E_0)|_{\rm OPE}}{\Gamma(E_0)|_{\rm OPE}}
\end{equation}
where $E_0$ is the lower cutoff on the photon energy, and $\Delta=m_b-2E_0$. This definition is such that the true decay rate $\Gamma(E_0)$ is obtained from the theoretical expression $\Gamma(E_0)|_{\rm OPE}$ obtained using a {\em local\/} operator product expansion by multiplying it with $[1+{\cal F}_E(\Delta)]$. Note that $\Gamma(E_0)|_{\rm OPE}$ refers to the formula used in previous calculations of the \bsg\ rate, see e.g.\ \cite{Misiak:2006zs}. The function ${\cal F}_E(\Delta)$ corresponds to the relative theoretical error made in these calculations due to the neglect of non-local power corrections from resolved photon contributions. 

To the order we are working, we obtain
\begin{equation}
\begin{aligned}
   {\cal F}_E(\Delta) 
   &= \frac{1}{m_b}\,\Bigg\{
    \left[ \bar F_{77}(\Delta,\mu) - \frac{C_F\alpha_s(\mu)}{4\pi}\,\Delta
    \left( 16\ln\frac{m_b}{\Delta} + 1 \right) \right] \\
   &\quad\mbox{}+ \frac{C_1(\mu)}{C_{7\gamma}(\mu)}
    \left[ \bar F_{17}(\Delta,\mu) + \frac{C_F\alpha_s(\mu)}{4\pi}\,
    \frac23\,\Delta + \frac{m_b\lambda_2}{9m_c^2} \right] \\
   &\quad\mbox{}+ \frac{C_{8g}(\mu)}{C_{7\gamma}(\mu)}
    \left[ \bar F_{78}(\Delta,\mu) - \frac{C_F\alpha_s(\mu)}{4\pi}\,
    \frac{10}{3}\,\Delta \right] \\
   &\quad\mbox{}+ \left( \frac{C_{8g}(\mu)}{C_{7\gamma}(\mu)} \right)^2
    \left[ \bar F_{88}(\Delta,\mu) - \frac{C_F\alpha_s(\mu)}{4\pi}\,
    \Delta \left( \frac29\ln\frac{m_b\Delta}{m_s^2} - \frac59 \right) 
    \right] \Bigg\} + \dots \,,
\end{aligned}
\end{equation}
where we have assumed that the Wilson coefficients are real (like in the Standard Model) and neglected effects proportional to $V_{ub}$. Note that the terms in the first line on the right-hand side vanish due to the relation (\ref{F77barres}). We can express the various other contributions in terms of suitably defined hadronic parameters of order $\Lambda_{\rm QCD}$, using the expressions for the quantities $\bar F_{ij}(\Delta,\mu)$ derived in the previous section under the assumption that $\Delta\gg\Lambda_{\rm QCD}$. Making explicit the dependence on the Wilson coefficients and factors of the strong coupling $g^2=4\pi\alpha_s$, we arrive at
\begin{equation}\label{finalform}
\begin{split}
   {\cal F}_E(\Delta) 
   &= \frac{C_1(\mu)}{C_{7\gamma}(\mu)}\,
    \frac{\Lambda_{17}(m_c^2/m_b,\mu)}{m_b}
    + \frac{C_{8g}(\mu)}{C_{7\gamma}(\mu)}\,
    4\pi\alpha_s(\mu)\,\frac{\Lambda_{78}^{\rm spec}(\mu)}{m_b} \\
   &\quad\mbox{}+ \left( \frac{C_{8g}(\mu)}{C_{7\gamma}(\mu)} \right)^2
    \left[ 4\pi\alpha_s(\mu)\,\frac{\Lambda_{88}(\Delta,\mu)}{m_b} 
    - \frac{C_F\alpha_s(\mu)}{9\pi}\,\frac{\Delta}{m_b}\,\ln\frac{\Delta}{m_s} \right]
    + \dots \,,
\end{split}
\end{equation}
where
\begin{eqnarray}\label{Lambdaijdef}
   \Lambda_{17}\Big(\frac{m_c^2}{m_b},\mu\Big)
   &=& e_c\,\mbox{Re} \int_{-\infty}^\infty \frac{d\omega_1}{\omega_1} 
    \left[ 1 - F\!\left( \frac{m_c^2-i\varepsilon}{m_b\,\omega_1} \right)
    + \frac{m_b\,\omega_1}{12m_c^2} \right] h_{17}(\omega_1,\mu) \,, \nonumber\\
   \Lambda_{78}^{\rm spec}(\mu) 
   &=& \mbox{Re}  
    \int_{-\infty}^{\infty} \frac{d\omega_1}{\omega_1+i\varepsilon} 
    \int_{-\infty}^{\infty} \frac{d\omega_2}{\omega_2-i\varepsilon}\,
    h_{78}^{(5)}(\omega_1,\omega_2,\mu) \,, \\
   \Lambda_{88}(\Delta,\mu)
   &=& e_s^2 \left[ \int_{-\infty}^{\Lambda_{\rm UV}}\!
    \frac{d\omega_1}{\omega_1+i\varepsilon} 
    \int_{-\infty}^{\Lambda_{\rm UV}}\!\frac{d\omega_2}{\omega_2-i\varepsilon}\,
    2 h_{88}^{\rm cut}(\Delta,\omega_1,\omega_2,\mu) 
    - \frac{C_F}{8\pi^2}\,\Delta \left( \ln\frac{\Lambda_{\rm UV}}{\Delta} - 1 \right)
    \right] . \nonumber
\end{eqnarray}
In the case of $\Lambda_{17}$ and $\Lambda_{88}$ we have factored out the appropriate powers of the quark electric charges. Because of the sum over light-quark flavors in (\ref{h785def}), the parameter $\Lambda_{78}^{\rm spec}$ receives contributions proportional to any one of the light-quark charges. The resulting hard breaking of isospin symmetry implies that its value will be different for charged and neutral $B$ mesons, even in the limit of exact isospin symmetry of the strong interaction. We will show in Section~\ref{subsec:78} that, in certain approximation schemes, $\Lambda_{78}^{\rm spec}$ is proportional to the electric charge of the spectator quark in the $B$ meson. 

Note that the parameters $m_c^2/m_b$ and $\Delta$ entering the arguments of $\Lambda_{17}$ and $\Lambda_{88}$ count as ${\cal O}(\Lambda_{\rm QCD})$. The dependence on the strange-quark mass in (\ref{finalform}) arises only because the function ${\cal F}_E(\Delta)$ is defined as the deviation from the partonic rate $\Gamma_{\rm part}(E_0)$. The true decay rate $\Gamma(E_0)$ in (\ref{calFdef}) is independent of $m_s$. Note also that the result for $\Lambda_{88}$ is formally independent of the UV cutoff $\Lambda_{\rm UV}$, and that it is the only hadronic parameter in (\ref{finalform}) that depends on the quantity $\Delta$. In the formal limit where the cut on the photon energy is removed, $\Delta\to m_b$, the linear growth (modulo logarithms) of the parameter $\Lambda_{88}$ with $\Delta$ implies that the corresponding contribution to ${\cal F}_E(\Delta)$ is promoted from a power-suppressed to a leading-order effect. Indeed, it is well known that in this limit there exists a leading-power, non-perturbative $Q_{8g}-Q_{8g}$ contribution related to the photon fragmentation off a strange quark or gluon \cite{Kapustin:1995fk}. For practical applications this observation is irrelevant. We will argue in Section~\ref{subsec:7.3} that, for realistic values of $E_0$ outside the endpoint region, the dependence of $\Lambda_{88}$ on $\Delta$ is very weak, and therefore the function ${\cal F}_E(\Delta)$ is almost equal to a constant.

Without further information about the soft functions, the $\Lambda_{ij}$ parameters are expected to be of order $\Lambda_{\rm QCD}$ apart from the electric charges factored out in (\ref{Lambdaijdef}). This would lead to very large effects of up to 30\% on the decay rate. Fortunately, it is possible to constrain the values of $\Lambda_{17}$ and $\Lambda_{78}^{\rm spec}$ by means of simple considerations, as we will now discuss. The input parameters used for the estimates in the following discussion are collected in Appendix~\ref{app:inputs}. The accuracy of our calculations is such that we are insensitive to the scale dependence of the subleading soft functions and the corresponding hadronic parameters. Even though we have indicated their $\mu$ dependence in the formulae given above, to properly control this dependence would require to extend our calculations to the next order in the expansion in powers of $\alpha_s(\mu)$.

\subsection{Analysis of the $\bm{Q_1^c-Q_{7\gamma}}$ contribution}

In order to obtain a reasonable estimate for the parameter $\Lambda_{17}$, we first collect everything we know about the function $h_{17}(\omega_1,\mu)$ defined below (\ref{eqn:h17}). As proved in Section~\ref{sec:Gtot}, this function must be real, and the symmetry relation (\ref{g17int}) then implies that it is an even function of $\omega_1$. It follows that all odd moments of $h_{17}$ vanish. Moreover, from (\ref{g17norm}) the normalization of $h_{17}$ is fixed to $2\lambda_2$. About the higher even moments nothing definite is known, but we can expect them to be proportional to an appropriate power of $\Lambda_{\rm QCD}$ times a not too large numerical factor. Finally, as a soft function, $h_{17}$ should not have any significant structures, such as peaks or zeros, outside the hadronic energy range.

The first functions that come to mind are an exponential and a Gaussian, 
\begin{equation}\label{h17simple}
   h_{17}(\omega_1,\mu) = \frac{\lambda_2}{\sigma}\,e^{-\frac{|\omega_1|}{\sigma}} \,,
    \quad \mbox{or} \quad
   h_{17}(\omega_1,\mu) = \frac{2\lambda_2}{\sqrt{2\pi}\sigma}\,
    e^{-\frac{\omega_1^2}{2\sigma^2}} \,,
\end{equation}
for which all even moments are finite. As long as $\sigma\ll 4m_c^2/m_b\approx 1.1$\,GeV, which with the power counting adopted in this paper is formally of order $\Lambda_{\rm QCD}$, then for all relevant $\omega_1$ values the argument of the penguin function $F(x)$ entering the definition of $\Lambda_{17}$ in (\ref{Lambdaijdef}) is much larger than 1/4, which is the radius of convergence for the Taylor expansion given in (\ref{Fexpand}). It is then a good approximation to expand the penguin function $[1-F(x)]$ to ${\cal O}(1/x^3)$. The first term in this expansion corresponds to the non-perturbative correction identified in \cite{Voloshin:1996gw}, which was already included in the partonic result and subtracted in (\ref{Lambdaijdef}). It therefore does not contribute to ${\cal F}_E(\Delta)$. The next term gives rise to an odd moment of $h_{17}$ and thus vanishes. The third term in the expansion contributes the amount
\begin{equation}\label{L17exp}
   \Lambda_{17}^{\rm expanded} 
   = - \frac{e_c}{280}\,\frac{m_b^3}{m_c^6}\,\lambda_2\,\langle\omega_1^2\rangle
\end{equation}
to $\Lambda_{17}$. Here $\langle\omega_1^2\rangle$ denotes the (normalized) variance of the function $h_{17}(\omega_1,\mu)$, which equals $2\sigma^2$ for the exponential form and 
$\sigma^2$ for the Gaussian. For a typical hadronic scale $\sigma=0.5$\,GeV this gives $\Lambda_{17}^{\rm expanded}=-6.9$\,MeV and $-3.4$\,MeV, respectively. Here and below we have used the input parameters collected in Appendix~\ref{app:inputs}. The corresponding contributions to the decay rate are very small, below 0.5\% in magnitude. 

It is interesting that, due to a weaker numerical suppression, certain $1/m_b$ corrections to $\Lambda_{17}$ can give a contribution of comparable size. They arise from the fact that the first moment of the function $g_{17}(\omega,\omega_1,\mu)$ with respect to $\omega$ does not vanish, see (\ref{g17mom1}). In order to calculate the resulting power-suppressed term, we replace the first relation in (\ref{Lambdaijdef}) by
\begin{equation}\label{L17power}
\begin{aligned}
   \Lambda_{17}\Big(\frac{m_c^2}{m_b},\mu\Big)
   &= e_c\,\mbox{Re} \int_{-\infty}^{\bar\Lambda}\!d\omega
    \int_{-\infty}^\infty \frac{d\omega_1}{\omega_1} \\
   &\quad\times \left\{ \left( \frac{m_b+\omega}{m_b} \right)^3 
    \left[ 1 - F\!\left( \frac{m_c^2-i\varepsilon}{(m_b+\omega)\,\omega_1} \right) \right]
    + \frac{m_b\,\omega_1}{12m_c^2} \right\} g_{17}(\omega,\omega_1,\mu) \,, 
\end{aligned}
\end{equation}
where the factor $(\frac{m_b+\omega}{m_b})^3$ appears because of the prefactor $E_\gamma^3$ in (\ref{HQexp}). Expanding now the penguin function to first order yields
\begin{equation}
   \Lambda_{17}\Big(\frac{m_c^2}{m_b},\mu\Big)
   = e_c\,\mbox{Re} \int_{-\infty}^{\bar\Lambda}\!d\omega
    \int_{-\infty}^\infty \frac{d\omega_1}{\omega_1} 
    \left\{ - \left( 1 + \frac{\omega}{m_b} \right)^4 \frac{m_b\,\omega_1}{12m_c^2} 
    + \frac{m_b\,\omega_1}{12m_c^2} + \dots \right\} g_{17}(\omega,\omega_1,\mu) \,,
\end{equation}
where the dots represent higher-order terms in the expansion of the penguin function, which in particular give rise to the contribution (\ref{L17exp}). The expression shown above yields a $1/m_b$-suppressed contribution to the parameter $\Lambda_{17}$, which we denote by $\delta\Lambda_{17}$. It is proportional to the normalized first moment of the function $g_{17}$ with respect to $\omega$, which according to (\ref{g17norm}) and (\ref{g17mom1}) is given by $\langle\omega\rangle=-\rho_{LS}^3/(6\lambda_2)\approx 0.24$\,GeV. We obtain 
\begin{equation}
   \delta\Lambda_{17} = \frac{2\rho_{LS}^3}{27m_c^2}
   \approx - (9.8\pm 5.2)\,\mbox{MeV} \,,
\end{equation}
which is formally a power correction proportional to $\Lambda_{\rm QCD}^2/m_b$ to the result in (\ref{L17exp}). Here $\rho_{LS}^3=3\rho_2$ corresponds to the spin-orbit term of the HQET Lagrangian introduced in (\ref{g17mom1}). 

In practice, it turns out that (\ref{L17exp}) provides a reasonable approximation only as long as $\sigma<0.3$\,GeV. Performing the convolution integral in (\ref{Lambdaijdef}) exactly, we find that for both model functions in (\ref{h17simple}) the resulting value of $|\Lambda_{17}|$ is maximized for certain values of $\sigma$, which depend on the functional form of $h_{17}$. Using the input parameters collected in Appendix~\ref{app:inputs}, we obtain $(\Lambda_{17}^{\rm exp})_{\rm max}=-4.6$\,MeV for $\sigma=0.51$\,GeV with the exponential model, and $(\Lambda_{17}^{\rm Gauss})_{\rm max}=-8.1$\,MeV for $\sigma=0.77$\,GeV with the Gaussian model. Note that the maximum values are smaller in magnitude than those one would derive from (\ref{L17exp}) with these values of $\sigma$.

\begin{figure}
\centering
\psfrag{x}[]{$\omega_1$~[GeV]}
\includegraphics[width=9cm]{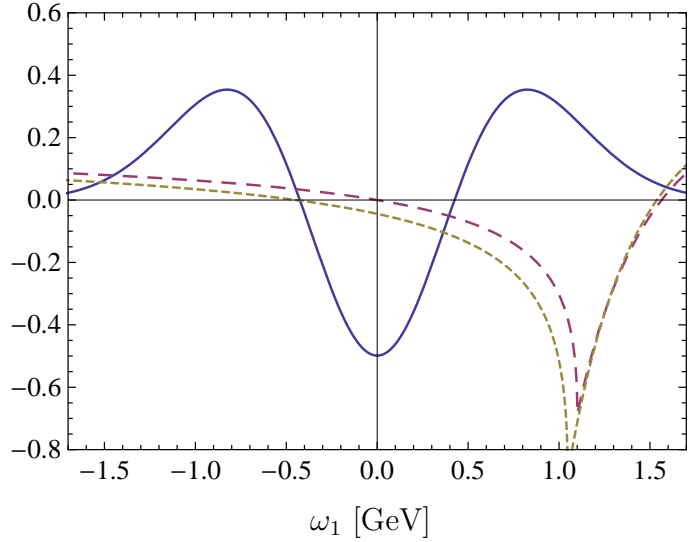}
\caption{\label{fig:h17}
Model function $h_{17}(\omega_1,\mu)$ from (\ref{h17model}) in units of GeV, with $\sigma=0.5$\,GeV and $\Lambda=0.425$\,GeV (solid line); weight function under the convolution integral in the definition of $\Lambda_{17}$ in (\ref{Lambdaijdef}) in units of GeV$^{-1}$ (long-dashed line); weight function including $1/m_b$ corrections, obtained by the substitution $\omega\to\langle\omega\rangle$ in (\ref{L17power}) (short-dashed line). See text for explanations.}
\end{figure}

The above estimates do not provide a conservative bound on the size of the hadronic parameter $\Lambda_{17}$. A significantly larger effect can be obtained if the soft function $g_{17}(\omega,\omega_1,\mu)$ exhibits a tail outside the region $|\omega_1|\ll 4m_c^2/m_b$. In analogy with the leading-order shape function, we expect that the function $g_{17}$ exhibits a radiative tail proportional to $1/\omega_1$ for large $\omega_1$. But even at the non-perturbative level, it is conceivable that a significant contribution to the integral results from the region of larger $\omega_1$ values. Consider, as an example, the model
\begin{equation}\label{h17model}
   h_{17}(\omega_1,\mu) 
   = \frac{2\lambda_2}{\sqrt{2\pi}\sigma}\,
    \frac{\omega_1^2-\Lambda^2}{\sigma^2-\Lambda^2}\,e^{-\frac{\omega_1^2}{2\sigma^2}} \,,
\end{equation}
which for $\Lambda$ and $\sigma$ of order $\Lambda_{\rm QCD}$ satisfies all requirements one would reasonably impose on the soft function. The solid curve in Figure~\ref{fig:h17} shows this function evaluated with $\sigma=0.5$\,GeV and $\Lambda=0.425$\,GeV. It features regions of positive and negative values and hence is less constrained at larger $\omega_1$ by the fact that the normalization is fixed to $2\lambda_2$. Having values of either sign is not problematic, because there is no probabilistic interpretation of the subleading soft functions. The long-dashed line in the figure shows the weight function under the convolution integral in the definition of $\Lambda_{17}$ in (\ref{Lambdaijdef}), including the charge factor $e_c$. With the above parameter choices for the soft function, we obtain $\Lambda_{17}=-42$\,MeV. By using another set of values, a correction with the opposite sign and of the same magnitude can be obtained. For example, taking $\sigma=0.5$\,GeV and $\Lambda=0.575$\,GeV we find $\Lambda_{17}=27$\,MeV. If we include the $1/m_b$ corrections as shown in (\ref{L17power}), using $(m_b+\omega)\to(m_b+\langle\omega\rangle)=(m_b-\rho_{LS}^3/6\lambda_2)$, we find $-62$\,MeV and 21\,MeV, respectively. Of course, these are just illustrative values, and one could obtain even larger negative or positive values by reducing the separation between $\sigma$ and $\Lambda$, which however will also increase the value of the soft function at $\omega_1=0$. Nevertheless, based on these considerations, it seems to us that
\begin{equation}\label{Lam17range}
   - 60\,\mbox{MeV} < \Lambda_{17} < 25\,\mbox{MeV}
\end{equation}
is a reasonably conservative range, which we will adopt for our analysis below. While this allows for a value significantly larger in magnitude than the naive estimate (\ref{L17exp}), it nevertheless strongly suggests that $\Lambda_{17}$ is considerably smaller in magnitude than $\Lambda_{\rm QCD}$. Note that the effect of a value of $\Lambda_{17}$ near the extreme values indicated above would be of the same magnitude as the effect of the leading-order, non-perturbative correction \cite{Voloshin:1996gw} resulting from the term proportional to $\lambda_2$ in the expression for $F_{17}^{\rm part}$ in (\ref{Fijpart}), which corresponds to $-m_b\lambda_2/(9m_c^2)\approx -48$\,MeV.

\subsection{Analysis of the $\bm{Q_{7\gamma}-Q_{8g}}$ contribution}
\label{subsec:78}

It is instructive to analyze this contribution using the language of flavor symmetry of the strong interaction. Due to the weighting by the quark electric charges, the relevant four-quark operator in (\ref{g78def}) is a pure $SU(3)$ octet, which can be decomposed into two parts corresponding to isospin $I=0,1$. The Wigner-Eckart theorem implies that  
\begin{equation}\label{decomp}
   \Lambda_{78}^{\rm spec} = \frac16\,\Lambda_{I=0}^{(8)} \pm \frac12\,\Lambda_{I=1}^{(8)}
   = \frac16 \left( \Lambda_{I=0}^{(8)} - \Lambda_{I=1}^{(8)} \right) 
    + e_{\rm spec}\,\Lambda_{I=1}^{(8)} \,,
\end{equation}
where the upper (lower) sign in the first equation refers to charged (neutral) $B$ mesons, and as before $e_{\rm spec}$ denotes the electric charge of the spectator quark in units of $e$. In the limit of unbroken $SU(3)$ flavor symmetry, it follows that $\Lambda_{I=0}^{(8)}=\Lambda_{I=1}^{(8)}$, since both parameters arise from the matrix element of the same $SU(3)$ octet operator. Hence, in this limit we obtain
\begin{equation}\label{eq:110n}
   \Lambda_{78}^{\rm spec}\big|_{SU(3)} = e_{\rm spec}\,\Lambda_{I=1}^{(8)} \,.
\end{equation}

Interestingly, the VIA discussed in Section~\ref{sec:Q7Q8} also predicts that $\Lambda_{78}^{\rm spec}$ is proportional to $e_{\rm spec}$ \cite{Lee:2006wn}, and we can use this fact to obtain a model estimate of the relevant $SU(3)$ reduced matrix element. From (\ref{VIA4ever}), we read off
\begin{equation}\label{eq:108}
   \Lambda_{I=1}^{(8)}\big|_{\rm VIA} = \Lambda_{I=0}^{(8)}\big|_{\rm VIA} 
   = - \left( 1 - \frac{1}{N_c^2} \right) \frac{F^2(\mu)}{8\lambda_B^2(\mu)} 
   \in [-386\,\mbox{MeV}, -35\,\mbox{MeV} ] \,,
\end{equation}
where in the last step we have used the parameter ranges discussed in Appendix~\ref{app:inputs}.

According to (\ref{decomp}), the isospin-averaged decay rate $[\Gamma(\bar B^0\to X_s\gamma)+\Gamma(B^-\to X_s\gamma)]/2$ depends only on $\Lambda_{I=0}^{(8)}$, while the isospin difference $[\Gamma(\bar B^0\to X_s\gamma)-\Gamma(B^-\to X_s\gamma)]$ is proportional to $\Lambda_{I=1}^{(8)}$. While {\em a priori\/} these two non-perturbative parameters are unrelated, we have just shown that they coincide both in the $SU(3)$ flavor-symmetry limit and in the VIA. It was first pointed out in \cite{Misiak:2009nr} that, in the limit of exact $SU(3)$ flavor symmetry, the isospin-averaged decay rate can be related to the isospin asymmetry,
\begin{equation}
   \Delta_{0-}
   = \frac{\Gamma(\bar B^0\to X_s\gamma)-\Gamma(B^-\to X_s\gamma)}%
          {\Gamma(\bar B^0\to X_s\gamma)+\Gamma(B^-\to X_s\gamma)} \,,
\end{equation}
without employing the VIA. This asymmetry has been measured by the BaBar Collaboration using two different experimental methods. For the ``sum-over-exclusive-modes method'' with $E_\gamma>1.9$\,GeV, they find $\Delta_{0-}=(-0.6\pm 5.8\pm 0.9\pm 2.4)\%$ \cite{Aubert:2005cua}, where the errors are statistical, systematic, and due to the production ratio $\bar B^0/B^-$, respectively. For the ``recoil method'' with $E_\gamma>2.2$\,GeV, they obtain instead $\Delta_{0-}=(-6\pm 15\pm 7)\%$ \cite{Aubert:2007my}, where the errors are statistical and systematic, respectively. The naive average of these two results is $\Delta_{0-}=(-1.3\pm 5.9)\%$. To the order we are working, the parameter $\Lambda_{I=1}^{(8)}$ is related to $\Delta_{0-}$ via
\begin{equation}\label{eq:110}
   \Lambda_{I=1}^{(8)}\big|_{\rm exp} = - \frac{C_{7\gamma}(\mu)}{C_{8g}(\mu)}\,
   \frac{m_b}{2\pi\alpha_s(\mu)}\,\Delta_{0-} 
   \approx (59\pm 268)\,\mbox{MeV} \,,
\end{equation}
where in the last step we have used the average experimental result with its large uncertainty given above. This value is consistent with the prediction (\ref{eq:108}) obtained in the VIA within errors, even though the central value has the opposite sign.

Allowing for $SU(3)$ flavor-symmetry breaking at the level of 30\%, i.e.\ $\Lambda_{I=0}^{(8)}=(1\pm 0.3)\,\Lambda_{I=1}^{(8)}$, we finally obtain
\begin{equation}
   \Lambda_{78}^{\rm spec} = (e_{\rm spec}\pm 0.05)\,\Lambda_{I=1}^{(8)} 
   \approx -4.5\,\mbox{GeV}\,(e_{\rm spec}\pm 0.05)\,\Delta_{0-} \,,
\end{equation}
which is meant as a range, not an error bar. This formula implies that, within the quoted uncertainty, the isospin asymmetry also determines the {\em flavor-averaged value\/} of $\Lambda_{78}^{\rm spec}$. For the corresponding contribution to the flavor-averaged value of the function ${\cal F}_E(\Delta)$, we obtain
\begin{equation}\label{eq:xx}
   {\cal F}_E^{\rm avg}(\Delta) \big|_{78} 
   = - (1\pm 0.3)\,\frac{\Delta_{0-}}{3} \,,
\end{equation}
which adds $SU(3)$-breaking effects to the estimate derived in \cite{Misiak:2009nr}. Note that this relation is independent of the values of the Wilson coefficients and other theoretical parameters. 

Due to the current large experimental uncertainties in the measurement of the isospin asymmetry, it is difficult to give a reliable estimate for $\Lambda_{78}^{\rm spec}$. Based on (\ref{eq:108}) and (\ref{eq:110}), we expect that the parameter $\Lambda_{I=1}^{(8)}$ is negative (assuming that the VIA is sufficiently reliable to predict the sign correctly), but since the experimental value allows for the entire range in (\ref{eq:108}) at the level of two standard deviations, we cannot restrict that range further at present. A future, more accurate measurement of $\Delta_{0-}$ could improve the situation. 

\subsection{Analysis of the $\bm{Q_{8g}-Q_{8g}}$ contribution}
\label{subsec:7.3}

Unfortunately, we have very little useful information about the soft function $h_{88}^{\rm cut}$ entering the definition of the hadronic parameter $\Lambda_{88}$ in (\ref{Lambdaijdef}). Its asymptotic behavior for large values of $\omega_1$ and $\omega_2$ can be derived from (\ref{g88pert}), and it ensures that $\Lambda_{88}$ is independent of the UV cutoff $\Lambda_{\rm UV}$. Note that the second term in the definition of $\Lambda_{88}$, which contains the logarithm of $\Lambda_{\rm UV}/\Delta$, is bound to give a very small contribution to $\Lambda_{88}$, because $(C_F e_s^2\Delta)/(8\pi^2)<3$\,MeV is very small for realistic values $E_0\ge 1.6$\,GeV. We thus expect that the hadronic parameter $\Lambda_{88}$ receives its dominant contributions from values $\omega_{1,2}={\cal O}(\Lambda_{\rm QCD})$, for which no useful constraints on the soft function $h_{88}^{\rm cut}$ exist. For the same reason, we expect that the linear growth of $\Lambda_{88}$ for large $\Delta$ is a numerically irrelevant effect. It then follows that the function $h_{88}^{\rm cut}(\Delta,\omega_1,\omega_2,\mu)$ is {\em approximately\/} equal to the function $h_{88}^{\rm cut}(\omega_1,\omega_2,\mu)$ shown in (\ref{h88naive}), even though this relation is not strictly valid. As mentioned earlier in the paragraph following that equation, this form would imply that the contribution to $\Lambda_{88}$ resulting from the double integral in (\ref{Lambdaijdef}) were strictly positive.

In summary, we expect that the hadronic parameter $\Lambda_{88}(\Delta,\mu)$ is, to a good approximation, independent of $\Delta$ and given by a positive, non-perturbative constant of order $e_s^2\,\Lambda_{\rm QCD}$:
\begin{equation}\label{L88mod}
   \Lambda_{88}(\Delta,\mu)\approx e_s^2\,\Lambda(\mu) \,, \qquad
   \Lambda(\mu) > 0 \,.
\end{equation}
We have backed up this expectation by using different models for the soft function $h_{88}^{\rm cut}$, for example by writing it as a product of two functions $f_1(\omega_1)\,f_2(\omega_2)$ and using various models such as exponentials or Gaussians. A particularly simple example is provided by functions $h_{88}^{\rm cut}(\omega_1,\omega_2,\mu)$ that are symmetric in both $\omega_i$ variables and have support for $\omega_i={\cal O}(\Lambda_{\rm QCD})$. In this case the third relation in (\ref{Lambdaijdef}) implies $\Lambda(\mu)\approx 2\pi^2 h_{88}^{\rm cut}(0,0,\mu)$, and the value at the origin scales like $h_{88}^{\rm cut}(0,0,\mu)\sim\Lambda_{\rm QCD}$. For our numerical analysis, we will consider the rather generous range $0<\Lambda(\mu)<1$\,GeV. Even for the largest value, the suppression by the charge factor $e_s^2=1/9$ in (\ref{L88mod}) renders the effect of this term on the decay rate to be very small.

\subsection{Summary of phenomenological estimates}

We are now in a position to study the implications of our analysis for the function ${\cal F}_E(\Delta)$ in (\ref{finalform}). Using the parameter values collected in Appendix~\ref{app:inputs}, we obtain from (\ref{Lam17range}) and (\ref{L88mod}) the contributions
\begin{equation}
\begin{aligned}
   {\cal F}_E\big|_{17} &\in [-1.7, +4.0]\,\% \,, \\
   {\cal F}_E\big|_{88} &\in [-0.3, +1.9]\,\% \,.
\end{aligned}
\end{equation}
The value of ${\cal F}_E|_{88}$ depends slightly on $\Delta$ and is obtained using $\Delta=1.45$\,GeV, corresponding to a cut at $E_0=1.6$\,GeV. For the case of ${\cal F}_E|_{78}$, we consider the charge-averaged contribution and quote separately the theoretical estimate obtained using the VIA and the experimental estimate derived from the measurement of the isospin asymmetry. In the latter case we allow for 30\% $SU(3)$ violation, as indicated in (\ref{eq:xx}), and take the 95\% confidence level experimental range. This yields
\begin{equation}
\begin{aligned}
   {\cal F}_E\big|_{78}^{\rm VIA} &\in [-2.8, -0.3]\,\% \,, \\
   {\cal F}_E\big|_{78}^{\rm exp} &\in [-4.4, +5.6]\,\% 
    \quad \mbox{(95\% CL)} \,.
\end{aligned}
\end{equation}

In order to obtain a conservative estimate of the combined theoretical analysis, we adopt a Baysean approach and add up the various contributions using the scanning method. In this way, we arrive at our final result
\begin{equation}\label{Fcurrent}
   -4.8\% < {\cal F}_E(\Delta) < +5.6\%
   \quad \mbox{(VIA for $\Lambda_{78}^{\rm spec}$)} \,,
\end{equation}
where we have used the theoretical estimate for ${\cal F}_E|_{78}$. When the experimental estimate is used instead, the range is expanded to 
\begin{equation}\label{Fcurrentexp}
   -6.4\% < {\cal F}_E(\Delta) < +11.5\%
   \quad \mbox{($\Lambda_{78}^{\rm spec}$ from $\Delta_{0-}$)} \,.
\end{equation}
We emphasize that the estimates in this sections should be considered as {\em ranges\/}, within which we expect the actual values of ${\cal F}_E$ to lie, without making a statement about the most likely values within these ranges.  

If in the future a more precise value of the isospin asymmetry can be measured, this could be used to reduce the uncertainty range somewhat. If, for example, we assume that the true isospin asymmetry lies in the center of the interval predicted by the VIA, $\Delta_{0-}=+4.6\%$, then in the absence of experimental uncertainties we would derive ${\cal F}_E|_{78}^{\rm exp}\in [-2.0, -1.1]\,\%$, where the remaining uncertainty stems from the unknown effects of $SU(3)$ breaking. In this ``ideal'' case, the combined result would be
\begin{equation}\label{final}
   -4.0\% < {\cal F}_E(\Delta) < +4.8\%
   \quad \mbox{(ideal case)} \,.
\end{equation}
We do not see a possibility to reduce this uncertainty in the foreseeable future, given that no theoretical tools exist to constrain the non-local matrix elements defining the soft functions entering the various resolved photon contributions studied in this paper. We therefore consider the range in (\ref{final}) as the {\em irreducible\/} theoretical uncertainty affecting any theoretical prediction of the \bsg\ branching ratio.

\section{Conclusions}

The inclusive radiative decay \bsg\ is used extensively in constraining extensions of the Standard Model. For example, it provides very stringent constraints on extended Higgs sectors in type-II 2-Higgs-doublet models and supersymmetric models. The theoretical prediction for the corresponding branching ratio is at a stage of precision where the remaining perturbative uncertainties are estimated to be of order 3\% \cite{Misiak:2006zs}. The limiting theoretical uncertainty arises from non-perturbative effects outside the realm of the local operator product expansion \cite{Lee:2006wn}. It is therefore important to analyze these effects in a systematic fashion. In this paper, we have for the first time provided a complete analysis of non-local $1/m_b$ corrections to the \bsg\ photon spectrum and decay rate, working at tree level in perturbation theory. Compared to inclusive semileptonic $B$ decays, non-perturbative effects in radiative decays are much more complicated to analyze. First of all, one must consider the contributions of many different operators in the effective weak Hamiltonian, not just one operator. More importantly, however, new types of non-local effects arise due to the hadronic substructure of the photon. Because photon conversion into light partons is a genuinely long-distance process, the decay \bsg\ is not a truly inclusive process, for which an expansion in local operators would apply. Indeed, from a conceptual point of view, it is as complicated as the semi-inclusive decay $\bar B\to X_s h$, with $h$ denoting a specific light hadron. No analogous effects arise in semileptonic processes, since the conversion of heavy $W$ bosons into light partons is a short-distance process.

Effective field theories, such as soft-collinear and heavy-quark effective theory, provide the necessary tools to analyze inclusive $B$ decays into light partons in the kinematical region of low hadronic invariant mass and large recoil energy, in which the hadronic final state is made up of a jet of collinear partons. For \bsg\ this is the endpoint region, where the photon has large energy $E_\gamma\approx m_b/2$ in the $B$-meson rest frame. Effective field theories are systematic, taking into account all possible contributions to a given decay amplitude and describing them in terms of well-defined, field-theoretic objects. This is especially important for radiative $B$ decays, where the diagrammatic approach used in the previous decade has missed the largest source of non-perturbative uncertainty \cite{Lee:2006wn}. 

In this paper, we have shown that the \bsg\ photon spectrum in the endpoint region obeys the novel factorization formula~(\ref{fact2}). The first term in this formula has the structure familiar from semileptonic $B$ decays. At each order in the $1/m_b$ expansion, it features products of hard functions $H_i$ and jet functions $J_i$ convoluted with soft functions $S_i$. We refer to this term as the direct photon contribution, since the photon couples directly to the weak vertex in a local interaction. The two remaining terms in the factorization formula describe resolved photon contributions, in which the photon couples indirectly to the weak vertex via conversion into light partons. The partonic substructure of the photon is described in terms of a new class of jet functions $\bar J_i$. These new terms appear first at order $1/m_b$ and arise from the contribution of operators other than $Q_{7\gamma}$ in the effective Hamiltonian. The new soft functions $S_i$ entering the resolved photon terms contain non-localities in two light-cone directions. Only one non-locality is removed when the photon spectrum is integrated over energy to obtain the total decay rate.\footnote{As before, by ``total'' rate we mean the rate defined with a lower cut $E_0$ on the photon energy that lies far outside the endpoint region, i.e., $m_b-2E_0\gg\Lambda_{\rm QCD}$.} 
As a result, we find that even the total decay rate receives non-local corrections of order $\Lambda_{\rm QCD}/m_b$. The new jet functions $\bar J_i$, which are defined in terms of propagators dressed by Wilson lines, are complex quantities carrying calculable, perturbative strong-interaction phases. The soft functions, on the other hand, were shown to be real in by the use of heavy-quark symmetry and the invariance of the strong interaction under parity and time reversal. The impact of the new strong phases on CP violation in \bsg\ decay will be considered in a future publication.

Phenomenologically the most important operators in the effective weak Hamiltonian are $Q_{7\gamma}$, $Q_{8g}$, and $Q_1^c$. We have explicitly evaluated the $1/m_b$ corrections to the \bsg\ photon spectrum arising from these operators, at tree-level in hard and hard-collinear interactions. This includes important contributions involving a hard-collinear gluon exchange, which carry a factor $g^2=4\pi\alpha_s$. Our results are summarized in relations (\ref{improved}), which replace the relations (\ref{Fijpart}) used in previous analyses of \bsg\ decay. The systematic methodology offered by the effective field-theory approach resolves a couple of puzzling features of the expressions (\ref{Fijpart}), such as the appearance of the strange-quark mass in the expression for $F_{88}^{\rm part}$, or of large logarithms in the expressions for $F_{77}^{\rm part}$ and $F_{88}^{\rm part}$. We point out that these features result from an improper separation of short- and long-distance physics. In our improved expressions (\ref{improved}), all long-distance physics is parameterized by well-defined hadronic matrix elements (the soft functions), while the logarithms entering the short-distance perturbative contributions contain ${\cal O}(1)$ ratios of scales. At order $1/m_b$, we find resolved photon contributions arising from the operator pairings $Q_{8g}-Q_{8g}$, $Q_{7\gamma}-Q_{8g}$, and $Q_1^c-Q_{7\gamma}$ in the squared decay amplitude. We also prove that the resolved photon contributions arising from the $Q_1^c-Q_1^c$ and $Q_1^c-Q_{8g}$ operator pairings are suppressed by two powers of $1/m_b$. Detailed analyses of the resolved photon contributions were presented in Section~\ref{sec:matching}, which constitutes the main technical part of the paper.

The non-perturbative soft functions, which are needed to describe the photon spectrum at order $1/m_b$ in the heavy-quark expansion, introduce new sources of hadronic uncertainties in the description of the photon spectrum in the endpoint region. These functions can neither be extracted from experiment, nor can they be computed using lattice gauge theory (since they involve operators containing fields separated by light-like distances), and unfortunately they are not much restricted by constraints on their normalization and moments. Hence, there is a vast freedom in constructing phenomenological models for the soft functions, which often depend on several convolution variables. The resulting uncertainties will impact any extraction of $|V_{ub}|$ via a combination of inclusive semileptonic and radiative decays. They will also affect the extraction of heavy-quark parameters such as $m_b$, $\bar\Lambda$, $\mu_\pi^2$ etc.\ from moments of the \bsg\ photon spectrum. A dedicated analysis of the resulting uncertainties will be presented elsewhere.

Our most important phenomenological result concerns the non-local power corrections to the \bsg\ decay rate defined with a cut $E_\gamma\ge E_0$, where $E_0$ is chosen to be far outside the endpoint region, $m_b-2E_0\gg\Lambda_{\rm QCD}$. In this region the direct photon contributions reduce to local matrix elements, and deviations from the naive model of a free heavy-quark decay start at order $1/m_c^2$ and $1/m_b^2$ and are calculable in terms of well-known heavy-quark parameters. The resolved photon contributions, on the other hand, are still expressed in terms of non-local operators, whose matrix elements are of order $1/m_b$. Their contributions to the integrated rate can be parameterized in terms of three non-perturbative parameters, $\Lambda_{17}$, $\Lambda_{78}^{\rm spec}$, and $\Lambda_{88}$, as shown in (\ref{finalform}). In (\ref{Lambdaijdef}), these parameters are expressed in terms of convolutions of calculable jet functions with non-perturbative soft functions. Needless to say, it is very difficult to estimate the values of these hadronic parameters. Nevertheless, we have provided arguments suggesting that all three parameters are much smaller than the naive expectation $\sim\Lambda_{\rm QCD}$. For the most important case of $\Lambda_{17}$, a detailed modeling of the corresponding soft function, taking into account the normalization conditions and moment relations we have derived in (\ref{g17norm}) and (\ref{g17mom1}), suggests that $\Lambda_{17}$ is significantly smaller in magnitude than $\Lambda_{\rm QCD}$, see (\ref{Lam17range}). For the second-most important case of $\Lambda_{78}^{\rm spec}$, we have provided two different arguments, based on the vacuum insertion approximation and on $SU(3)$ flavor symmetry, suggesting that to a good approximation this hadronic parameter is proportional to the electric charge of the light spectator inside the $B$ meson, see~(\ref{eq:110n}). While the parameter $\Lambda_{I=1}^{(8)}$ entering in this relation can indeed be of order $\Lambda_{\rm QCD}$ -- see (\ref{eq:108}) and (\ref{eq:110}) -- the weighting by the spectator charge reduces the corresponding contribution to the isospin-averaged decay rate by a factor $(e_u+e_d)/2=1/6$. Finally, as shown in (\ref{L88mod}), the parameter $\Lambda_{88}$ is suppressed by a charge factor $e_s^2=1/9$, and its value is therefore bound to be much smaller than $\Lambda_{\rm QCD}$. Our final estimates for the hadronic uncertainty from non-local $1/m_b$ corrections in the theoretical prediction for the \bsg\ decay rate defined with the cut $E_\gamma>1.6$\,GeV has been given in (\ref{Fcurrent}) and (\ref{Fcurrentexp}). It depends on whether the contribution from $\Lambda_{78}^{\rm spec}$ is estimated using the vacuum insertion approximation or the current experimental value of the isospin asymmetry $\Delta_{0-}$. We have emphasized that even a precision measurement of the isospin asymmetry would not help to reduce the uncertainty by a significant amount. Relation (\ref{final}) shows that in this ideal case an irreducible uncertainty of about 4--5\% remains. At present, we do not see any hope to reduce this error using well-controlled theoretical methods. 

The analysis presented in this paper applies without alteration, apart from some obvious substitutions, to the decay $\bar B\to X_d\gamma$. First, one needs to change the definition of $\lambda_q$ to $V_{qb} V_{qd}^*$ and replace $m_s$ in (\ref{Fms}) and (\ref{jsubm}) by $m_d\approx 0$. Second, one has to replace the $s$-quark fields by $d$-quark fields in the definitions of the soft functions $f_u$ and $f_v$ contributing to $F_{77}^{\rm SSF}(E_\gamma,\mu)$ in (\ref{F77SSF}), and in the definitions of $\bar g_{78}$ and $\bar g_{78}^{\rm cut}$ contributing to $F_{78}^{(b)}(E_\gamma,\mu)$ in (\ref{F78c}). Notice, however, that none of these functions contribute to the integrated decay rate.

\vspace{0.5cm}\noindent 
{\em Acknowledgments:} 
We would like to thank Thomas Becher, Uli Haisch, Richard Hill, Tobias Hurth, Paul Langacker, and Satoshi Mishima for useful discussions. M.N.\ gratefully acknowledges the hospitality and support from the Institute of Advanced Study in Princeton and the University of Chicago during various stages of this work. S.J.L.\ and G.P.\ gratefully acknowledge the hospitality and support from the Johannes Gutenberg University of Mainz during various stages of this work. The research of M.N.\ is supported in part by a Jensen Professorship from the Klaus Tschira Foundation. The research of M.B.\ and M.N.\ is supported in part by the German Federal Ministry for Education and Research grant 05H09UME ({\em Precision Calculations for Collider and Flavour Physics at the LHC}), and by the Research Centre {\em Elementary Forces and Mathematical Foundations\/} funded by the Excellence Initiative of the State of Rhineland-Palatinate. The work of G.P.\ is supported by the Department of Energy grant DE-FG02-90ER40560.

\begin{appendix}

\section{Effective weak Hamiltonian}
\label{app:Heff}
\renewcommand{\theequation}{A.\arabic{equation}}
\setcounter{equation}{0}

We use the form of the effective weak Hamiltonian for \bsg\ decay as
presented in \cite{Beneke:2001ev}, i.e.\
\begin{equation}\label{eq:Weak Hamiltonian}
   {\cal H}_{\rm eff} = \frac{G_F}{\sqrt{2}} \sum_{q=u,c} \lambda_q\,
   \bigg( C_1\,Q_1^q + C_2\,Q_2^q + \sum_{i=3,...,6} C_i\,Q_i 
   + C_{7\gamma}\,Q_{7\gamma} + C_{8g}\,Q_{8g} \bigg) \,,
\end{equation}
where $\lambda_q=V_{qb} V_{qs}^*$, and the Wilson coefficients depend on the scale $\mu$ at which the operators are renormalized. The explicit form of the operator basis is
\begin{equation}
\begin{aligned}
   Q_1^q &= (\bar q b)_{V-A}\,(\bar s q)_{V-A} \,, &\qquad
    Q_2^q &= (\bar q_i b_j)_{V-A}\,(\bar s_j q_i)_{V-A} \,, \\
   Q_3 &= (\bar s b)_{V-A} \sum{}_q\,(\bar q q)_{V-A} \,, &\qquad
    Q_4 &= (\bar s_i b_j)_{V-A} \sum{}_q\,(\bar q_j q_i)_{V-A} \,, \\
   Q_5 &= (\bar s b)_{V-A} \sum{}_q\,(\bar q q)_{V+A} \,, &\qquad
    Q_6 &= (\bar s_i b_j)_{V-A} \sum{}_q\,(\bar q_j q_i)_{V+A} \,, \\
   Q_{7\gamma} &= \frac{-e m_b}{8\pi^2}\,
    \bar s\sigma_{\mu\nu}(1+\gamma_5)F^{\mu\nu}b \,, &\quad
    Q_{8g} &= \frac{-g m_b}{8\pi^2}\,
     \bar s\sigma_{\mu\nu}(1+\gamma_5) G^{\mu\nu} b \,,
\end{aligned}
\end{equation}
where $i$ and $j$ are color indices, and for the penguin operators a summation over quark flavors $q=u,d,s,c,b$ is implied. We use the short-hand notation $(\bar q b)_{V\mp A}\equiv\bar q\gamma^\mu(1\mp\gamma_5) b$ etc. Our sign convention is such that $iD_\mu=i\partial_\mu+g\,T^a A_\mu^a+e\,e_q A_\mu$, where $T^a$ are the $SU(3)$ color generators, and $e_q$ are the quark electric charges in units of $e$.

\section{Input parameters}
\label{app:inputs}
\renewcommand{\theequation}{B.\arabic{equation}}
\setcounter{equation}{0}

Here we collect the input parameter values used in the numerical analysis in Section~\ref{sec:conc}. As our default choice for the factorization scale $\mu$ entering the Wilson coefficients, the strong coupling constant, and the various hadronic quantities, we take the hard-collinear scale $\mu=1.5$\,GeV, which is indeed a scale of order $\sqrt{m_b\Lambda_{\rm QCD}}$. This is an appropriate scale choice, given that we neglect RG evolution effects. 

The $b$-quark mass enters our expressions either via the photon energy ($E_\gamma\approx m_b/2$ near the peak of the spectrum) or as the heavy-quark expansion parameter. It is therefore appropriate to adopt a low-scale subtracted heavy-quark mass, such as the mass defined in the shape-function scheme \cite{Bosch:2004th}. Specifically, we use $m_b=4.65$\,GeV. The charm-quark mass enters as a running mass in charm-penguin diagrams with a soft gluon emission, which are characterized by a hard-collinear virtuality. We therefore use $m_c=\overline{m}_c(\mu)$ defined in the $\overline{\rm MS}$ scheme, with $\mu=1.5$\,GeV fixed as described above. This corresponds to the choice adopted in \cite{Misiak:2006zs}, and following these authors we set $\overline{m}_c(\mu)=1.131$\,GeV. Finally, for the strange-quark mass we take $m_s=m_b/50$, which is the value commonly adopted in the literature on \bsg\ decay.

We also need input values for some HQET matrix elements. The parameters $\lambda_2$ and $\rho_{LS}^3$ are extracted from a global fit to $\bar B\to X_c l\,\bar\nu$ experimental data by the Heavy Flavor Averaging Group (HFAG) \cite{Barberio:2008fa}. Unfortunately, in many cases these and other parameters are extracted from a combined fit to $\bar B\to X_c l\,\bar\nu$ and \bsg, an approach that was criticized in \cite {Neubert:2008cp}. Only recently HFAG has started quoting also values obtained using only semileptonic data. The most recent results are $\lambda_2=(0.12\pm 0.02)$\,GeV$^2$ and $\rho_{LS}^3=(-0.17\pm 0.09)\,\mbox{GeV}^3$ \cite{HFAG:fpcp09}. For simplicity, we always use the central values for these quantities. For the first inverse moment of the $B$-meson light-cone distribution amplitude, we take the range $250\,\mbox{MeV}<\lambda_B<750\,\mbox{MeV}$, which covers predictions obtained using QCD sum rules and other methods \cite{Grozin:1996pq,Ball:2003fq,Braun:2003wx,Lee:2005gza,Le Yaouanc:2007mz,Kawamura:2010tj}. Finally, to the level of accuracy of our calculations, the parameter $F$ can be extracted from the relation $F=f_B\sqrt{M_B}$, and using $f_B=(193\pm 10)$\,MeV \cite{Laiho:2009eu} we obtain $0.177\,{\rm GeV}^3<F^2<0.217\,{\rm GeV}^3$.

\section{\boldmath NNLO matching of $\cal{H}_{\rm eff}$ to SCET\unboldmath}
\label{app:SCETops}
\renewcommand{\theequation}{C.\arabic{equation}}
\setcounter{equation}{0}

In this appendix we present the matching of the effective weak Hamiltonian operators $Q_{7\gamma}$, $Q_{8g},$ and $Q_1^{c,u}$ onto SCET up to NNLO in the expansion parameter $\sqrt{\lambda}$, with $\lambda\sim\Lambda_{\rm QCD}/m_b$. Although there is a large number of possible operators, only some of them are needed in practice. One subset, which was presented already in Section~\ref{subsec:SCETmatching}, is needed for the study of the resolved photon contributions at tree level. Another subset is needed for the analysis of the power corrections to the direct photon contributions at ${\cal O}(\alpha_s)$. In the first part of this Appendix we perform the matching at tree level. In the second part we include also the contribution of one-loop quantum fluctuations.

\subsection{Tree-level matching}

We begin with the current-type operators $Q_{7\gamma}$ and $Q_{8g}$. At LO, with a scaling of $\lambda^{5/2}$, the operator $Q_{7\gamma}$ is the only one which gives a contribution to
\bsg. The contribution of the operator $Q_{8g}$ begins at NLO, i.e.\ ${\cal O}(\lambda^{3})$. We then perform the tree-level matching of $Q_1^{c,u}$, whose contribution begins at the NNLO, ${\cal O}(\lambda^{7/2})$. For simplicity, we denote LO, NLO, and NNLO operators by
superscripts (0), (1), and (2), respectively.

\subsubsection{Matching of $\bm{Q_{7\gamma}}$}

The operator $Q_{7\gamma}$ in the weak Hamiltonian is given by
\begin{equation}
   Q_{7\gamma}=-\frac{em_b}{8\pi^2}\bar s
   \sigma_{\mu\nu}(1+\gamma_5)F^{\mu\nu}b= -\frac{em_b}{4\pi^2}\bar
   s\,[i\delslash  \Aslash^{\rm em}_{\perp}]\,(1+\gamma_5)b \,,
\end{equation}
where it is assumed that the photon is real, i.e., it is transversely polarized. The tree-level matching of $Q_{7\gamma}$ can be read off from \cite{Beneke:2002ph}. As was done there, we separate $Q_{7\gamma}$ into ``A'' and ``B'' terms, according to whether they
contain hard-collinear gluon fields or not. Suppressing the $-\frac{e m_b}{4\pi^2}\,e^{-im_b\,v\cdot x}$ factor, $Q_{7\gamma}$ is matched onto the operators
\begin{equation}
\begin{aligned}
Q_{7\gamma A}^{(0)} &= \hcbar\,\frac{\nbslash}{2}\, [i n\cdot \partial \Aslash^{\rm em}_{\perp}]\,(1+\gamma_5)  h  \,, \\
Q_{7\gamma A}^{(1)} &=
\hcbar\,\frac{\nbslash}{2}\, [i n\cdot \partial \Aslash^{\rm em}_{\perp}]\,(1+\gamma_5)\,x_\perp^\mu D_\mu h
+ \hcbar\,[i\delslash_{\perp}  \Aslash^{\rm em}_{\perp}]\,(1+\gamma_5) h  \,, \\
Q_{7\gamma A}^{(2)} &= \hcbar\,\frac{\nbslash}{2}\, [i n\cdot \partial \Aslash^{\rm em}_{\perp}]\,(1+\gamma_5)
\left[\frac{n\cdot x}{2}  \bar n\cdot D h + \frac{x_\perp^\mu x_\perp^\nu}{2} D_\mu D_\nu h + \frac{i\Dslash}{2m_b}\,h \right] \\
&\quad\mbox{}+ \hcbar\,[i\delslash_{\perp} \Aslash^{\rm em}_{\perp}]\,(1+\gamma_5)\,x_\perp^\mu D_\mu h
+\hcbar\,\frac{\nbslash}{2}\,\frac{i\!\overleftarrow{\delslash}\!_\perp}{i\bar n\cdot\overleftarrow{\partial}}\, [i\delslash_{\perp} \Aslash^{\rm em}_{\perp}]\,(1+\gamma_5)\,h \,,
\end{aligned}
\end{equation}
and
\begin{equation}
\begin{aligned}
   Q_{7\gamma B}^{(1)} &=
    \hcbar\,\frac{[i n\cdot \partial \Aslash^{\rm em}_{\perp}]}{m_b}\, g\,\calAslash_{hc \perp} (1+\gamma_5) h  \,, \\
   Q_{7\gamma B}^{(2)} &=
- \hcbar\,\frac{\nbslash}{2}\,[i n\cdot \partial \Aslash^{\rm em}_{\perp}]\,(1+\gamma_5)\,\frac{1}{i\bar n\cdot \partial}\,g\, n\cdot {\EuScript A}_{hc} h +
         \hcbar\,\frac{[i n\cdot \partial \Aslash^{\rm em}_{\perp}]}{m_b}\,g\,n\cdot {\EuScript A}_{hc} (1-\gamma_5) h  \\
&\quad\mbox{}+ \hcbar  \frac{[i n\cdot \partial \Aslash^{\rm em}_{\perp}]}{m_b}\,g\,\calAslash_{hc \perp}(1+\gamma_5) x_\perp^\mu D_\mu h \\
&\quad\mbox{}- \hcbar\,\frac{\nbslash}{2}\,[i n\cdot \partial \Aslash^{\rm em}_{\perp}]\,(1+\gamma_5)\,\frac{1}{i\bar n\cdot\partial}\,
           \frac{(i\delslash_{\perp}\,g\,\calAslash_{hc \perp})}{m_b} h \\
&\quad\mbox{}- \hcbar\,\frac{\nbslash}{2}\,g\,\calAslash_{hc \perp}\,\frac{1}{i\bar n\cdot\overleftarrow{\partial}}\,[i\delslash_{\perp} \Aslash^{\rm em}_{\perp}]\,(1+\gamma_5) h \,,
\end{aligned}
\end{equation}
where the covariant derivative $D^\mu$ only contains soft fields. Note that the hard-collinear fields are not sterile (see the discussion in Section~\ref{sec:fact}), so they still couple to soft gluons.

\subsubsection{Matching of $\bm{Q_{8g}}$ \label{sec:Q8}}

The operator $Q_{8g}$ in the weak Hamiltonian is given by
\begin{equation}
   Q_{8g}=-\frac{gm_b}{8\pi^2}\bar s
   \sigma_{\mu\nu}(1+\gamma_5)G^{\mu\nu}b \,.
\end{equation}
We can discard the non-abelian part of $G^{\mu\nu}$, since we only work to first order in $g$. A priori, we can match the $s$ quark onto either a hard-collinear or anti-hard-collinear quark and the gluon onto a hard-collinear or anti-hard-collinear gluon. We do not consider matching the $s$ quark onto a soft particle, since it will only give rise to a contribution beyond NNLO. Also, the gluon and the $s$ quark cannot both be anti-hard-collinear, since the
necessary conversions would lead to a suppression beyond NNLO. The three remaining cases will be considered in turn.

\begin{figure}[ht]
\newcommand{\jwt}{0.042\textwidth}
\begin{center}
\begin{tabular}{ccccc}
SCET&&${\cal{L}}_{\rm{SCET}},{\cal O}(\lambda)$&&{}\\
\epsfig{file=operator1.ps,width=3cm} &
\raisebox{\jwt}{$+$} &
\epsfig{file=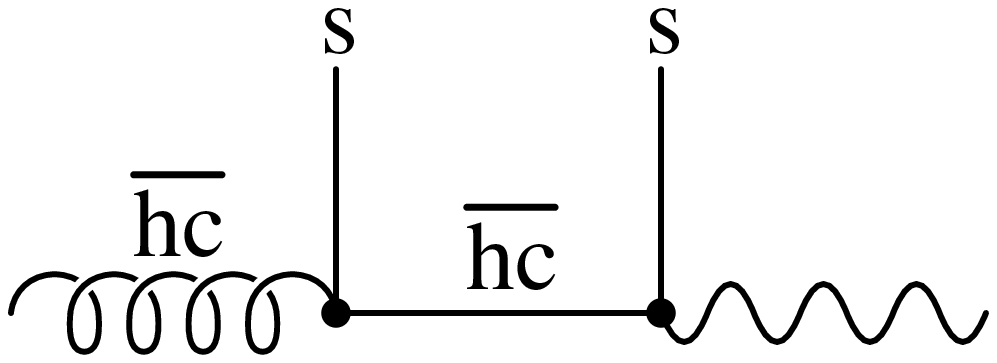,width=4cm} &
\raisebox{\jwt}{$\longrightarrow$} &
\epsfig{file= 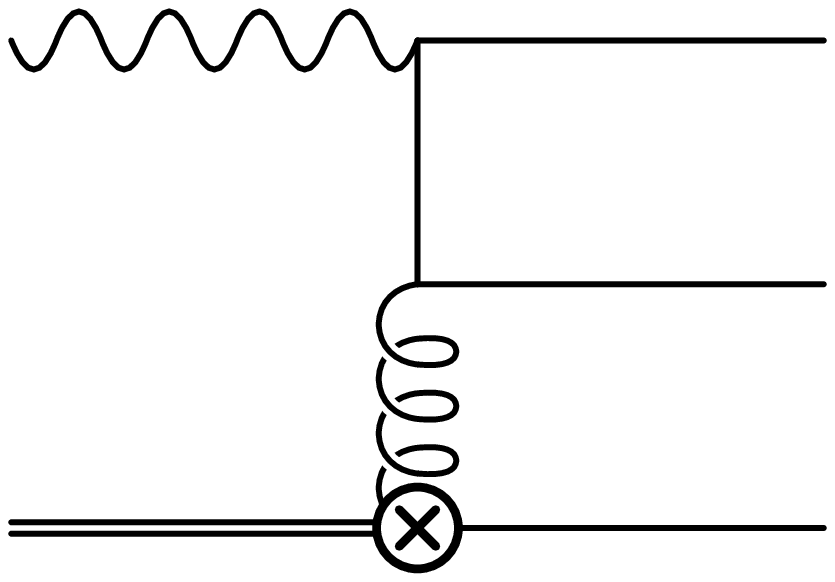,width=3cm}
\end{tabular}
\caption{\label{fig:Q8gMatching0}
Graphical illustration of the tree-level matching procedure for a $Q_{8g}$ contribution, showing how a resolved photon contribution arises after a SCET Lagrangian insertion.}
\end{center}
\end{figure}

For the case of an anti-hard-collinear gluon and hard-collinear $s$ quark in $Q_{8g}$, the conversion of the anti-hard-collinear gluon onto a photon is ${\cal O}(\lambda)$ in power counting. This process is illustrated in Figure~\ref{fig:Q8gMatching0}. In this case $Q_{8g}$ is matched onto a single operator (again suppressing the overall $-\frac{g\,m_b}{4\pi^2}e^{-im_b\,v\cdot x}$ factor):
\begin{equation}
   Q_{8g,\,{\overline{hc}\,\,{\rm gluon}}}^{(2)}=
     \hcbar\,\frac{\nbslash}{2}\,\left[in\cdot\partial\calAslash_{\overline{hc} \perp}\right](1+\gamma_5)h \,, \quad \mbox{followed by}\;{\cal O}(\lambda)\; {\rm conversion.}
\end{equation}
This is the first example of an operator that gives rise to a resolved photon contribution. The conversion is displaced along the light cone by the anti-hard-collinear propagator.

For the case of a hard-collinear gluon in $Q_{8g}$, we have to distinguish between the different scaling of the components of $G^{\mu\nu}$. We will need the following components of $\sigma_{\mu\nu}\,G^{\mu\nu}$:
\begin{equation}
\begin{aligned}
   {\cal O}(\lambda^{1/2}): &\quad
    2\,\frac{\nslash}{2}\,[i\bar{n}\cdot\partial\Aslash_{\perp}] \\
   {\cal O}(\lambda): &\quad
   2\,[i\delslash_{\perp}\Aslash_{\perp}-i\partial_{\perp}\cdot
   A_{\perp}] \,+\,
   \left(\frac{\nslash}{2}\frac{\nbslash}{2}-\frac{\nbslash}{2}\frac{\nslash}{2}\right)
   [i\bar{n}\cdot\partial \,n\cdot A]
\end{aligned}
\end{equation}

\begin{figure}[ht]
\newcommand{\jw}{0.059\textwidth}
\begin{center}
\begin{tabular}{ccccc}
{}&&{}&&SCET\\
\epsfig{file=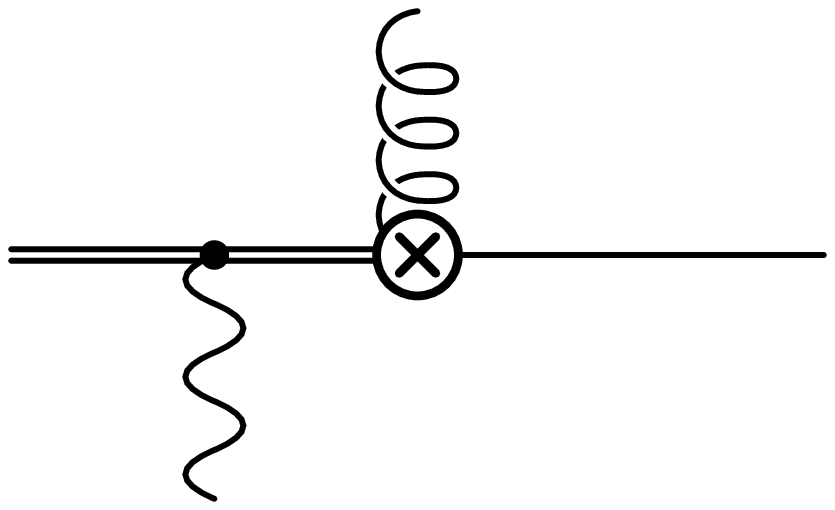,width=3cm} &
\raisebox{\jw}{$+$} &
\epsfig{file=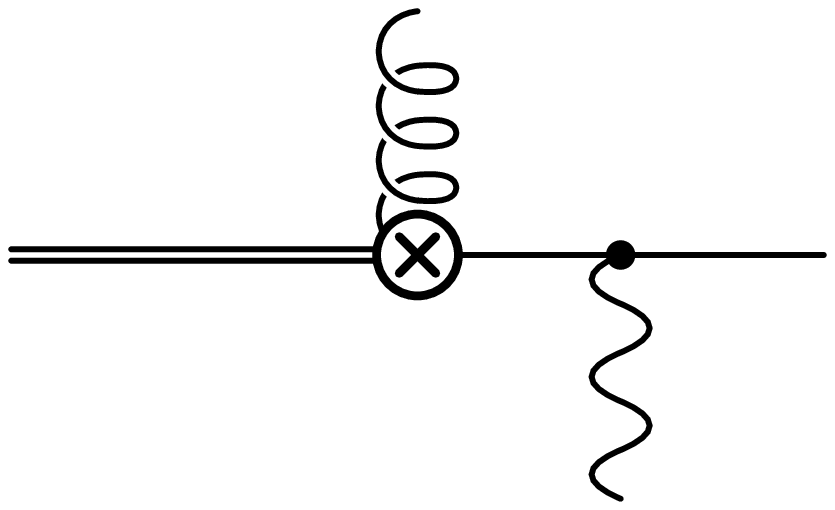,width=3cm} &
\raisebox{\jw}{$\longrightarrow$} &
\epsfig{file= 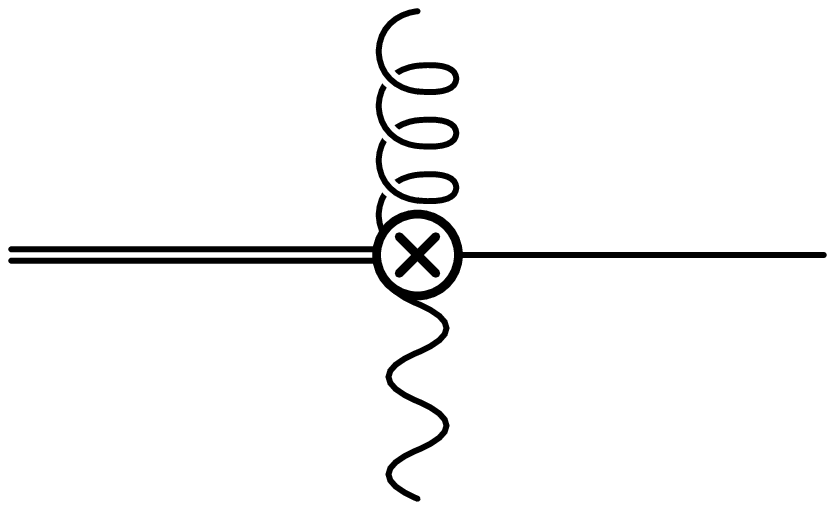,width=3cm}
\end{tabular}
\caption{\label{fig:Q8gMatching} Graphical illustration of the tree-level matching procedure for the operator $Q_{8g}$ for the case when the $s$ quark is matched onto a hard-collinear field.}
\end{center}
\end{figure}

If the $s$ quark is matched onto a hard-collinear field, we need to have the anti-hard-collinear photon emitted from the $b$ or $s$ quark lines, as shown in Figure~\ref{fig:Q8gMatching}. This would match onto SCET operators that contain both a hard-collinear gluon and an anti-hard-collinear photon suppressed by $m_b$ for the $b$-quark line, and by $\bar n\cdot\partial$ for the $s$-quark line. We also need to consider the three possibilities for $G^{\mu\nu}$ and the multipole expansion of the heavy-quark field. In total we have
\begin{equation}\label{eq:Q8Q8hcq}
\begin{aligned}
Q_{8g,\,{\rm hc\,quark}}^{(1)}&=
\hcbar\frac{1}{i\bar n\cdot\overleftarrow{\partial}}e_d e\Aslash^{\rm em}_{\perp}[i\bar{n}\cdot\partial\calAslash_{hc \perp}]\,(1+\gamma_5) h \,, \\
Q_{8g,\,{\rm hc\, quark}} ^{(2)}&=
\hcbar\frac{1}{i\bar n\cdot\overleftarrow{\partial}}e_d e\Aslash^{\rm em}_{\perp}[i\bar{n}\cdot\partial\calAslash_{hc \perp}]\,(1+\gamma_5)x_\perp^\mu D_\mu h \\
&\quad\mbox{}-\hcbar\frac{i\overleftarrow{\delslash}_{\perp}}{i\bar n\cdot\overleftarrow{\partial}}\frac{\nbslash}{2m_b}\,[i\bar{n}\cdot\partial\calAslash_{hc \perp}]
e_d e\Aslash^{\rm em}_{\perp} (1+\gamma_5)h \\
&\quad\mbox{}+\hcbar\frac{1}{i\bar n\cdot\overleftarrow{\partial}}e_d e\Aslash^{\rm em}_{\perp}\frac{\nbslash}{2}\,
[i\delslash_{\perp}\calAslash_{hc \perp}-i\partial_{\perp}\cdot  {\EuScript A}_{hc \perp}]\,(1+\gamma_5) h \\
&\quad\mbox{}-\hcbar\,[i\delslash_{\perp}\calAslash_{hc \perp}-i\partial_{\perp}\cdot  {\EuScript A}_{hc \perp}]\,
\frac{\nbslash}{2m_b}e_d e\Aslash^{\rm em}_{\perp}(1+\gamma_5) h \\
&\quad\mbox{}+\hcbar\frac{1}{i\bar n\cdot\overleftarrow{\partial}}e_d e\Aslash^{\rm em}_{\perp}\frac{\nbslash}{2}
\left[\frac{i}{2}\bar n\cdot \partial \,n\cdot {\EuScript A}_{hc}\right] (1+\gamma_5) h \\
&\quad\mbox{}+\hcbar\frac{\nbslash}{2m_b}\left[\frac{i}{2}\bar n\cdot \partial \,n\cdot {\EuScript A}_{hc}\right]e_d e\Aslash^{\rm em}_{\perp} (1+\gamma_5) h \,.
\end{aligned}
\end{equation}

If the $s$ quark is matched onto an anti-hard-collinear field, it can only be converted to an anti-hard-collinear photon and a soft $s$ quark. The conversion costs us $\lambda^{1/2}$, so the lowest-order operator possible is ${\cal O}(\lambda^3)$. Considering all the possible structures for $G^{\mu\nu}$ and the multipole expansion, we find
\begin{eqnarray}\label{eq:Q8Q8ahcq}
Q_{8g,\,{\rm \overline{hc}\, quark }}^{(1)}&=&
\ahcbar \frac{\nslash}{2}\,[i\bar{n}\cdot\partial\calAslash_{hc \perp}]\,(1+\gamma_5) h \,,
\quad{\rm followed\;by}\;{\cal O}(\lambda^{1/2})\; {\rm conversion,} \nonumber\\
Q_{8g,\,{\rm \overline{hc}\, quark}} ^{(2)}&=&
\ahcbar \frac{\nslash}{2}\,[i\bar{n}\cdot\partial\calAslash_{hc \perp}]\,(1+\gamma_5)x_\perp^\mu   D_\mu h \,,
\quad{\rm followed\;by}\;{\cal O}(\lambda^{1/2})\; {\rm conversion,} \\
&&\mbox{}+\ahcbar\, [i\delslash_{\perp}\calAslash_{hc \perp}-i\partial_{\perp}\cdot {\EuScript A}_{hc \perp}]\,(1+\gamma_5) h \,,
\quad{\rm followed\;by}\;{\cal O}(\lambda^{1/2})\; {\rm conversion,} \nonumber\\
&&\mbox{}+\ahcbar\, \frac{\nslash}{2}\left[\frac{i}{2}\bar{n}\cdot\partial \,n\cdot {\EuScript A}_{hc}\right](1+\gamma_5) h \,,
\quad{\rm followed\;by}\;{\cal O}(\lambda^{1/2})\; {\rm conversion.} \nonumber
\end{eqnarray}

\subsubsection{Matching of $Q_1^{q}$}

To simplify the notation we write the operator as $Q_1^q=\bar s\,\Gamma_1 q\,\bar q\,\Gamma_2 b$, where $\Gamma_1\otimes\Gamma_2=\gamma^\mu(1-\gamma_5)\otimes\gamma_\mu(1-\gamma_5)$. At tree level, the light quarks can only be matched onto hard-collinear or anti-hard-collinear fields. A matching of any of the light quarks onto even one soft field would lead to a suppression of ${\cal O}(\lambda^{4})$. As a result, $Q_1$ is matched at NLO, before taking into account any conversions. When there is more than one anti-hard-collinear quark field, no conversion is allowed at tree-level. Hence, we are left with the following cases (suppressing the $e^{-im_b\,v\cdot x}$ factor).

\paragraph{\boldmath $\bar{s}=\hcbar,\;q=\hc,\;\bar{q}=\hcbar$:}
We need to consider an attachment of one anti-hard-collinear photon emitted either from the heavy or one of the hard-collinear quark lines. This leads to four possible ${\cal O}(\lambda^{7/2})$ operators:
\begin{equation}
\begin{aligned}
&-\hcbar\Gamma_1\frac{\nbslash}{2m_b}e_d e\Aslash^{\rm em}_{\perp}
h\,\hcbar\Gamma_{2}\hc
+\hcbar\,e_d e\Aslash^{\rm em}_{\perp}\frac{\nbslash}{2}\frac{1}{i\bar n\cdot\overleftarrow{\partial}}\Gamma_1 h\, \hcbar\Gamma_{2}\hc \\
&\mbox{}+\hcbar\Gamma_1 h\,\hcbar\,e_d e\Aslash^{\rm
em}_{\perp}\frac{\nbslash}{2}\frac{1}{i\bar
n\cdot\overleftarrow{\partial}}\Gamma_{2\mu}\hc -\hcbar\Gamma_1
h\,\hcbar\Gamma_{2}\frac{\nbslash}{2}\,e_d e\Aslash^{\rm
em}_{\perp}\frac{1}{i\bar n\cdot\overrightarrow{\partial}}\hc\,.
\end{aligned}
\end{equation}

\paragraph{\boldmath $\bar{s}=\hcbar,\;q=\ahc,\;\bar{q}=\hcbar$ {\rm and} $\bar{s}=\hcbar,\;q=\hc,\;\bar{q}=\ahcbar$:} 
In this case we need to convert the anti-hard-collinear quark to a photon via $\ahc\to A^{\rm em}_\perp+q$ and $\ahcbar\to A^{\rm em}_\perp+\bar{q}$, which is ${\cal O}(\lambda^{1/2})$. Therefore, we can only have
\begin{equation}
\begin{aligned}
&\hcbar\Gamma_1 h\,\ahcbar\Gamma_{2}\hc \,,
\quad{\rm followed\;by}\;{\cal O}(\lambda^{1/2})\; {\rm conversion,} \\
&\hcbar\Gamma_1 h\,\hcbar\Gamma_{2}\ahc \,, \quad{\rm
followed\;by}\;{\cal O}(\lambda^{1/2})\; {\rm conversion.}
\end{aligned}
\end{equation}

\paragraph{\boldmath $\bar{s}=\ahcbar,\;q=\hc,\;\bar{q}=\hcbar$:}
Supplemented with SCET Lagrangian for $\ahcbar\to A^{\rm em}_\perp+\bar{q}$, only one NNLO operator is possible:
\begin{equation}
\ahcbar\Gamma_1 h\,\hcbar\Gamma_{2}\hc \,,\quad{\rm
followed\;by}\;{\cal O}(\lambda^{1/2})\; {\rm conversion.}
\end{equation}

\subsection{Loop matching}

We now perform the matching including the contributions of loops. These are only relevant for $Q_1$, where the two up-type quarks are contracted and a number of gauge bosons are emitted from the internal lines. The contribution of three gauge bosons would lead to
further power or loop suppression, so we only need to consider one or two bosons.

The one gauge boson loops are easy to analyze. In the NDR scheme only $Q_5$ and $Q_6$ give a non zero contribution. Furthermore, this contribution only modifies the coefficients $Q_{7\gamma}$ and $Q_{8g}$ to $C_i^{\rm eff}$ \cite{Buras:1993xp}, with
\begin{equation}
C_{7\gamma}^{\rm eff}=C_{7\gamma}+\sum_{i=1}^6
y_iC_{i} \,,\qquad C_{8g}^{\rm eff}=C_{8g}+\sum_{i=1}^6
z_iC_{i} \,,
\end{equation}
where $y_i,z_i$ depend on the scheme. They vanish in the 't~Hooft-Veltman scheme, while in the NDR scheme the non zero ones are $y_5=-1/3$, $y_6=-1$, $z_5=1$. The effective coefficients are regularization-scheme invariant. The contribution of the one-boson
loop would therefore be to change $C_{7\gamma,8g}$ to $C_{7\gamma,8g}^{\rm eff}$.

A more involved contribution arises for the loops with two external bosons, which is the main focus in this subsection. These can only be one photon and one gluon. Two external gluons would lead to further power or loop suppression.

As usual, the $b$ quark is matched onto the heavy quark field $h$. Since there is already a photon in the operator, the $s$ quark cannot be anti-hard-collinear. On the other hand, a soft $s$ quark would lead to power suppression, so it must be hard-collinear. There
are two possibilities for the gluon emitted from the loop: it can either be hard-collinear or it can be soft. If the gluon is hard-collinear, the loop momentum is hard. If the gluon is soft, the loop momentum is anti-hard-collinear. For the first case, a photon and a hard-collinear gluon, we need to calculate the loop diagram in QCD in order to perform the matching. For the second case, a photon and and a soft gluon, one would need the tree-level
matching of $Q_1$ onto the operator of (\ref{Q1q}). The conversion of the two anti-hard collinear quarks to a photon and a soft gluon would be calculated in SCET. Alternatively, one can calculate the process in QCD and use the fact that the two calculations are equivalent, since only one momentum region, anti-hard-collinear, contributes in this case.

We have explicitly calculated the appropriate QCD one-loop diagram, arising from the four-quark operator $\left(\bar{s}_i\Gamma_2q_j\right)\left(\bar q_k \Gamma_1 b_l\right)$. Alternatively, the result can be read off from the two gluon calculation in \cite{Beneke:2002jn}, where one of the gluons is replaced by a photon. Using the notation of \cite{Beneke:2002jn} the amplitude is given by
\begin{equation}
\label{eq:notrace} {\cal A}=\frac {e\,e_q} {4\pi}\frac g {4\pi}
(T^a)_{mn} \bar s_m\Gamma_2 A_{\mu\nu} \Gamma_1 b_n
\epsilon_1^{*\mu}(q_1)\epsilon_2^{*a\,\nu}(q_2)
\delta_{ij}\delta_{kl} \,,
\end{equation}
where
\begin{equation}\label{eq:Amunu}
\begin{aligned}
A_{\mu\nu}&=\left[(4r-1)F(r)-4r\right]\left[\frac{m_q}{2r}\,g_{\mu\nu}-\frac
{q_\mu q_\nu} {m_q}
\right]+\left[F(r)-1\right]i\epsilon_{\rho\alpha\mu\nu}\left(q_1^\rho-q_2^\rho\right)\gamma^\alpha\gamma^5 \\
&\quad\mbox{}+ \frac{2r}{m_q^2}\left[F(r)-1\right]\left[q_\mu
i\epsilon_{\rho\sigma\alpha\nu}q_1^\rho q_2^\sigma-q_\nu
i\epsilon_{\rho\sigma\alpha\mu}q_1^\rho
q_2^\sigma\right]\gamma^\alpha\gamma^5
-\frac{F(r)}{m_q}i\epsilon_{\rho\sigma\mu\nu}q_1^\rho
q_2^\sigma\gamma^5 \,.
\end{aligned}
\end{equation}
Here $m_q$ and $e_q$ are the mass and charge of the quark in the loop, $q_1,\epsilon_1$ ($q_2,\epsilon_2$) are the momentum and polarization of the photon (gluon), $r=m_q^2/q^2-i\epsilon$, $q^2=(q_1+q_2)^2$, and $F(r)$ is the penguin function defined in (\ref{eq:F(r)}). Alternatively, one could write in a gauge-invariant notation
\begin{equation}
\begin{aligned}
A_{\mu\nu}\epsilon_1^{*\mu}(q_1)\epsilon_2^{*a\,\nu}(q_2) &= \left[(1-4r)F(r)+4r\right]\frac{1}{2m_q}F^{\mu\nu}G_{\mu\nu}^a
+\frac{F(r)}{2m_q}\,\tilde G_{\rho\mu}^aF^{\rho\mu}i\gamma_5 \\
&\quad\mbox{}+ [1-F(r)]\,\frac{2}{q^2}\left(G^a_{\mu\alpha}\tilde{F}^{\mu\beta}+
F_{\mu\alpha}\tilde{G}^{a\,\mu\beta}\right)iq^\alpha\gamma_\beta\gamma_5 \,.
\end{aligned}
\end{equation}
Here we are using the convention
\begin{equation}
\tilde{F}^{\mu\nu}=-\frac12\epsilon^{\mu\nu\alpha\beta}F_{\alpha\beta}\quad
(\epsilon^{0123}=-1)
\end{equation}
and the fact that for an external gluon $q_2\cdot\epsilon_2^{*a}=0$.

\subsubsection{Matching with $A^{\rm em}_{\perp{\overline{hc}}}$ and $A^{\rm gluon}_s$}

The diagram with an anti-hard-collinear photon and a soft gluon emitted from the internal line already contributes at ${\cal O}(\lambda^{7/2})$. Since $A^{\rm s}_{{\rm gluon}}$ scales homogeneously $\sim\left(\lambda,\lambda,\lambda\right)$, it is natural to use the gauge invariant form for the resulting operators, rather than decomposing the gauge fields into their light-cone components. Furthermore, only the axial part of $A_{\mu\nu}$ in
(\ref{eq:Amunu}) yields a non-zero result, as the $\Gamma_i$ in (\ref{eq:notrace}) are the usual $V-A$ Dirac structures, when matching $Q_1$. Therefore we only need
\begin{equation}
\begin{aligned}
A_{\mu\nu}\epsilon_1^{*\mu}(q_1)\epsilon_2^{*a\,\nu}(q_2)
&= \frac{2}{q^2}\left(G^a_{\mu\alpha}\tilde{F}^{\mu\beta}+F_{\mu\alpha}\tilde{G}^{a\,\mu\beta}\right)iq^\alpha\gamma_\beta\gamma_5[1-F(r)] \\
&\approx \frac{2}{q^2}\left(G^a_{\mu\alpha}\tilde{F}^{\mu\beta}\right)iq^\alpha\gamma_\beta\gamma_5[1-F(r)] \,,
\end{aligned}
\end{equation}
which follows from the fact, that $q^\alpha F_{\mu\alpha}$ vanishes at the lowest order in $\lambda$.

For $Q_1$ the loop can consist of any up-type quark that is not integrated out in the weak effective Lagrangian. When matching onto SCET the $u$ quark should be taken to be massless, so we can replace $F(m_u^2/q^2)$ by 0. For charmed quarks we have $m_c^2\sim m_b\Lambda_{\rm QCD}$, which is of the same order as $q^2$. As a result, $F(m_c^2/q^2)$ should not be expanded for $c$ quarks.

In position space, we find that at NNLO the $Q_1^q$ operators are matched onto 
\begin{eqnarray}
Q_1^{u\,(2)}(x) &=&
\left(\frac{ee_u}{4\pi^2}\right)\hcbar(x)\,T^a\gamma^\beta(1-\gamma_5)
h(x)e^{-im_bv\cdot x} \nonumber\\
&&\times
\int\frac{d^4q_1}{(2\pi)^4}\frac{d^4q_2}{(2\pi)^4}e^{i(q_1+q_2)x}\frac
1{(q_1+q_2)^2+i\epsilon}i
(q_1^\alpha+q_2^\alpha)gG^a_{\mu\alpha}(q_2)\,\epsilon^{\mu\beta\rho
\sigma} F_{\rho\sigma}(q_1) \,, \nonumber\\
Q_1^{c\,(2)}(x) &=& \left(\frac{ee_c}{4\pi^2}\right)\hcbar(x)\,T^a\gamma^\beta(1-\gamma_5)h(y)e^{-im_bv\cdot x}\int
\frac{d^4q_1}{(2\pi)^4}\frac{d^4q_2}{(2\pi)^4}e^{i(q_1+q_2)x} \\
&&\times
\frac{1}{(q_1+q_2)^2+i\epsilon}\left[1-F\left(\frac{m_c^2}{(q_1+q_2)^2}-i\epsilon\right)\right]i
(q_1^\alpha+q_2^\alpha)gG^a_{\mu\alpha}(q_2)\,\epsilon^{\mu\beta\rho\sigma}
F_{\rho\sigma}(q_1) \,, \nonumber
\end{eqnarray}
where we show explicitly the dependence of the momentum-space $F_{\rho\sigma}$ and $G^a_{\mu\alpha}$ on $q_1$ and $q_2$.

\subsubsection{Matching with $A^{\rm em}_{\perp{\overline{hc}}}$ and $A^{\rm gluon}_{\perp hc}$}

In this case $q^2$ is hard, i.e.\ $q^2\sim m_b^2$. Therefore $F(m_q^2/q^2)$ can be expanded around zero for charm quarks as well as for up quarks. The first order correction resulting from this expansion gives a power suppressed contribution, so we can just set $F(m_q^2/q^2)$ to zero. 
Depending on the polarization of the gluon field we can get either an NLO or an NNLO operator. We need to also include corrections from the multipole expansion of the heavy-quark field and subleading matching on the hard-collinear quark. In total we find that $Q^q_1$ is matched onto (suppressing an overall factor $\frac{ee_q g}{4\pi^2}\,e^{-im_b\,v\cdot x}$)
\begin{equation}\label{eq-Qi1}
\begin{aligned}
Q^{q\,(1)}_1 &= \hcbar \frac{\nbslash}{2}\,
\epsilon_{\perp\mu\nu}[{\EuScript A}^{\nu}_{hc \perp}\,n\cdot \partial A^{{\rm em}\;\mu}_\perp]\, (1-\gamma_5)h \,, \\
Q^{q\,(2)}_1 &= \hcbar\,\epsilon_{\perp\mu\nu}[A^{\rm em}_\perp\cdot\partial_\perp {\EuScript A}^{\nu}_{hc \perp}]\, \gamma^\mu_\perp(1-\gamma_5)h \\
&\quad\mbox{}+ \frac{1}{2}\,\hcbar\,\epsilon_{\perp\mu\nu}[A^{{\rm em}\;\nu}_\perp \bar n\cdot\partial\, n\cdot {\EuScript A}_{hc}]\, \gamma^\mu_\perp(1-\gamma_5)h \\
&\quad\mbox{}- \hcbar\frac{i\overleftarrow{\delslash}_{\perp}}{i\bar n\cdot\overleftarrow{\partial}} \,\epsilon_{\perp\mu\nu}
[A^{{\rm em}\;\mu}_\perp\bar{n}\cdot \partial {\EuScript A}^{\nu}_{hc \perp}]\, (1-\gamma_5)h \\
&\quad\mbox{}+ \hcbar \frac{\nbslash}{2}\,\epsilon_{\perp\mu\nu}
[{\EuScript A}^{\nu}_{hc \perp} n\cdot \partial A^{{\rm em}\;\mu}_\perp]\, (1-\gamma_5) x_\perp^\mu D_\mu h \,,
\end{aligned}
\end{equation}
where $\epsilon_{\perp\mu\nu}=\frac{1}{2}\epsilon_{\alpha\beta\mu\nu}\bar n^\alpha n^\beta$.

\end{appendix}

\end{document}